\documentclass[prd,floatfix,twocolumn,showpacs]{revtex4}
\usepackage{epsfig}
\usepackage{graphicx}
\usepackage{amsmath}
\usepackage{color}
\usepackage{longtable}

\newcommand{\sigmaT}{\sigma_{\rm T}}
\newcommand{\nuD}{\nu_{\rm D}}
\newcommand{\Tm}{T_{\rm m}}
\newcommand{\Tr}{T_{\rm r}}
\newcommand{\kB}{k_{\rm B}}
\newcommand{\nmax}{n_{\rm max}}
\newcommand{\nH}{n_{\rm H}}

\newcommand{\xHI}{x_{\rm HI}}
\newcommand{\xHII}{x_{\rm HII}}

\newcommand{\etaC}{\eta_{\rm c}}
\newcommand{\tauLL}{\tau_{\rm LL}}
\newcommand{\tauS}{\tau_{\rm S}}
\newcommand{\phaL}{\pha_{\rm L}}
\newcommand{\phaLO}{\pha_{\rm L}^{(0)}}
\newcommand{\phaC}{\pha_{\rm C}}
\newcommand{\Pesc}{P_{\rm esc}}
\newcommand{\PescIV}{P_{\rm esc}^{\rm(IV)}}
\newcommand{\Pprd}{P_{\rm esc}}
\newcommand{\PS}{P_{\rm S}}
\newcommand{\PC}{P_{\rm C}}
\newcommand{\PMC}{P_{\rm MC}}
\newcommand{\nuin}{\nu_{\rm in}}
\newcommand{\nuout}{\nu_{\rm out}}
\newcommand{\taucoh}{\tau_{\rm coh}}
\newcommand{\tauinc}{\tau_{\rm inc}}
\newcommand{\Hethree}{$^3{\mathrm {He}}$}
\newcommand{\Hefour}{$^4{\mathrm {He}}$}

\newcommand{\HI}{H{\sc ~i}}
\newcommand{\HII}{H{\sc ~ii}}
\newcommand{\HeI}{He{\sc ~i}}
\newcommand{\HeII}{He{\sc ~ii}}
\newcommand{\HeIII}{He{\sc ~iii}}

\newcommand{\LyaHe}{$2^1P^o-1^1S$}
\newcommand{\LybHe}{$3^1P^o-1^1S$}
\newcommand{\LyHe}{$n^1P^o-1^1S$}
\newcommand{\InteraHe}{$2^3P^o-1^1S$}
\newcommand{\InterHe}{$n^3P^o-1^1S$}

\newcommand{\QuadHe}{$n^1D-1^1S$}
\newcommand{\TPHe}{$2^1S-1^1S$}

\newcommand{\beq}{\begin{equation}}
\newcommand{\eeq}{\end{equation}}
\newcommand{\beqa}{\begin{eqnarray}}
\newcommand{\eeqa}{\end{eqnarray}}

\newcommand{\pha}{\mathcal{N}}
\newcommand{\barr}{\begin{array}}
\newcommand{\earr}{\end{array}}
\newcommand{\dphadnu}{\frac{\partial \pha}{\partial \nu}}

\newcommand{\comment}[1]{}

\begin{document}
\title{Primordial helium recombination I: feedback, line transfer, and continuum opacity}

\author{Eric R. Switzer}
\email{switzer@princeton.edu}
\affiliation{Department of Physics, Princeton University, Princeton, New Jersey 08544, USA}

\author{Christopher M. Hirata}
\affiliation{School of Natural Sciences, Institute for Advanced Study, Einstein Drive, Princeton, New Jersey 08540, USA}

\date{\today}

\begin{abstract}
Precision measurements of the cosmic microwave background temperature anisotropy on scales $\ell > 500$ will be available in the near future.  
Successful interpretation of these data is dependent on a detailed understanding of the damping tail and cosmological recombination of both hydrogen 
and helium.  This paper and two companion papers are devoted to a precise calculation of helium recombination.  We discuss several aspects of the 
standard recombination picture, and then include feedback, radiative transfer in \HeI\ lines with partial redistribution, and continuum opacity from 
\HI\ photoionization.  In agreement with past calculations, we find that \HeII\ recombination proceeds in Saha equilibrium, whereas \HeI\ recombination 
is delayed relative to Saha due to the low rates connecting excited states of \HeI\ to the ground state.  However, we find that at $z<2200$ the 
continuum absorption by the rapidly increasing \HI\ population becomes effective at destroying photons in the \HeI\ \LyaHe\ line, causing \HeI\ 
recombination to finish around $z\approx 1800$, much earlier than previously estimated.
\end{abstract}

\maketitle
 
\section{Introduction}
\label{sec:introduction}

Cosmological recombination occurs when the photon gas in the early universe has cooled sufficiently for bound atoms to form.  As the free electrons become locked 
in the ground states of these atoms, the opacity from Thomson scattering drops, and the signatures of thermal inhomogeneities in the recombination plasma begin to 
stream freely across the universe.  These signatures reach us today in the cosmic microwave background (CMB). The best limits on the CMB temperature anisotropy over 
the sky, down to $\sim 0.4^\circ$ are provided by the Wilkinson Microwave Anisotropy Probe (WMAP) \cite{2006astro.ph..3449S}.  In conjunction with other surveys, these 
data tightly constrain cosmological model parameters \cite{2004PhRvD..69j3501T,2006astro.ph..3449S}.
 
Most future and recent CMB experimental efforts aim to measure the polarization or temperature anisotropy \cite{2001PhRvD..63d2001L, 
2001PhRvL..86.3475J, 2002ApJ...568...46P, 2002MNRAS.334...11A, 2004AdSpR..34..491T, 2003NewAR..47..939K, 2004SPIE.5498...11R,2004ApJ...600...32K, 
2001ApJ...561L...7S, 2002ApJ...571..604N, 2002ApJ...568...38H, 2003ApJ...591..556P, 2002MNRAS.329..890G, 2005MNRAS.363...79G} on smaller scales than 
those measured by WMAP.  Small-scale CMB temperature anisotropy measurements further constrain the matter and baryonic matter fractions, $\Omega_mh^2$ 
and $\Omega_bh^2$ \cite{2001ApJ...555...88W}, the spectral index of the primordial scalar power spectrum, $n_s$, and its possible running, $\alpha_s$.  
Measurements of $n_s$ will be integral to the viability or eventual rejection of a wide variety of inflation models \cite{1997PhRvD..56.3207D, 
2003ApJS..148..213P}.

The most conspicuous features in the CMB temperature anisotropy on small scales are acoustic oscillations 
\cite{1970Ap&SS...7....3S,1970ApJ...162..815P, 1987MNRAS.226..655B, 1997Natur.386...37H} and Silk damping 
\cite{1968ApJ...151..459S,1997ApJ...479..568H}.  Silk damping is determined directly by the free electron abundance, which is set by the recombination 
history.  Here, photons in the ionized gas will diffuse over a characteristic scale $k_D \propto \sqrt{n_e}$ (where $n_e$ is the free electron number 
density and $k_D$ is the damping wavenumber), exponentially damping photon perturbations with wavenumbers $k>k_D$.  If the free electron density is 
overpredicted by recombination models, then $k_D$ is also overpredicted and Boltzmann codes will predict too much power on small scales.  Measurements 
of $n_s$ using CMB data including the damping tail region will then be biased downward. In light of the WMAP 3-year analysis \cite{2006astro.ph..3449S}, it is crucial to 
understand how $n_s$ differs from 1.  While the differences among recently published recombination histories are too small to be important for the WMAP 
$n_s$ measurement, the corrections are expected to be significant (at the $\Delta n_s\sim 0.02$ level) for Planck \cite{2006MNRAS.373..561L}, and 
presumably also for high-$\ell$ experiments such as ACT \cite{2003NewAR..47..939K} and SPT \cite{2004SPIE.5498...11R}.  
Successful interpretations of data from the next generation of small-scale CMB anisotropy experiments will depend on a solid understanding of recombination.

Fundamentally, the problem in cosmological recombination is to solve consistently for the evolution of the atomic level occupations and
the radiation field (which has both a thermal piece, and a non-thermal piece from the radiation of the atoms themselves) in an expanding background.  
The highly-excited states are kept close to equilibrium by the high rates interconnecting them, and 
the rate of formation of the ground state in both helium and hydrogen is dominated by the occupation of the $n=2$ states and the rates connecting the $n=2$
states to the ground state.  In both helium and hydrogen, the decay channels to the ground state through the allowed transitions are dramatically
suppressed relative to the vacuum rates by the optical depth in the gas.  Indeed, in both systems, the two-photon decay rate from the $n=2$ $S$ state (singlet 
in the case of \HeI) is comparable to the rate in allowed decay channels.  This is the so-called ``$n=2$ bottleneck'' \cite{2000ApJS..128..407S}.

The primordial recombination was first investigated theoretically in the 1960s \cite{1968ApJ...153....1P, 1968ZhETF..55..278Z} using a 
simple ``three-level atom'' (TLA) approximation.  The TLA tracks the abundance of ground state hydrogen atoms (\HI\ $1s$), excited 
hydrogen atoms (assumed to be in Boltzmann equilibrium), and free electrons.  The TLA allows for recombination to and photoionization 
from excited \HI\ levels, and allows excited atoms to decay to the ground state by Ly$\alpha$ ($2p\rightarrow 1s$) or two-photon 
($2s\rightarrow 1s$) emission.  Subsequently, a substantial literature developed, testing some of the assumptions of the TLA and extending 
it to the problem of helium recombination \cite{1969PThPh..42..219M, 1971PThPh..46..416M, 1983A&A...123..171L, 1984ApJ...280..465L, 
1991A&A...246....1F, 1998A&A...335..403G}.  Seager et al.\ \cite{2000ApJS..128..407S} provide the current benchmark precision 
recombination calculation by simulating $300$ levels in \HI, 200 levels in \HeI, 100 levels in \HeII, interactions with the radiation 
field in the Sobolev approximation, basic hydrogen chemistry, and matter temperature evolution.  They found that the three level model 
\cite{1968ApJ...153....1P, 1968ZhETF..55..278Z} with a ``fudge factor'' inserted to speed up \HI\ recombination was an accurate 
approximation to their full multilevel atom solution.  Their recombination code, {\sc Recfast} \cite{1999ApJ...523L...1S}, is packaged 
into most of the CMB anisotropy codes in common use, and underlies the cosmological constraints from the CMB, including those recently 
reported by WMAP.

\begin{figure}[!ht]
\epsfxsize=3.0in
\begin{center}
\epsffile{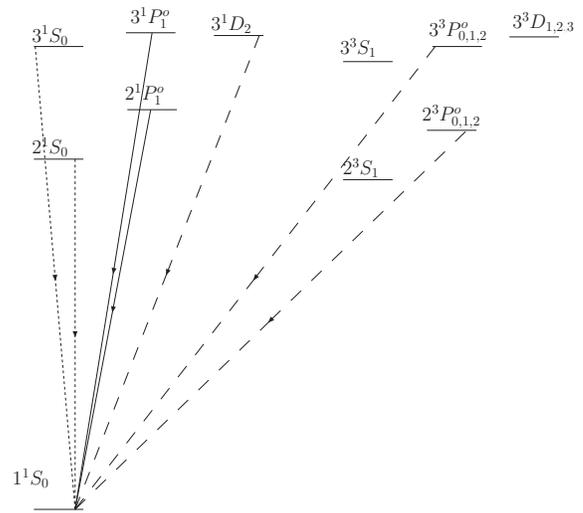}
\end{center}
\caption{Formation of neutral helium: a Grotrian diagram (up to $n=3$) for \HeI.  Singlet ($S=0$ parahelium) and triplet 
($S=1$ orthohelium) levels and higher-order transitions give a rich system of low-lying transitions.  Marked are two-photon transition 
(light dashed lines from $2^1S_0$ and $3^1S_0$), (allowed) electric dipole transitions (solid lines from $2^1P_1^o$ and $3^1P_1^o$, like the Lyman series in 
\HI), the intercombination lines (heavy dashed lines from $2^3P_1^o$ and $3^3P_1^o$), and quadrupole transitions (from $3^1D_2$).  The 
dipole transitions are treated in Sec.~\ref{ss:mcmethods} using a Monte Carlo method with partial redistribution, forbidden 
one-photon lines are treated in Sec.~\ref{sec:anacompletecont} using an analytic method for complete redistribution.  (The energy levels are 
not drawn to scale.)}
\label{figs:heigrotrian}
\end{figure}

There are several reasons why it is now timely to revisit the cosmological recombination. Recent work \cite{2006A&A...446...39C, 2005AstL...31..359D, 
2004MNRAS.349..632L, 2006astro.ph.10691W, 2006AstL...32..795K, 2006astro.ph.12322W} has led to suspicion that some pieces of the recombination problem, such as matter 
temperature evolution \cite{2004MNRAS.349..632L,2006astro.ph.12322W}, two-photon transitions from high-lying states \cite{2005AstL...31..359D}, the effect of \HeI\ 
intercombination lines \cite{2005AstL...31..359D,2006astro.ph.10691W}, departures of the $l$ sublevels of hydrogen from their statistical population ratios 
\cite{2007MNRAS.374.1310C}, and stimulated two-photon transitions \cite{2006A&A...446...39C, 2006AstL...32..795K} are not completely understood or were 
absent in past work.  Two-photon transitions and intercombination rates speed \HeI\ recombination by facilitating the formation of the ground state, and as
described in \cite{2005AstL...31..359D} these could constitute a serious correction \HeI\ recombination, pushing \HeI\ much closer to equilibrium than 
in traditional models \cite{2000ApJS..128..407S}.  These considerations led to significant changes in cosmological model parameter estimates \cite{2006MNRAS.373..561L} and 
warrant further consideration.  (The allowed, intercombination, quadrupole, and 
two-photon transitions up to $n=3$ are shown in the Grotrian diagram, Fig.~\ref{figs:heigrotrian}.)  
In addition, some of the issues considered during the 1990s, such as the effect of \HI\ 
continuum opacity on \HeI\ recombination, were not fully resolved; we will show here that this problem is much more complicated than previously believed.  
Finally, the imminent prospect of precision CMB data at high $\ell$ from ACT, SPT, and Planck has ``raised the bar'' for theorists and demands a much better
understanding of recombination than was needed a decade ago.  

This is the first in a series of three papers devoted to the subject of helium recombination; the companion papers (Hirata and Switzer astro-ph/0702144 and 
Switzer and Hirata astro-ph/0702145) will be referred to as 
Paper II and Paper III. We have not yet completed the solution of hydrogen recombination because the radiative transfer is more complicated, the 
treatment of the high-$n$ levels is more important, and the required accuracy is much greater.  We do not believe our existing code is accurate 
enough for that application (see the discussion in Paper III).  This paper describes the standard multi-level atom code, as well as the most important 
improvements we have made, namely inclusion of feedback from spectral distortions, semiforbidden and forbidden lines, \HI\ bound-free opacity, and 
radiative transfer in the \HeI\ lines in the cases of both complete and partial redistribution.  Paper II describes several effects that we find to 
have only a small influence on recombination: two-photon transitions in \HeI, interfering damping wings, and photons in the \HeI\ continuum.  Paper III 
discusses the effect of the isotope shift between \Hethree\ and \Hefour\ resonance lines, Thomson scattering, rare processes, 
collisional process, and peculiar velocities.  Although we find these effects to be small, it is necessary to investigate them in order to be sure of this.  
Paper III also contains a summary of the remaining uncertainty in helium recombination and its implications for the CMB power spectrum.

One unresolved issue that was discussed extensively in the 1990s is whether hydrogen photoionization can destroy photons in the \HeI\ 
\LyHe\ resonances (in particular, \LyaHe) and the \HeI\ $1^1S$ continuum, pushing \HeI\ recombination closer to Saha equilibrium.  This effect was 
ignored in the earliest work on helium recombination \cite{1969PThPh..42..219M}, however Hu et al.\ \cite{1995PhRvD..52.5498H} argued that the effect 
was strong enough to force helium into Saha equilibrium.  The argument is essentially that the timescale for the absorption of these photons is much 
less than the Hubble time.  On the other hand, Seager et al.\ \cite{2000ApJS..128..407S} argued that hydrogen photoionization was negligible because 
the rate of \HeI\ line excitation was much faster than the rate of \HI\ photoionization.  We show here that both of these arguments are too simple: the 
effectiveness of \HI\ opacity depends not only on the optical depth per Hubble time from \HI, but also on the width of the \HeI\ line and the 
redistribution of photons within the line. We have therefore supplemented our level code with a radiative transfer analysis of the \HeI\ lines that 
takes account of \HI\ continuum opacity and emission/absorption of resonance radiation, including coherent scattering.  The result is that \HI\ opacity 
has little effect at $z>2200$, because even though the \HI\ optical depth per Hubble time can be $\gg 1$, the \HI\ optical depth within the width of any 
individual \HeI\ line is small, so the standard Sobolev escape rate is unmodified and recombination proceeds similarly to Seager et al. 
\cite{2000ApJS..128..407S}.  At lower redshifts the exponentially increasing \HI\ abundance begins to absorb photons out of the \HeI\ \LyaHe\ line, 
rapidly accelerating \HeI\ recombination and leading to its completion by $z\sim 1800$.

Leung et al. \cite{2004MNRAS.349..632L} argued that the heating due to the $13.6\,$eV energy release per \HI\ recombination would increase the matter 
temperature and thus slightly delay recombination.  However this energy release couples inefficiently to the matter because it is delivered almost 
entirely to the photons, which are not absorbed by the matter \cite{2006astro.ph.12322W}.  This energy actually goes into the spectral distortion to 
the CMB, not the matter.  In order to properly follow the matter temperature during recombination it is necessary to directly keep track of the changes 
in the total energy in translational degrees of freedom, and the number of such degrees of freedom (which decreases during recombination); the ratio of 
these quantities is $\kB\Tm/2$.  This was done by Seager et~al. \cite{2000ApJS..128..407S}, and we do not believe any modification of the basic 
methodology is necessary.

Finally, the feedback of radiation field distortions from resonances onto lower-lying resonances is expected to slow recombination rates 
\cite{2000ApJS..128..407S}.  For example, escaped radiation from Ly$\beta$ in \HI\ could redshift onto the Ly$\alpha$ transition and excite atoms; analogous 
effects occur in \HeI\ (and in \HeII, although this case is less important for CMB anisotropies).  This is included iteratively in the level code 
developed here.

\comment{
Dubrovich and Grachev \cite{2005AstL...31..359D} investigate the effect of two-photon transitions in \HI\ and para-\HeI\ from high-lying states 
($n>2$), and of radiative transfer in the intercombination lines \HeI\ \InteraHe.  These are found to push \HeI\ recombination closer to Saha 
equilibrium, dropping the density of free electrons before hydrogen recombination.  We add the quadrupole series (\QuadHe), and the intercombination 
series (\InterHe) to the recombination level code.  Two-photon transitions from $n>2$ levels are investigated in Paper II and are found to be less 
significant 
than anticipated \cite{2005AstL...31..359D} in \HeI\ because of destructive interference between various contributions to the decay amplitude, and 
because the resonances in the two-photon decay rates from $n>2$ are already included in the 
one-photon transitions in the level code.  Paper II also considers stimulated two-photon emission \cite{2006A&A...446...39C}, 
two-photon absorption of radiation field distortions \cite{2006AstL...32..795K}, and Raman scattering.
}

In this paper, we will develop the effects through several sections.  The multi-level atom code is described in Sec.~\ref{sec:newlevelcode}.  We 
introduce \HI\ continuum opacity and its effects on radiative transfer between \HeI\ lines in Sec.~\ref{sec:heibbbfpihi}.  
Sec.~\ref{sec:anacompletecont} discusses an analytic approach to transport with complete redistribution and continuous (\HI) opacity.  
Sec.~\ref{sec:partialred} expands this treatment to include partial redistribution using a Monte Carlo simulation.  We conclude in 
Sec.~\ref{sec:conclusion}.  In Appendix~\ref{sec:atomicappendix} we discuss the atomic data used here; Appendix~\ref{ss:integraldetails} 
describes issues with transport under complete redistribution; and Appendix~\ref{sec:mcimplementation} describes our implementation of the Monte Carlo 
method for solving line profiles with coherent scattering.  Appendix~\ref{sec:mcsobolev} examines the limiting cases in which the Monte Carlo 
method reduces to the method of Sec.~\ref{sec:anacompletecont} and to the widely used Sobolev escape probability method.  Finally, 
Appendix~\ref{ss:heic} describes the handling of photons above the \HeI\ photoionization threshold (24.6 eV) in our code.

\section{A new level code with radiative feedback}
\label{sec:newlevelcode}

The standard recombination scenario discussed in this paper is a homogeneous, interacting gas of protons, helium nuclei, electrons, and photons in an 
expanding background.  The background dynamics (i.e. the Friedmann equation) are allowed to include other homogeneously distributed components (e.g. 
neutrinos, dark matter, and dark energy) but in our treatment we only consider their effect on $H(z)$, neglecting any direct interaction with the 
baryons and photons.  Several extensions to this standard scenario have been proposed but will not be considered here: energy injection from 
self-annihilating or decaying dark matter, primordial magnetic fields, small-scale inhomogeneities or peculiar velocities, and others 
\cite{2005MNRAS.363..521G, 2001PhRvD..63d3505B, 2005MNRAS.364....2M, 2005PhRvD..72b3508P, 2000ApJ...539L...1P, 2006MNRAS.369.1719M, 
2001NewA....6...17H, 2004PhRvL..92c1301P, 2000PhRvD..62l3508A, 1986A&A...157L...7B}.  The underlying physics in the standard problem is known, but the 
solution is complicated by the interactions between radiation and atomic level occupations, and the variety of atomic rates.  Many atomic rates 
involved are time-consuming to calculate, or not known well (for example, collisional and intercombination rates in \HeI.)  Collisional processes 
contribute to a lesser extent because of the high photon to baryon ratio during recombination.  These are discussed in Paper III, 
where we argue that they are unimportant for helium recombination; in this paper we restrict ourselves to radiative processes.

Recombination calculations must ultimately meet the practical constraint that they be fast enough (running in a few seconds) to act as inputs to CMB 
anisotropy codes.  The general procedure for developing a recombination code that meets this constraint is to first ``over-simulate'' the recombination 
history, and prioritize the effects.  One then develops a practical method that encapsulates the important physics.  Additional effects can be included 
as parameterized corrections to the model.  The TLA approximation has been practical for CMB studies to date, and a constant correction factor is 
sufficient to bring it into agreement with more complete methods \cite{2000ApJS..128..407S}.  CMB researchers have been lucky that such a fast 
approximation is possible, but in the post-WMAP era it is not clear whether the TLA approximation will be sufficiently accurate. Tighter experimental 
constraints on the high CMB multipoles in the Silk damping region will need to be matched by confidence that the underlying recombination physics is 
well understood.  Here, we use a multilevel atom code similar to Seager et al.\ \cite{2000ApJS..128..407S} to assess the significance of several new 
processes relative to a base model with one set of cosmological parameters.  This level of detail is essential to developing an understanding of new 
physical effects.  Unfortunately, it leads to a code that is much too slow for direct incorporation into Boltzmann codes.  Developing and extension
to the TLA approximation (or similarly fast model) for a range of cosmological parameters will be the subject of later work.

Many of the important details in the calculations are wrapped up in the notation, so we present the major symbols used here in Table~\ref{tab:table1}.

\begin{table*}
\caption{\label{tab:table1}Symbols used in this paper. Units of ``1'' mean the quantity is dimensionless.}
\begin{tabular}{cccclcl}
\hline\hline
Symbol & & Units & & Description & & Equation \\
\hline
$a$ & & 1 & & Voigt unitless width parameter & & Eq.~(\ref{eq:voigt}) \\
$A_{i \rightarrow j}$ & & s$^{-1}$ & & Einstein spontaneous one-photon decay rate for $i$ to $j$ & & \\
$f_{\rm coh}$ & & 1	& & fraction of line photon absorptions resulting in coherent scattering & & \\
$f_i$ & & 1 & & fraction of line photon absorptions followed by transition to level $i$ & & \\
$f_{\rm inc}$ & & 1	& & fraction of line photon absorptions resulting in incoherent processes & & $f_{\rm inc}=1-f_{\rm coh}$ \\
$g_i$ & & 1 & & degeneracy of level $i$, not including nuclear spin & & \\
$H$ & & s$^{-1}$ & & Hubble rate & & \\
$K$ 	& & cm$^3\,$s	& & Peebles $K$-factor \cite{1968ApJ...153....1P} & & $K=\lambda_{\rm line}^3/8\pi H$ \\ 
$n_e$ & & cm$^{-3}$ & & electron density & & \\
$n_i$ & & cm$^{-3}$ & & density of atoms in level $i$ & & \\
$\nH$ & & cm$^{-3}$ & & total density of all hydrogen nuclei & & \\
$\pha$	& & 1	& & photon phase space density & & Eq.~(\ref{eqn:photonphase}) \\
$\phaC$     & & 1   & & $\pha$ in equilibrium with the continuum opacity & & Eq.~(\ref{eqn:phacontinuum}) \\
$\phaL$ 	& & 1	& & $\pha$ in equilibrium with the line opacity & & Eq.~(\ref{eqn:phaline}) \\
$\phaLO$ 	& & 1	& & modification of $\phaL$ used with coherent scattering & & Eq.~(\ref{eq:phalo-def}) \\
$\pha_{\rm Pl}$ & & 1 & & photon phase space density for blackbody distribution & & $\pha_{\rm Pl} = 1/(e^{h\nu/\kB\Tr}-1)$ \\
$\pha_\pm$ 	& & 1	& & photon phase space density on the blue ($+$) or red ($-$) side of a line & & Eq.~(\ref{eqn:linedistortion}) \\
$\bar \pha$ & & 1	& & photon phase space density averaged over the profile & & Eq.~(\ref{eqn:phaintegral}) \\
$\PC$ & & 1 & & probability of photon emitted in line being absorbed by \HI & & Eq.~(\ref{eq:pc}) \\
$\Pesc$ & & 1 & & escape probability from the line (equal to $\PS$ in Sobolev approximation) & & \\
$\PMC$ & & 1 & & prob. of photon in MC being lost by \HI\ absorption or redshifting & & \\
$\PS$ & & 1 & & Sobolev escape probability & & Eq.~(\ref{eqn:sobolevescprob}) \\
${\mathcal Q}$ & & erg$\,$s$^{-1}$ & & heating per hydrogen nucleus per unit time & & \\
$R_{ij}$ & & s$^{-1}$ & & transition rate from level $i$ to level $j$ & & \\
$\Tm$, $\Tr$ & & K & & matter ($\Tm$) or radiation ($\Tr$) temperature \\
$x$ & & 1	& & frequency relative to line center in Doppler units & & $x=(\nu-\nu_{\rm line})/\Delta\nuD$ \\
$x_e$ & & 1 & & abundance of electrons relative to total hydrogen nuclei & & $x_e=n_e/\nH$ \\
$x_i$ & & 1		& & abundance of the state $i$ relative to total hydrogen nuclei	& & $x_i=n_i/\nH$ \\
$x_{\rm tot}$ & & 1 & & total number of free particles per hydrogen nucleus & & \\
$\alpha_i$ & & cm$^3\,$s$^{-1}$ & & recombination coefficient to level $i$ & & \\
$\beta_i$ & & s$^{-1}$ & & photoionization rate from level $i$ & & \\
$\Gamma_i$ & & s$^{-1}$ & & width of level $i$ & & \\
$\Gamma_{\rm line}$ & & s$^{-1}$ & & Lorentz width of the line & & \\ 
$\Gamma_{\cal Q}$ & & erg$\,$cm$^{-3}\,$s$^{-1}$ & & heating rate per hydrogen nucleus & & \\
$\Delta\nuD$ & & Hz	& & Doppler width of line & &  \\
$\etaC$ & & Hz$^{-1}$	& & differential optical depth from continuous opacity & & Eq.~(\ref{eqn:gammacdef})\\
$\Lambda_{i \rightarrow j}$ & & s$^{-1}$ & & spontaneous two-photon decay rate from $i$ to $j$ & & \\
$\Lambda_{\cal Q}$ & & erg$\,$cm$^{-3}\,$s$^{-1}$ & & cooling rate per hydrogen nucleus & & \\
$\nu_{\rm line}$, $\nu_{ul}$	& & Hz	& & frequency of line center; $\nu_{ul}$ for specific upper and lower levels & & \\ 
$\nu_{{\rm th},i}$ & & Hz & & photoionization threshold frequency from level $i$ & & \\
$\xi$ & & 1 & & rescaled photon phase space density & & Eq.~(\ref{eq:xi-pha}) \\
$\Pi$ & & Hz$^{-1}$ & & probability distribution of photon frequency in Monte Carlo & & \\
$\sigma_{ic}$ & & cm$^2$ & & photoionization cross section from level $i$ & & \\
$\sigmaT$ & & cm$^2$ & & Thomson cross section & & \\
$\tau_{\rm coh}$ 	& & 1 & & Sobolev optical depth from coherent scattering & & $\tau_{\rm coh}=f_{\rm coh}\tauS$ \\
$\tau_{\rm inc}$ 	& & 1 & & Sobolev optical depth from incoherent processes & & $\tau_{\rm inc}=f_{\rm inc}\tauS$ \\
$\tauLL$ & & 1       & & optical depth from continuous opacity between lines & & Eq.~(\ref{eqn:linelinedepth}) \\
$\tauS$ 		& & 1 & & total Sobolev optical depth & & Eq.~(\ref{eqn:sobolevdepth}) \\
$\phi$ & & Hz$^{-1}$	& & atomic line profile & & \\
$\chi$ & & 1	& & photon-atom scattering angle & & Eq.~(\ref{eqn:cohfreqshift}) \\
\hline\hline
\end{tabular}
\end{table*}

\subsection{Summary of the method}
\label{ss:summary}

For brevity, we will only highlight the physical arguments of the multilevel code, and differences from Seager et al \cite{2000ApJS..128..407S}.  The 
multilevel atom code tracks levels up to a maximum principal quantum number $\nmax$.  In this paper, we use a smaller model for \HI\ with 245 levels 
(up to $\nmax=200$), \HeI, 289 levels ($\nmax=100$), and for \HeII, 145 levels ($\nmax=100$), resolving $l$ sublevels up to $n=10$, and including 
quadrupole and intercombination transitions in the \HeI\ rates.  (A detailed discussion of atomic data can be found in 
Appendix~\ref{sec:atomicappendix}.  In Paper III, we investigate the effect of changing $\nmax$ and find that these can be neglected here.)  Unless 
stated otherwise, we assume a $\Lambda$CDM 
cosmology with $\Omega_b=0.04592$, $\Omega_m= 0.27$, $\Omega_r=8.23\times10^{-5}$, zero spatial curvature, massless neutrinos, a Hubble parameter of 
$h=0.71$, and present-day radiation temperature $\Tr(z=0) = 2.728$~K. The fiducial fractional helium abundance (i.e. ratio of helium to hydrogen 
nuclei) is $f_{\rm He}=0.079$.  The Hubble rate in such a cosmology is
\beq
H(z)=H_0 \sqrt{\Omega_\Lambda + \Omega_m(1+z)^3 + \Omega_r(1+z)^4}.
\label{eqn:backgroundcosmology} 
\eeq 
The number density of hydrogen nuclei is given by \cite{2000ApJS..128..407S, 1999ApJ...523L...1S}
\beq \nH(z) = 1.123\times10^{-5}\frac{\Omega_b h^2}{1+3.9715 f_{\rm He}} (1+z)^3
\,{\rm cm}^{-3}, \label{eqn:hydrogendensity} \eeq
where $3.9715$ is the ratio of atomic masses of $^4$He and $^1$H.  We will use the photon phase space density $\pha(\nu)$ to track the radiation 
spectrum instead of the specific intensity $J_\nu$, because the former is 
conserved along a trajectory in free space whereas $J_\nu$ decreases as the universe expands.  The relation between these is
\beq
\pha(\nu) = \frac{c^2}{2h\nu^3} J_\nu.
\label{eqn:photonphase}
\eeq

Photoionization and re\-com\-bination con\-tri\-bu\-tions to atomic level population dynamics are discussed in Seager et al.\ \cite{2000ApJS..128..407S}, and we follow their treatment here.  The rate of change of the average occupation of an atomic level is given by a series of bound-bound and bound-free rate equations.  The photoionization rate from some level $i$ is given by
\beq \beta_i = \frac{8 \pi }{ c^2} \int_{\nu_{{\rm th},i}}^\infty \sigma_{ic} (\nu) \nu^2 \pha(\nu) d \nu, \label{eqn:photorate} \eeq  
where $\pha(\nu)$ is the photon phase space density, $\nu_{th}$ is the photoionization threshold frequency from level $i$, and $\sigma_{ic}$ is the photoionization cross-section from that level, as a function of frequency (implicit in subsequent equations).  The photorecombination rate density for spontaneous and 
stimulated processes is
\beq \alpha_i = \frac{8 \pi }{ c^2} \left ( \frac{ n_i }{ n_e n_c} \right )^{\rm LTE}
\! \int_{\nu_{{\rm th},i}}^\infty \sigma_{ic} \nu^2 [1+\pha(\nu)] e^{-h\nu/\kB 
\Tm} d \nu, \label{eqn:recombrate} \eeq
where $\Tm$ is the matter temperature.  The prefactor is the Saha ratio of the occupation of the level $i$ (in local thermal equilibrium, LTE) to the free electron density times the 
continuum state density,
\beq \left ( \frac{ n_i }{ n_e n_c} \right )^{\rm LTE} = \left ( \frac{ h^2 }{ 2 \pi m_e \kB \Tm } \right )^{3/2} \frac{g_i }{ 2 g_c} e^{h \nu_{\rm th} /\kB \Tm}, 
\label{eqn:saharatio} \eeq
where $i$ labels a bound state of a species, and $c$ labels the continuum state (eg. for hydrogen this is \HII) of the species; $g_i$ and $g_c$ are the 
state degeneracies.  The bound-free rate for a level $i$ is then
\beq \frac{dx_i }{ dt} = \alpha_i n_e x_c - \beta_i x_i. \label{eqn:bfdiffeq} \eeq

Except for several transitions in \HeI\ where transport is calculated separately to include new effects, single photon bound-bound rates are calculated 
in the standard Sobolev approximation with complete redistribution,
\beq \frac{d x }{ dt } = A_{u \rightarrow l} \PS [ x_u (1+\pha_+) - \frac{g_u }{ g_l} x_l \pha_+], \label{eqn:sobolevapprox} \eeq
where $A_{u\rightarrow l}$ is the Einstein rate coefficient connecting an upper bound state $u$ to a lower bound state $l$, and $\pha_+$ is the phase space density of radiation on the blue side of the line.  In the ``original'' version of the code (i.e. before feedback is included) this is taken to be the Planck distribution, $\pha_+=\pha_{\rm Pl}=(e^{h\nu/\kB\Tr}-1)^{-1}$.  In the Sobolev approximation, the rates are modulated by the probability that a photon will escape from the resonance, allowing the average occupation 
state of the gas to change.  The probability is associated with the Sobolev effective optical depth,
\beq \tauS = \frac{A_{u\rightarrow l} c^3 }{ 8 \pi H(z) \nu_{ul}^3 } n_{\rm H} \left ( x_l \frac{g_u }{ g_l} - x_u \right ), \label{eqn:sobolevdepth} 
\eeq
so that
\beq \PS = \frac{1 - e^{-\tauS} }{ \tauS}. \label{eqn:sobolevescprob} \eeq

\subsection{Matter temperature}
\label{ss:tmat}

The matter temperature, $\Tm$, departs from the radiation temperature due to adiabatic expansion, Compton cooling, free-free, 
bound-free, and bound-bound processes \cite{2000ApJS..128..407S}.  Here, we write the heat exchange terms to emphasize the individual
processes, kinetic degrees of freedom in the gas, and we expand the description of free-free processes in \cite{2000ApJS..128..407S}.  
Nothing is fundamentally new or unexpected in this treatment, and, indeed, we find that the departure of $T_m$ from $T_r$ is negligible 
during \HeI\ recombination.  The rate of change of the matter temperature is related to the heating/cooling processes and adiabatic expansion 
in the gas through
\beq
{\dot T}_{\rm m}  = \frac{2 }{ 3 \kB x_{\rm tot}} \left ( \mathcal{Q}_{\rm es} + \mathcal{Q}_{\rm ff} + \mathcal{Q}_{\rm bf} + \mathcal{Q}_{\rm bb} 
\right ) - 2 H \Tm, 
\label{eqn:coolingdiffeq} \eeq
where the ``es'' subscript denotes Compton cooling and the others denote radiative atomic processes between bound and free levels.  
An individual heat exchange process $\mathcal{Q}$ has the general form
\beq \mathcal{Q} = \frac{\Gamma_{\cal Q}-\Lambda_{\cal Q} }{ n_{\rm H}} - \frac{3 }{ 2} \dot x_{\rm tot} \kB \Tm, \eeq
where $\Gamma_{\cal Q}$ and $\Lambda_{\cal Q}$ are the heating and cooling rates (in $\mathrm{erg} \, \mathrm{cm}^{-3}\, \mathrm{s}^{-1}$), and $\dot 
x_{\rm tot}$ accounts for processes that modify the number of kinetic particles in the gas.
The Compton (electron scattering) cooling term is
\beq \mathcal{Q}_{\rm es} = -4 x_e c \sigmaT a_{\rm R} \Tr^4 \frac{ \kB (\Tr -\Tm) }{ m_e c^2 },  \eeq
where we have used the radiation constant $a_{\rm R} = \pi^2 \kB^4 / 15 c^3\hbar^3$.
The free-free contribution to $\dot T_{\rm m}$ is the integral of the free-free opacity over the radiation temperature blackbody distribution (heating) minus 
the matter temperature blackbody distribution (cooling),
\beq \mathcal{Q}_{\rm ff} =
\frac{ 8 \pi }{ n_H c^2} \int \nu^2 [\pha_{\rm Pl}(\nu,\Tr) - \pha_{\rm Pl}(\nu,\Tm)] h \nu \alpha_{\rm ff} (\nu)d\nu, \eeq
where $\pha_{\rm Pl}$ is the blackbody distribution, and the length absorption coefficient $\alpha_{\rm ff}(\nu)$ in units of cm$^{-1}$ is 
\cite{1986rpa..book.....R}
\begin{eqnarray} \alpha_{\rm ff}(\nu) &=& \frac{4 e^6 }{ 3 m_e h c} \sqrt{\frac{ 2 \pi }{ 3 \kB m_e }} \Tm^{-1/2} \nu^{-3}
n_e n_{\rm HII}  \nonumber \\ 
&&\times (1- e^{-h\nu/\kB \Tm} ) \bar g_{\rm ff}. \end{eqnarray}
The thermally averaged Gaunt factor $\bar g_{\rm ff}$ is given by Sutherland \cite{1998MNRAS.300..321S}.
In principle there is an additional correction due to free-free radiation from electron-\HeII\ and electron-\HeIII\ collisions, however these other 
species are an order of magnitude less abundant than \HII, and deviations from $\Tm=\Tr$ are negligible for helium recombination, regardless.  Therefore we 
have not included helium free-free radiation in our matter temperature evolution.

The bound-free contribution to $\dot T_{\rm m}$ is from energy exchanged in photorecombination and photoionization, and due to the change in $x_{\rm tot}$,
\begin{eqnarray}
\mathcal{Q}_{\rm bf} &=&
\sum_i \biggl [ \left (\frac{\Gamma_{{\cal Q},i} }{ n_H} - \frac{3 }{ 2} \beta_i x_i \kB \Tm \right ) \nonumber \\ &&- \left (\frac{\Lambda_{{\cal Q},i} }{ n_H} - \frac{3 }{ 2} 
\alpha_i n_e x_c \kB \Tm \right ) \biggr ]. \end{eqnarray}
The energy exchanged per bound-free process includes the heating due to photoionization,
\beq \frac{\Gamma_{{\cal Q},i} }{ n_H}= x_i \frac{ 8 \pi }{ c^2 } \int_{\nu_{{\rm th},i}}^\infty \sigma_{ic} \nu^2 \pha(\nu) h (\nu-\nu_{{\rm th},i}) d \nu, \eeq
and the cooling due to recombination,
\begin{eqnarray} \frac{\Lambda_{{\cal Q},i} }{ n_H} \!\! &=& \!\! n_e x_c \frac{ 8 \pi }{ c^2 }\left ( \frac{ n_i }{ n_e n_c } \right )^{\rm LTE} 
\nonumber 
\\ &&\!\! \times \int_{\nu_{{\rm th},i}}^\infty \sigma_{ic} \nu^2 [1 + \pha(\nu)] h (\nu - \nu_{{\rm th},i}) d\nu.
\end{eqnarray}

We have computed these results using the blackbody function for ${\cal N}(\nu)$, since it is the thermal CMB photons that are responsible for the 
photoionizations and stimulated recombinations in the Balmer, Paschen, etc. continua of \HI, and the analogous continua of \HeI.  The potential 
pitfall in this assumption is that some of the photons emitted in the \HeI\ 21.2 eV line ($2^1P^o$--$1^1S$) photoionize \HI\ directly from the $1s$ 
level.  This provides an additional source of heating during helium recombination that is not included in our code.  However a simple calculation shows 
that it is negligible.  The maximum amount of thermal energy that can be injected by such photons is $21.2-13.6=7.6\,$eV per photon.  Since a fraction
$f_{\rm He}/x_{\rm tot}\sim 0.04$ of the matter particles during this era are heliums (\HeI\ or \HeII), this implies an injected energy of 
$\sim 0.3\,$eV per particle, even assuming every helium recombination injects the maximum amount of energy.  For comparison the total energy in 
translational modes of the matter is $(3/2)\kB\Tm=0.6\,$eV per particle.  Even at the very end of helium recombination, the Compton cooling time is
\beqa
t_{\rm Compton} &=& \frac{(3/2)x_{\rm tot}\kB(\Tm-\Tr)}{{\cal Q}_{\rm es}}
\nonumber \\
&=& \frac{3x_{\rm tot}m_ec}{8x_e\sigma_{\rm T}a_{\rm R}T_{\rm r}^4}
\approx 7\times 10^6\,{\rm s};
\eeqa
in comparison, the timescale for helium to recombine is several times $10^{12}\,$s.  The maximum fractional change in the matter temperature is then 
$0.3\,$eV/$0.6\,$eV multiplied by the ratio of the Compton cooling to recombination time, which is of order $10^{-6}$.  This can be neglected.

The contribution from bound-bound transitions (${\cal Q}_{\rm bb}$) is negligible through a similar argument.  Bound-bound processes do not change 
$x_{\rm tot}$, so the term ${\cal Q}_{\rm bb}$ is determined by the average amount of energy delivered to the gas by a photon that is injected into 
(or redshifts onto) a resonance line.  In general, the net heating rate can only be calculated by a detailed radiative transfer analysis \cite{2004ApJ...602....1C}. 
However, as we saw in the previous paragraph, the injection of energy changes $\Tm$ at the $10^{-6}$ level if each \HeI\ recombination releases 7.6 eV of 
energy into the gas.  Since there is a total of 24.6~eV available per \HeI\ recombination, $\Tm$ would change by only several parts in $10^6$ even 
if {\em all} of this energy were released into the gas by resonance line scattering.  Therefore we can neglect the bound-bound contribution to $\Tm$.

Time derivatives for the occupations and temperatures are converted to redshift derivatives according to
\beq \frac{d z }{ dt } = - (1+z)H(z). \eeq
Equations~(\ref{eqn:bfdiffeq}), (\ref{eqn:sobolevapprox}), and (\ref{eqn:coolingdiffeq}) give a set of stiff equations for the level occupations, as in 
Seager et~al. \cite{2000ApJS..128..407S} that can be solved using the semi-implicit extrapolation (Bader-Deuflhard) method 
\cite{1992nrca.book.....P}.

\begin{figure}[!ht]
\epsfxsize=3.1in
\begin{center}
\epsffile{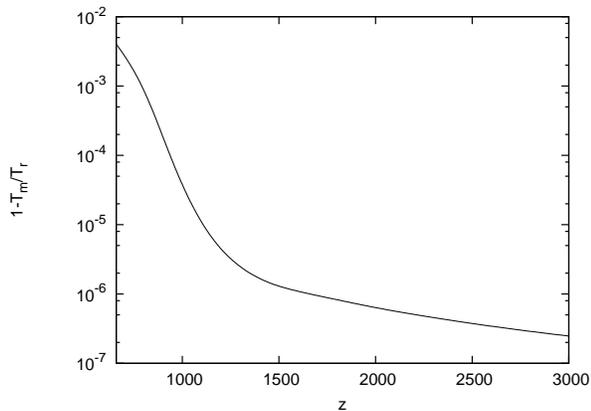}
\end{center}
\caption{\label{figs:tmatter}
The fractional difference in the matter temperature relative to the radiation temperature, $1-\Tm/\Tr$.  Note the sign: matter is cooler than 
radiation at late times because of its different adiabatic index ($5/3$ versus $4/3$).  During \HeI\ recombination for $1600 < z < 3000$, the 
fractional difference is $< 10^{-6}$.}
\end{figure}

Our matter temperature evolution is shown in Fig.~\ref{figs:tmatter}.  The important conclusion is that the fractional temperature difference $(\Tr-\Tm)/\Tr$ is 
negligible throughout helium recombination ($z\ge 1600$).  This is in accord with the methods and results of Seager et al.\ \cite{2000ApJS..128..407S}.
At low redshifts when hydrogen recombines the matter and radiation temperatures fall out of equilibrium, but that era is not the subject of this paper.

\subsection{Feedback} 
\label{ss:feedback}

Photons that escape a high-lying resonance will readily redshift to a lower-lying line owing to the (nearly) negligible optical depth between lines.
This will excite the lower-lying state, and suppress the overall formation rate of the ground state.  This radiation is manifest as a spectral distortion to the 
thermal spectrum of radiation incident on the blue side of the lines.  (The calculation shares some similarity to \cite{2005PhRvD..72h3002S} where we 
showed that the relic recombination radiation is enough to ionize most of the neutral lithium population in the era following recombination,
except here the radiation is fed back onto recombination.)

Feedback between levels can be calculated in a number of ways.  The most direct way is to simulate a frequency-discretized radiation field along with 
the atomic levels.  In this method, the evolution of the radiation field bins and the evolution of the occupation states of the atoms are solved for 
simultaneously.  This is numerically very difficult.  Because of the huge range of rate scales in the system and the large number of radiation bins 
that it is necessary to track, this method quickly becomes difficult to manage and ill-conditioned.  For this reason, we use an iterative method to 
include feedback in the level code.  This is both practical to implement, and accurate.

The level code calculates the radiation distortions generated by transitions from excited states to the ground state in a first pass, for each redshift 
step, for each species.  The first pass assumes a black-body radiation with local distortions in the Sobolev approximation.  In a second pass, we 
transport the distortion generated by the $(i+1)$th transition to the $i$th transition to the ground state of the same species.  That is, we only 
transport the radiation from the next higher-lying ground-excited transition, in the same species.  (Inter-species feedback between \HeI\ and \HI\ is 
discussed in depth in subsequent sections.)  The distortion is recalculated for each level in the second pass, and these are transported to the lower 
levels and applied to a third pass of the level code.  The iteration continues in this way.  Because the iteration step accounts for much of the total 
feedback effect, subsequent iterations give progressively smaller corrections, and the procedure converges rapidly\comment{; see Fig.~\ref{figs:feedback}}.  The 
typical fractional contribution of the fifth iteration is $| \Delta x_e | \approx 2 \times 10^{-4}$ -- we stop there.  

For hydrogen, the distortion is determined by a simple vacuum transport equation for one resonance.  In helium, the problem is complicated by 
continuous opacity from \HI\ photoionization.  Here, distortion photons are absorbed by \HI\ in flight, and may not redshift down to the lower level.  
This is treated in Sec.~\ref{ss:fbwithcon}.\comment{, and the effects are summarized in Fig.~\ref{figs:tll}.}

The radiation phase space density is convenient for cosmological calculations because it is conserved along a ray, yielding the bound-bound transport 
equation for complete redistribution (here for \HI),
\beqa \dphadnu &=& \tauS \phi(\nu) [ \pha(\nu) - \phaL ] \nonumber \\
        &\approx&  \tauS \phi(\nu) \left[ \pha(\nu) - \frac{x_u g_l }{ x_l g_u}  \right], 
\label{eqn:linetransport} \eeqa
where $\phaL$ is the radiation phase space density that would be in equilibrium with the line,
\beq \phaL = \frac{ x_u }{ x_l (g_u/g_l) - x_u } \approx \frac{ x_u g_l }{ x_l g _u}. \label{eqn:phaline} \eeq
[This is obtained by setting the de-excitation rate $A_{u\rightarrow l}(1+\pha)$ equal to the excitation rate $g_uA_{u\rightarrow l}\pha/g_l$.]
Here we have used the approximation that the upper level is significantly less occupied than the lower level, which is valid when the lower level is the ground state since the energy of the first excitation in \HI, \HeI, or \HeII\ is many times $\kB\Tr$.
The transport equation has the solution
\beq \pha(\nu) = \frac{x_u g_l }{ x_l g_u} - C \exp \left[ -\tauS \int_\nu^\infty \phi(\nu) d \nu \right], \label{eqn:fblinetransportsoln} \eeq
where $C$ is the constant of integration.  This constant is obtained from the initial condition that the far blue side of the line has some phase space density $\pha_+$, which implies
\beq
C = \frac{x_ug_l}{x_lg_u}-\pha_+.
\eeq
Thus the phase space density $\pha_-$ on the red side of the line is
\beq \pha_- = \pha_++\left (\frac{x_u g_l }{ x_l g_u} - \pha_+ \right ) (1-e^{-\tauS}). \label{eqn:linedistortion} \eeq

Notice that as $\tauS$ becomes very large, then the phase space density on the red side of the line approaches $\phaL$.  Because $\pha$ is conserved 
during the transport between \HI\ and \HeII\ resonances (i.e. there is negligible continuum absorption or emission), the phase space density on the 
blue side of the \HI\ Ly$n$ resonance (i.e. $1s$--$np$) is simply the phase space density on the red side of the Ly$(n+1)$ resonance at an earlier 
time:
\beqa
\pha_+\left({\rm Ly}n,z\right) &=& \pha_-\left[{\rm Ly}(n+1), z'\right],
\nonumber \\
z' &=& \frac{1-(n+1)^{-2}}{1-n^{-2}}(1+z)-1,
\eeqa
where $[1-(n+1)^{-2}]/[1-n^{-2}]$ is the ratio of line frequencies.  A similar result holds for \HeII.  Because of the existence of \HI\ continuum 
opacity during \HeI\ recombination, this result does not apply to \HeI; we will treat the \HeI\ problem in Sec.~\ref{sec:heibbbfpihi}.

\section{Hydrogen continuum opacity and helium recombination}
\label{sec:heibbbfpihi}

One of the major issues in recombination physics during the 1990s was whether helium recombines in Saha equilibrium, or is delayed. On 
the one hand, studies by Seager et al.\ \cite{2000ApJS..128..407S} and Matsuda et al.\ \cite{1969PThPh..42..219M,1971PThPh..46..416M} found that there 
is an ``$n=2$ bottleneck'' in which \HeI\ recombines slowly because the two-photon \TPHe\ transition is slow and the \LyaHe\ line is extremely 
optically thick; hence an excited helium atom has only a low probability of reaching the ground state and is most likely to be reionized by the CMB.  
These studies thus found a slower-than-Case B recombination for \HeI\ (where the \LyHe\ transitions are optically thick \cite{1938ApJ....88...52B} and 
processes that depopulate a level through \LyHe\ are ``blocked'' by absorption of the same quanta, on average, thus greatly suppressing electron capture
followed by direct cascade to the ground state).  On the other hand, there is some neutral hydrogen present during the helium 
recombination era, and Hu et al.\ \cite{1995PhRvD..52.5498H} argued that this can speed up helium recombination by absorbing resonance-line and 
continuum photons that would otherwise excite or ionize \HeI.  There appears to be no satisfactory explanation in the literature for why the more 
recent works do not agree, but the difference in CMB spectra is several percent at high $\ell$ \cite{2000ApJS..128..407S} so it is essential that the 
issue be resolved.  This section, as well as Secs.~\ref{sec:anacompletecont} and \ref{sec:partialred}, are devoted to this issue.

There are fundamentally three ways that \HI\ continuous opacity could speed up \HeI\ recombination:
\newcounter{LC}
\begin{list}{\arabic{LC}.\ }{\usecounter{LC}}
\item\label{it:1} Hydrogen can suppress feedback in the \HeI\ lines by absorbing \HeI\ line radiation before it redshifts down to the next line and 
excites a helium atom.
\item\label{it:3} If the \HI\ opacity is very large, it could directly absorb \HeI\ resonance line photons, thus increasing the effective line escape 
probability above its Sobolev value.
\item\label{it:2} Sometimes a helium atom recombines directly to the ground state.  In the case of hydrogen, such recombinations are ineffective 
because the 
emitted photon immediately ionizes another atom.  However in the case of helium, it is possible to produce a neutral helium atom but have the emitted 
photon ionize an \HI\ atom instead of \HeI.  This results in a net recombination of helium.
\end{list}
We treat mechanism \#\ref{it:1} in Secs.~\ref{ss:cont} and 
\ref{ss:fbwithcon}.  The problem of absorption of \HeI\ line photons by \HI\ 
(mechanism \#\ref{it:3}) is more complicated.  The physical picture is outlined in Sec.~\ref{ss:linewithcon}, and it is split into two 
cases.  For the helium intercombination and quadrupole lines, there is negligible coherent scattering within the line since the upper 
level has allowed decays (and allowed pathways to other states) whereas re-emission of the photon to the ground state of \HeI\ is semiforbidden or forbidden.  This case is the 
simplest to consider and it is treated analytically in Sec.~\ref{sec:anacompletecont}.  The other case is that of the allowed \HeI\ 
\LyHe\ lines, in which coherent scattering plays a key role alongside incoherent absorption/emission processes and \HI\ opacity in 
determining the line profile and net decay rate.  This situation is treated via Monte Carlo simulation in Sec.~\ref{sec:partialred}.
Mechanism \#\ref{it:2} produces a negligible effect and so we discuss it in Appendix~\ref{ss:heic}.  We will develop the distinction between 
coherent scattering and incoherent processes much more carefully in Sec.~\ref{ss:linewithcon}~and~Sec.~\ref{ss:setup}.

\subsection{Continuum opacity} 
\label{ss:cont}

The photons emitted in resonant transitions in \HeI\ from excited states to the ground state have energies above the \HI\ photoionization threshold. 
The opacity from photoionization influences transport both within and between \HeI\ lines.  In this section, we describe how the continuum opacity is 
calculated, and Secs.~\ref{sec:anacompletecont} and \ref{sec:partialred} describe details of transport subject to continuous opacity.  Throughout, we 
use $\etaC$ to represent the continuum depth per unit frequency,
\beq \etaC  = \frac{d \tau  }{ d \nu} \biggl |_{\rm continuum} = \frac{ \nH x_{1s} \sigma_{c1} c }{ H \nu}, \label{eqn:gammacdef} \eeq 
where $\sigma_{c1}$ is the photoionization cross-section of neutral hydrogen, and $x_{1s}$ is the ground state occupation fraction (the excited 
states have much lower occupation numbers and lower photoionization cross sections, so we neglect them).  Stimulated 
recombination to H$(1s)$ can be neglected.  The continuum optical depth is also slowly varying as a function of frequency for the \HeI\ ground-resonance 
transitions because the energies are above the \HI\ photoionization threshold and single-electron atoms such as \HI\ posess no multiple-excitation resonances in their cross section.

\begin{figure}[!ht]
\epsfxsize=3.1in
\begin{center}
\epsffile{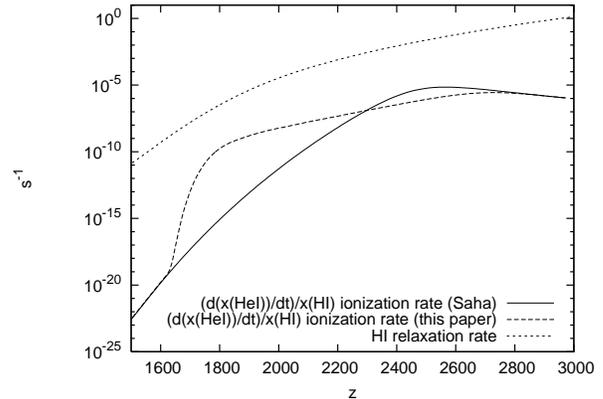}
\end{center}
\caption{The ionization/relaxation timescales of \HI\ during the period of \HeI\ recombination, assuming each \HeI\ recombination generates a photon 
that photoionizes a hydrogen atom.  Here we consider two \HeI\ recombination histories: one in equilibrium and the history derived here 
(Fig.~\ref{figs:mcmodified}).  The ``\HI\ relaxation rate'' is the inverse-timescale $t_{\rm Saha}^{-1}$ (Eq.~\ref{eqn:saharelaxtime}) for \HI\ to 
return to Saha equilibrium if its abundance is perturbed.  In either \HeI\ history, the ionizing radiation from \HeI\ is not sufficient to push \HI\ 
evolution out of Saha equilibrium.}
\label{figs:saha_times}
\end{figure}

In standard recombination theory, the neutral hydrogen population is well-described by the Saha distribution at early times ($z>1700$):
\beq \xHI \approx x_e \xHII \nH \left ( \frac{ h^2 }{ 2 \pi m_e \kB \Tm} \right )^{3/2} e^{\chi_{\rm HI}/\kB \Tm}, \eeq
where $\xHII=1-\xHI\approx 1$.  One concern in using this equation to estimate the effect on helium recombination is that radiation from helium recombination could knock the neutral hydrogen population out of equilibrium.  This would tend to lower the neutral hydrogen population, and decrease the continuum opacity.  In the worst case, each \HeI\ recombination photoionizes a hydrogen atom, giving the characteristic time scale for ionization, $t_{\rm ionize} \approx \xHI/\dot x_{\rm HeI}$.  The Saha relaxation time for this perturbation to fall back to Saha equilibrium is well-approximated by the time 
scale for perturbations to decay in the TLA approximation (neglecting the \HI\ recombination rate at these high redshifts),
\begin{eqnarray}
t_{\rm Saha} &\approx& \frac{\beta_H + \Lambda_{2s\rightarrow1s} + (K\nH\xHI)^{-1} }{ \Lambda_{2s\rightarrow1s} + (K\nH\xHI)^{-1}} \nonumber \\ 
&&\times \frac{1 }{ \beta_H} e^{E_{2p\rightarrow 1s}/\kB T},
\label{eqn:saharelaxtime} 
\end{eqnarray}
where $K=\lambda_{{\rm Ly}\alpha}^3/8\pi H$ is the Peebles $K$-factor \cite{1968ApJ...153....1P} and $\beta_H$ is the effective hydrogen photoionization coefficient from $n=2$ in the three-level approximation.  These are calculated, and shown for 
comparison in Fig.~\ref{figs:saha_times}.  It is evident that a perturbation to the neutral hydrogen population caused by \HeI\ photoionizing radiation 
quickly relaxes back to the Saha evolution.

\begin{figure}[!ht]
\epsfxsize=2.9in
\begin{center}
\epsffile{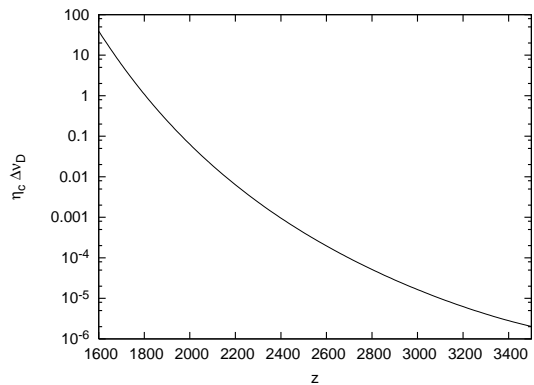}
\end{center}
\caption{The continuum optical depth $d\tau_c/d\nu= \etaC$ times the Doppler width of \HeI\ \LyaHe\ as a function of redshift.}
\label{figs:gammacdelta}
\end{figure}

The unitless quantity $\etaC \Delta \nu_D$ (where $\Delta \nu_D$ is the the \Hefour\ Doppler width) during the \HeI\ recombination era is shown in 
Fig.~\ref{figs:gammacdelta}.

The energy separations between transitions from excited states to the ground state in \HeI\ are typically much greater than the optically thick line 
width.  Thus, radiative transport in \HeI\ can be thought of as taking place through two phases.  In the first, 
continuum processes influence transport within a line.  This modifies the transition rates, and sets the escape probability from a 
given resonance, which may exceed the Sobolev value.  This is described in Sec.~\ref{ss:linewithcon}.  In a second phase, continuum processes influence 
the transport of radiation between resonances, as described in the next section.

\subsection{Feedback from transport between \HeI\ lines and continuous opacity}
\label{ss:fbwithcon}

Section~\ref{ss:feedback} addressed the feedback of a radiation distortion produced by a higher resonance on lower-lying resonances in \HI\ and \HeII.  For \HI\ and \HeII, this transport is in free space in the approximation that the resonances are spaced more widely than their widths.  (This approximation is not entirely valid for \HI\ at late times, however this is not relevant to helium recombination and will be deferred to a later paper.)  In \HeI, the picture is more complicated because transport is subject to opacity from the photoionization of neutral hydrogen.  There is a sufficient neutral population that, when integrated over the photon's trajectory, the feedback between levels can be significantly suppressed.  This is true especially near the end of the \HeI\ recombination and beginning of \HI\ recombination.

The algorithm presented earlier to include feedback iteratively can be easily modified to include feedback suppression between lines: calculate the 
distortions for all resonances, and in the next iteration, multiply them by a suppression factor before applying them to the lower line.  Changing 
variables to redshift, the total depth of the continuum between lines is
\beq \tauLL  = \int_{z_{abs}}^{z_{em}} \frac{ n_{\rm H} x_{1s} \sigma_{c1} c  }{ H} \frac{dz }{ 1+z}. \label{eqn:linelinedepth} \eeq

Let $\pha_i$ be the radiation field on the blue side of the lower line assuming there is no line-line optical depth $\tau_{LL}$.  Then 
the nonthermal distortion produced by the higher state is $\pha_i$ minus the Planck spectrum (since at times earlier than $z \sim 1600$ \HI\ is in Saha 
equilibrium, to a good approximation), which is suppressed by the line-line depth.  The final radiation field just above the frequency of the lower-lying line is 
then
\beq \pha_+ = \left ( \pha_i - \pha_{\rm Pl} \right ) e^{-\tauLL} + \pha_{\rm Pl},
\label{eqn:linelinetransportsoln} \eeq
where the Planck spectrum is $\pha_{\rm Pl}=1/(e^{h\nu/\kB\Tr}-1)$.

\begin{figure}[!ht]
\epsfxsize=3.2in
\begin{center}
\epsffile{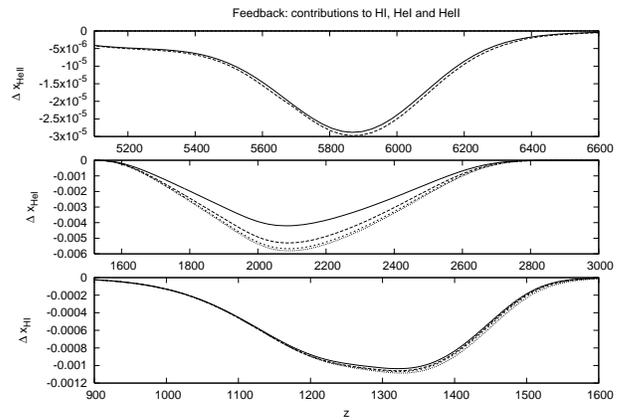}
\end{center}
\caption{A comparison of the effect of feedback of a spectral distortion produced by higher-lying states on lower-lying states in the same species after several iterations, 
relative to the reference model, for \HeII, \HeI, and \HI, where $\Delta x$ is the abundance with feedback minus without feedback.  Here, continuous opacity from 
hydrogen photoionization between \HeI\ is included.  (discussed in Fig.~\ref{figs:tll}).  In all cases, feedback has the effect of retarding the formation of the neutral species.  
A larger number of iterations ($\sim 5$) are needed for \HeI\ recombination than for \HI\ or \HeII.  Here, the uppermost line is the first iteration, moving down with further 
iterations and better convergence.}
\label{figs:feedback}
\end{figure}

\begin{figure}[!ht]
\epsfxsize=3.0in
\begin{center}
\epsffile{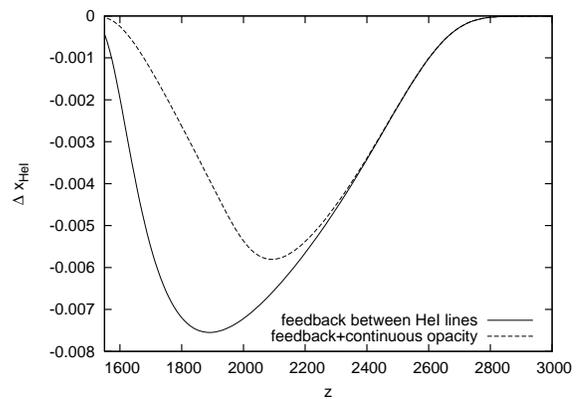}
\end{center}
\caption{The effect of hydrogen continuum absorption on the feedback between transitions to the ground state in \HeI.  Feedback slows \HeI\ 
recombination, but becomes increasingly less significant as more hydrogen recombines, increasing the bound-free opacity and absorbing the distortion.}
\label{figs:tll}
\end{figure}

The effect of feedback is shown in Fig.~\ref{figs:feedback}, and the effect of including $\tau_{\rm LL}$ is shown in Fig.~\ref{figs:tll}.

\subsection{Transport within \HeI\ lines and continuous opacity: physical argument} 
\label{ss:linewithcon}

Previous analyses \cite{1995PhRvD..52.5498H, 2000ApJS..128..407S} identify neutral hydrogen as a potential catalyst that 
could cause \HeI\ recombination to proceed closer to Saha equilibrium.  This is because photons locked in the optically thick \HeI\ 
\LyHe\ lines can ionize neutral hydrogen.  This removes the photons and thus prevents them from re-exciting a helium atom:
\begin{eqnarray}
\mathrm{He}(n^1P^o) &\rightarrow& \mathrm{He}(1^1S) + \gamma, \nonumber \\
\mathrm{H}(1s) + \gamma &\rightarrow& \mathrm{H}^+ + e^-.
\end{eqnarray}
The purpose of this section is to investigate this process and related processes in the \HeI\ intercombination and quadrupole lines.  We will only 
discuss the physical situation here, leaving the detailed calculation to Sec.~\ref{sec:anacompletecont} (for intercombination and quadrupole lines) and 
\ref{sec:partialred} (for permitted lines).  In those sections the objective will be to evaluate the increase in effective escape probability $P_{\rm 
esc}$ (Eq.~\ref{eqn:sobolevescprob}) due to \HI.  Before we do so, however, we must discuss the issue of coherent scattering and its implications for 
what is meant by the ``width'' of a \HeI\ line; we also provide a simple physical explanation for why \HI\ opacity becomes important when it does (i.e. 
at $z\sim 2200$).

\subsubsection{Coherent vs. incoherent processes}

Optically thick lines in cosmological recombination are usually treated by the Sobolev method (see e.g. Sec. 2.3.3 of Ref.~\cite{2000ApJS..128..407S} 
for a good discussion).  This method describes decays from an excited level $u$ of atoms to a lower level $l$ without regard to how the 
atom reached the excited state.  This a good approximation in the case of forbidden lines, and it is also good for very optically thick lines in the 
absence of other sources of opacity (in this case the radiation field in the optically thick part of the line comes to equilibrium with the population 
ratio of $u$ and $l$, and thereafter the details of the radiative transfer matter little).  However in the case of \HeI\ recombination, where \HeI\ and 
\HI\ compete for photons, no equilibrium is established and the resonance line profile matters.  In particular, the frequency distribution of photons 
emitted in e.g. \LyaHe\ need not be a Voigt profile and indeed need not be the same as the absorption profile $\phi(\nu)$.  The reason is that 
sometimes level $u$ is populated by resonant absorption of a photon from level $l$, which subsequently decays back to $l$ without any intermediate 
interactions.  In this case, conservation of energy requires the frequency of the final photon to be the same as the frequency of the initial photon 
in the atom's rest frame, and hence such events will be called ``coherent scattering.'' In the comoving frame the photon's frequency undergoes a small 
fractional shift of order $v/c$, where $v$ is the atomic velocity.  This small frequency shift has a minor influence on recombination,
and will be included in the calculation of Sec.~\ref{sec:partialred}.  For emphasis we will continue to call these scattering events coherent 
with the understanding that coherence is meant to be exact in the atom's frame; this should not lead to any confusion since there is {\em no} type of 
scattering that is coherent in the comoving frame instead of the scatterer's frame.  (Note, however, that the term ``coherent scattering'' applied in the 
comoving frame does appear in literature.)

The process described above differs from ``incoherent scattering'' \cite{1989ApJ...338..594K, 1990ApJ...353...21K}, in which the excited atom undergoes 
other interactions (almost always involving one or more photons) before returning to $l$.  It is only in the latter case that the final photon's 
frequency distribution can be described by a Voigt profile (complete redistribution).  It is also possible for the atom in the excited state $u$ to become ionized, which we will 
consider to be an incoherent process since if the electron later recombines (probably onto another atom) the spectrum of emitted radiation will bear no 
memory of the frequency of the photon that initially excited the atom.  Based on this distinction, one may split the Sobolev optical depth for a line 
into pieces: $\tauS=\tau_{\rm coh}+\tau_{\rm inc}$, depending on the fractions $f_{\rm coh}$ and $f_{\rm inc}$ of photon absorptions in the line that 
go to coherent or incoherent processes, respectively.  Note that by classifying ionization from the excited state as an incoherent process, we ensure 
$f_{\rm inc}+f_{\rm coh}=1$.

The practical implication of this distinction is that when solving for the phase space density $\pha(\nu)$ across an optically thick line, incoherent 
processes can play a dominant role even if $f_{\rm inc}\ll 1$, as is the case for the \HeI\ $2^1P^o$--$1^1S$ line.  This is because incoherent 
scattering can transport a photon from the line center to a far damping wing (or vice versa) in one scattering event, and an excitation followed by 
ionization can remove a photon from the resonance line.  In contrast, coherent scattering can only move a photon far into the damping wings by taking 
many ``baby steps'' using the Doppler shift from the atom's velocity, and it by itself cannot create or destroy a line photon.  We distinguish two 
physically distinct cases here: one where the coherent scattering is negligible (to be studied in detail in Sec.~\ref{sec:anacompletecont}) and a more 
complicated case of partial redistribution where it is not (Sec.~\ref{sec:partialred}).

\subsubsection{When is continuum opacity important?}

\begin{figure}[!ht] 
\epsfxsize=3.4in 
\begin{center} 
\epsffile{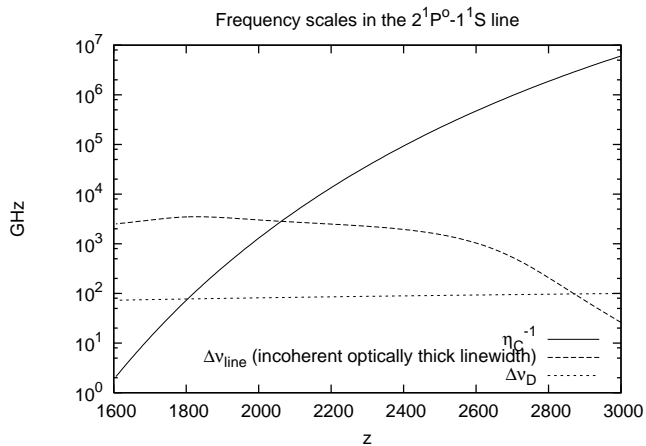} 
\end{center} 
\caption{\label{figs:hydrogen_depth} Inverse of the differential optical depth $\etaC$ from hydrogen photoionization as a function
 of redshift, compared to the optically thick line-width due to incoherent processes in the \LyaHe\ line.  Continuum 
processes start to become important over scales inside the (incoherent) optically thick part within the line around $z=2100$.  Also plotted is the 
Doppler width of the line, emphasizing that the line is optically thick out into the wings.
Continuum processes do not act on scales smaller than the Doppler core until $z<1800$.}
\end{figure}

The \HeI\ \LyaHe\ line is in the extreme UV at $21.2$ eV, where the optical depth from neutral hydrogen photoionization is 
nearly constant in the neighborhood of the resonance.  The continuous opacity from neutral hydrogen is also present in radiation 
transport within the \HeI\ \LyHe\ series, [\HeI] \QuadHe, and \HeI] \InterHe.  Seager et al.\ \cite{2000ApJS..128..407S} concluded that 
this effect is negligible for \LyaHe\ because the neutral hydrogen photoionization rate from photons in the \LyaHe\ resonance is orders 
of magnitude lower than \LyaHe\ photoexcitation rate during the \HeI\ recombination history, due to the sparse population of neutral 
hydrogen. A better criterion for the significance of neutral hydrogen is whether or not continuous opacity affects radiative transport 
within the line.  The two natural scales in the transport problem are: (i) the frequency $\etaC^{-1}$ that a photon can be expected to 
traverse by redshifting before it is absorbed in a hydrogen photoionization event, and (ii) the range of frequencies $\Delta\nu_{\rm 
line}$ over which the line is thick to incoherent scattering/absorption.  If $\etaC^{-1}\gg\Delta\nu_{\rm line}$ then helium atoms 
will re-absorb the resonance radiation from other helium atoms, without regard to the \HI\ abundance.  If, on the other hand, 
$\etaC^{-1}\le\Delta\nu_{\rm line}$, then \HI\ can destroy photons that would otherwise have re-excited helium atoms, and thereby 
accelerate \HeI\ recombination.  We thus care about the continuum optical depth within a line,
\beq
\tau_{\rm C} = \etaC \Delta \nu_{\rm line}.
\label{eqn:tauin}
\eeq
In the case of lines that are optically thick into the damping wings, such as the \HeI\ \LyHe\ lines, $\Delta\nu_{\rm line}$ may be calculated by integrating the asymptotic line profile $\phi(\nu)\sim \Gamma_{\rm line}/4\pi^2\Delta\nu^2$ until the optical depth in the wing becomes unity:
\beq
\Delta\nu_{\rm line} = \frac{\Gamma_{\rm line} \tau_{\rm inc} }{ 4 \pi^2}.
\eeq
Here $\Gamma_{\rm line}$ is the Lorentz width of the line, $\tau_{\rm inc}$ is the Sobolev optical depth through the line from 
incoherent processes (i.e. all absorption processes other than coherent scattering; this will be calculated in 
Sec.~\ref{sec:partialred}), and $\etaC$ is the differential continuum optical depth, 
Eq.~(\ref{eqn:gammacdef}).  Fig.~\ref{figs:hydrogen_depth} compares the optically thick linewidth (from incoherent processes) 
to the inverse differential optical depth $\etaC^{-1}$ and the Doppler width for the \LyaHe\ transition in \Hefour.
The line may be optically thick in the Sobolev sense out to much larger frequency 
separations due to coherent scattering, but a coherent scattering event results in no net change in the atomic level 
populations and hence does not directly affect recombination.  (It only has an indirect effect by changing the radiation spectrum.)

The general problem of the escape probability including the continuum opacity and coherent scattering as well as incoherent 
emission/absorption processes is quite complicated.  Therefore we will solve it in two steps.  In Sec.~\ref{sec:anacompletecont} we will 
solve the problem {\em without} coherent scattering.  This is a conceptually simple problem -- all one has to do is compute the 
probability of a photon either redshifting out of the line or being absorbed by \HI\ before being re-absorbed by \HeI\ -- and the 
machinery for solving it has already been developed for the theory of line transfer in stellar winds \cite{1985ApJ...293..258H}.  
Despite its simplicity, the solution in Sec.~\ref{sec:anacompletecont} is an accurate description of the intercombination and quadrupole 
lines because there is negligible coherent scattering in these lines.  (The reason is that if an atom absorbs a photon in these lines and reaches a 
$n^3P^o$ or $n^1D$ level, its next step is almost always to undergo one of the allowed decays rather than to emit a photon in the intercombination or 
quadrupole line and go back to the ground state.)  In Sec.~\ref{sec:partialred} we address the problem {\em with} coherent scattering as well as 
incoherent processes and continuum opacity, and describe its solution via a Monte Carlo simulation.

\section{The modified escape probability without coherent scattering} 
\label{sec:anacompletecont}

This section is concerned with calculating the net decay rate in \HeI\ lines with continuous opacity (from \HI) but no coherent 
scattering.  This is a line radiative transfer problem with ``complete redistribution'' in the sense that a photon emitted in the line 
has a frequency distribution given by the intrinsic line profile, independent of the spectrum of incident radiation:
\beq p(\nuout  | \nuin ) = \phi(\nuout ). \eeq

There is negligible coherent scattering in the \HeI] \InterHe\ and [\HeI] \QuadHe\ lines (see Sec.~\ref{ss:linewithcon}), and results in this section 
are readily applied to those lines.  The macroscopic picture is that in emission, a line in the gas always has the same shape, regardless of the 
excitation field.  Throughout, $\phi$ is Voigt-distributed and accounts for the natural resonance width of the line and Doppler broadening from thermal 
motion, set by the matter temperature.  In Sec.~\ref{sec:partialred} we discuss the case of a line in a gas where some fraction of the photons are 
re-emitted coherently (with the same frequency as the incoming photon in the rest frame of the atom) and some are reemitted incoherently.

The fundamental quantity that determines the transition rate between a lower level $l$ and upper level $u$ is the radiation phase space 
density integrated over the atomic absorption profile, $\bar \pha (\nu_{ul})$.  In the case of complete redistribution through incoherent 
scattering in the presence of continuous opacity, the radiation phase space density evolves as
\beq
\dot\pha = \dot\pha_{\rm Hubble}
+ \dot\pha_{\rm cont}
+ \dot\pha_{\rm inc}.
\eeq
(We assume that the line width is small compared to $\nu_{ul}$ so that factors of $\nu/\nu_{ul}$ appearing in the transport equation can be dropped.)
In this section, we will develop each of these terms and solve the transport equations for the radiation profile over the line and the escape probability.
Of these terms, the Hubble redshifting term is,
\beq
\dot\pha_{\rm Hubble}=H\nu_{ul}\dphadnu,
\label{eq:r1}
\eeq
We will work in the steady-state approximation, where $\dot \pha = 0$; possible corrections to this are considered in Paper II.  Then, frequency provides 
a convenient domain over which to solve for $\pha$, moving the Hubble term and dividing by $H \nu_{ul}$.  Breaking up $\dot\pha_{\rm cont}$ and $\dot\pha_{\rm inc}$
into emission and absorption pieces, there are four terms that appear in the steady-state equation for the phase space density: absorption and emission by continuum processes and 
emission and absorption by incoherent processes in the line.  We may write these as
\beqa
\dphadnu
&=& \left(\dphadnu \right)_{\rm cont-abs}
+ \left(\dphadnu \right)_{\rm inc-abs}
\nonumber \\&&
+ \left(\dphadnu \right)_{\rm cont-em}
+ \left(\dphadnu \right)_{\rm inc-em}.
\eeqa
The continuum absorption term has already been determined (Eq.~\ref{eqn:gammacdef}) and is $\etaC\pha(\nu)$.  The line absorption term 
depends on the line profile and is $\tauS\phi(\nu)\pha(\nu)$ since $\tauS\phi(\nu)$ is the optical depth per unit frequency.  The continuum emission term is
\beq
\left(\dphadnu \right)_{\rm cont-em} = -\etaC\pha_{\rm C},
\eeq
where $\phaC$ is the phase space density of photons that would be required for the reaction
\beq
{\rm H}(1s) + \gamma \leftrightarrow {\rm H}^++e^-
\eeq
to be in equilibrium.  The usual equilibrium constant argument shows that $\pha_{\rm C}\propto n_{\rm HII}n_e/n_{\rm HI}$; the proportionality constant can be determined from the principle of detailed balance as
\beq \phaC(\nu) = \left ( \frac{n_e n_c }{ n_i} \right ) \left ( \frac{ n_i }{ n_e n_c} \right )^{\rm LTE} e^{-h\nu/\kB \Tm} .
\label{eqn:phacontinuum} \eeq
Since we have seen that hydrogen is in Saha equilibrium during the relevant era, and that $\Tm=\Tr$, we have $\phaC(\nu)=e^{-h\nu/\kB\Tr}$.  The \HeI] \InterHe\ and 
[\HeI] \QuadHe\ lines are at very high energies ($\ge 20.6\,$eV), so the photon phase space density $\pha\ll1$ and we can neglect stimulated emission 
and other consequences of the photon's bosonic nature.  The line emission term is
\beq
\left(\dphadnu \right)_{\rm inc-em} = -\tauS\phi(\nu)\pha_{\rm L},
\eeq
where $\pha_{\rm L}$ is the phase space density that would exist if only the line processes were important.  This can also be determined 
by setting the excitation rate $x_l(g_u/g_l)A_{u\rightarrow l}\pha_{\rm L}$ equal to the de-excitation rate $x_uA_{u\rightarrow l}(1+\phaL)$:
\beq \phaL = \frac{ x_u }{ x_l (g_u/g_l) - x_u}. \label{eqn:phalinefreq} \eeq
The transport equation is thus
\beq \dphadnu = \etaC (\pha-\phaC)+\tauS \phi(\nu) (\pha-\phaL). \label{eqn:inccontransportproblem} \eeq
It has the general solution:
\beq \pha(\nu)=e^{\Phi_1(\nu)} \left \{ C - \int_{\nu_2}^\nu [\phaC \etaC + \phaL \tauS \phi(\tilde \nu)] e^{-\Phi_1(\tilde \nu)} d \tilde \nu \right \}, \label{eqn:inccontransportsoln} \eeq
where
\beq \Phi_1(\nu) = \int_{\nu_1}^\nu [\etaC + \tauS \phi(\tilde \nu)]d \tilde \nu. \label{eqn:bigphiintegral} \eeq
Here $\nu_1$ is an arbitrary but fixed frequency, and $\nu_2$ is the frequency at which we set the initial condition.
It is convenient to expand Eq.~(\ref{eqn:inccontransportsoln}) into the pieces that depend linearly on the constant $C$ and on $\phaC$ and $\phaL$:
\beq
\pha(\nu) = CI_i(\nu) + \phaC I_C(\nu) + \phaL I_L(\nu),
\eeq
where the individual profiles are
\beqa
I_i(\nu) &=& e^{\Phi_1(\nu)},
\nonumber \\
I_C(\nu) &=& -e^{\Phi_1(\nu)} \int_{\nu_2}^\nu \etaC e^{-\Phi_1(\tilde \nu)} d \tilde \nu , {\rm ~and}
\nonumber \\
I_L(\nu) &=& -e^{\Phi_1(\nu)} \int_{\nu_2}^\nu \tauS \phi(\tilde \nu) e^{-\Phi_1(\tilde \nu)} d \tilde \nu.
\eeqa

\begin{figure}[!ht]
\epsfxsize=3.4in
\begin{center}
\epsffile{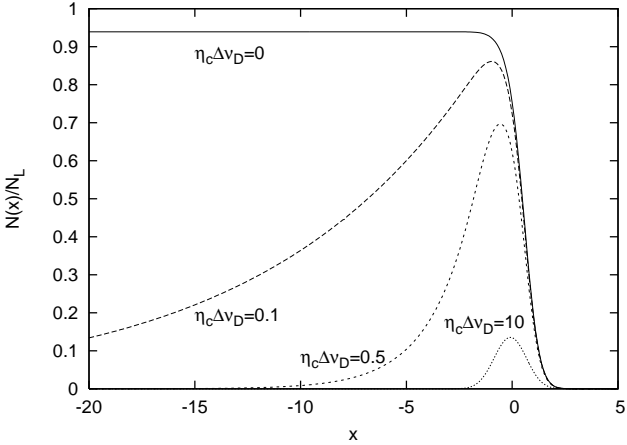}
\end{center}
\caption{The radiation profile near the intercombination line $1^1S_0 
\rightarrow 2^3P^o_1$ resonance with Voigt parameter $a=\Gamma_{\rm line}/(4 \pi 
\Delta\nuD) = 10^{-5}$ for the Sobolev optical depth $\tauS=2.8$ and $\phaC=0$, for several 
sample continuum optical depths $\etaC \Delta\nuD$, showing the effect of 
continuous opacity.  Because of the low optical depth and small natural 
line width, very little radiation extends more than three doppler widths 
above the line, and the effect of the continuum is significant in relaxing 
the radiation phase space density near line-center.}
\label{figs:intercomlineprofile}
\end{figure}

We integrate this phase space density over the profile, and break the integral into three pieces that emphasize the physical processes:
\beq \bar \pha = C \bar I_i + \phaC \bar I_C + \phaL \bar I_L. \label{eqn:barphafromintegrals} \eeq
Here the overbar denotes averaging over the line profile, e.g.
\beq \bar \pha = \int_{-\infty}^{\infty} \phi(\nu) \pha(\nu) d \nu. \label{eqn:phaintegral} \eeq
Bringing the outer exponent under the integral in the expression of $\bar I_L$:
\begin{eqnarray}
\bar I_L &=& -\int_{-\infty}^{\infty} \phi (\nu) \int_{\nu_2}^\nu \tauS \phi(\tilde \nu) \nonumber \\
 &&\times \exp \left \{ - \int_\nu^{\tilde \nu} [\etaC + \tauS \phi(y)]dy \right \}  d \tilde \nu  d\nu.
\label{eqn:lineintegralexpression}
\end{eqnarray}
We will take the starting frequency to be on the far blue side of the line ($\nu_2 \rightarrow \infty$), appropriate for expanding media 
\cite{1992ApJ...387..248H}.  With this choice, it is convenient to switch the order of integration and absorb the leading minus sign.  
The $\bar I_L$ line integral is related to the original Sobolev problem by setting $\etaC=0$,
\beq \bar I_L \biggr |_{\etaC =0} = 1- \frac{1-e^{-\tauS} }{ \tauS}= 1- \PS. \label{eqn:lineintegralsobolev} \eeq
Simply, $\PS = 1 - \bar I_L (\etaC = 0)$.  We follow Rybicki and Hummer \cite{1985ApJ...293..258H} in then calculating the difference in the line integral with 
and without continuum absorption (which is related to the probability of absorption between incoherent scattering events, $\PC$):
\begin{eqnarray}
\Delta \bar I_L \!\!&=&\!\! \bar I_L - \bar I_L \biggr |_{\etaC =0} = -\int_{-\infty}^{\infty} \phi(\nu) \int_{\nu_2}^\nu \tauS \phi(\tilde 
\nu) \label{eqn:absorptionprob} \nonumber \\ &&\!\!\times \exp \left \{ -\tauS \int_\nu^{\tilde \nu} \phi(y)dy \right \} \left [ 
e^{-\etaC (\tilde \nu - \nu)} -1 \right ] d\tilde \nu d\nu.
\nonumber \\ &&
\label{eq:pc}
\end{eqnarray}
We integrate this numerically over a Voigt profile to find the probability of absorption by a continuum process.  The Voigt profile and 
its integral have a wide dynamic range--details of the evaluating the integral are described in Appendix~\ref{ss:integraldetails}.  The main goal is to find the integral of the radiation phase space density over the line profile ($\bar 
\pha$ in Eq.~\ref{eqn:barphafromintegrals}), and the modified escape probability.  Examples of line radiation profiles with incoherent scattering and continuum opacity are shown in Fig.~\ref{figs:intercomlineprofile}.

The overall value of $\bar\pha$ is required in order to compute excitation and de-excitation rates.  With the boundary $\nu_2\rightarrow\infty$, we have $\Phi_1(\nu_2)\rightarrow\infty$ and hence $C\rightarrow 0$, while $\bar I_i$ remains finite at fixed $\nu_1$.  Therefore
\beq
\bar\pha = \phaC\bar I_C+\phaL\bar I_L.
\eeq
Now, if we had $\phaC=\phaL$, the solution to Eq.~(\ref{eqn:inccontransportproblem}) would be simply $\pha=\phaL$ and hence $\bar\pha=\phaL$.  Thus
$\bar I_C+\bar I_L=1$ and hence
\beqa
\bar\pha &=& \phaC\bar I_C + \phaL(1-\PS+\Delta\bar I_L)
\nonumber \\ &=& \phaC(\PS-\Delta\bar I_L) + \phaL(1-\PS+\Delta\bar I_L).
\label{eqn:barphaprob}
\eeqa

The net downward transition rate from $u$ to $l$ is
\beq
\dot x|_{\rm line} = A_{u\rightarrow l}\left[ x_u(1+\bar\pha) - \frac{g_u}{g_l}x_l\bar\pha\right].
\eeq
For $x_u\ll x_l$ and $\bar\pha\ll 1$, this can be re-expressed in terms of $\phaL$ as
\beq
\dot x|_{\rm line} = A_{u\rightarrow l} \frac{g_u}{g_l}x_l (\phaL-\bar \pha),
\eeq
from which we find
\beqa \dot x|_{\rm line} \!\!&=&\!\!
A_{u\rightarrow l}\frac{g_u}{g_l}x_l(\PS-\Delta\bar I_L)(\phaL-\phaC)
\nonumber \\ &=&\!\! \Pesc A_{u\rightarrow l} \left( x_u - \frac{g_u }{
g_l} x_l \phaC \right), \label{eqn:incconrateequation2} \eeqa
where $\Pesc = \PS - \Delta\bar I_L$.  Note that $\Delta\bar I_L$ is manifestly negative according to Eq.~(\ref{eq:pc}), so continuum opacity enhances the escape probability.  The 
transition rate from from level $u$ to $l$ is then set by the occupations 
of the states, and by the radiative escape probability, $\Pesc$.  

Before continuing, we note one subtlety of our analysis.  We have assumed that the continuum on the blue side of the line is optically thick to \HI\ photoionization, or equivalently, we neglected feedback from unabsorbed spectral distortions from a higher-frequency line. [Formally this was done when we argued that $\Phi_1(\nu_2)\rightarrow\infty$.]  This is not as restrictive an assumption as it seems (even though it is violated at the beginning of \HeI\ recombination), because the widths of the \HeI\ lines are small compared to their separation.  Thus if the continuum opacity between lines is not large, then the continuum opacity {\em within} a line can be neglected, in which case we know that the correct escape probability is the Sobolev result, $\Pesc=\PS$ (and the converse, where opacity is important within the line only once with transport of the distortion between lines is suppressed). Since our solution here for $\Pesc$ reduces to the Sobolev optical depth in the limit where continuum opacity within the line is turned off, it follows that using this $\Pesc$ in the level code in the form
\beq \dot x|_{\rm line} = \Pesc A_{u\rightarrow l} \left( x_u - \frac{g_u }{
g_l} x_l \pha_+ \right), \label{eqn:incconrateequation} \eeq
i.e. replacing $\phaC$ in Eq.~(\ref{eqn:incconrateequation2}) with $\pha_+$, will recover the correct solution in both the case where continuum opacity within the line is important (where we have $\pha_+=\phaC$) and the case where it is not.

The total escape probability is too time consuming to calculate on a 
case-by-case basis, as part of the level code.  Instead, we calculate a 
lookup table of modified escape probabilities as a function of the \HeI\ 
ground state fraction and redshift, assuming that the neutral hydrogen 
population is in Saha equilibrium during \HeI\ recombination.  
The details of this calculation are given in Appendix~\ref{ss:integraldetails}.  
The Monte Carlo methods developed in Sec.~\ref{ss:mcmethods} also 
generate a grid of modified escape probabilities as a function of redshift 
and the occupation of the \HeI\ ground state.  These are log-interpolated 
and applied in to the recombination level code, as described in Sec.~\ref{ss:levelcoderes}.

\section{The modified escape probability including coherent scattering}
\label{sec:partialred}

In the previous section, we considered the problem of calculating the decay rate of \HeI\ lines with no coherent scattering.  This section aims to 
include coherent scattering in the problem, which is necessary in order to handle the Lyman-like series of lines, \HeI\ \LyHe.  Coherent scattering 
complicates the problem because the frequency distribution of photons emitted in the line depends on the existing spectrum of radiation (photons are 
not completely redistributed across the line).  Our approach to this problem will be to realize that a helium atom reaches the upper level $n^1P_1^o$ 
by emitting or absorbing a photon, and then it leaves this level by emitting or absorbing a photon.  Therefore, all processes involving this level can 
be regarded as a form of resonant two-photon absorption, resonant two-photon emission, resonant Raman scattering, and resonant Rayleigh scattering.  
Physically one should write down the rates for these processes and solve the relevant level population/radiative transfer problem.

\subsection{Setup}
\label{ss:setup}

The goal here is to replace the usual treatment of the line through one-photon processes by an inherently two-photon treatment (two-photon absorption, two-photon emission, and scattering) in which $u$ represents the intermediate state.  In the case of each \HeI\ \LyHe\ line, we will denote the lower level by $l=1^1S$ and the upper level by $u=n^1P^o$.  The rates for two-photon processes in the vicinity of resonance can be expressed in terms of the constituent one-photon absorption and emission processes, and depend in particular on the branching fractions that determine the fate of a helium atom in level $u$.  Throughout, we will make several assumptions about the radiation field and rates during recombination:
\begin{enumerate}
\item $\pha(\nu_{ul})\ll 1$ implies that stimulated emission in \HeI\ $n^1P^o$--$1^1S$ lines can be neglected.  (This is because $h\nu_{ul}\gg\kB\Tr$ 
so the Wien curve has $\pha\ll1$, and the spectral distortion raises $\pha$ to at most $\sim x_{n^1P^o}/3x_{1^1S}\ll 1$.)
\item The transitions from $u$ to other excited states (or the continuum) see approximately a blackbody spectrum and are Sobolev optically thin.
\item The time for $x_u$ to change significantly is much longer than the lifetime of the state, so we can work in a ``steady state".
\end{enumerate}

\begin{figure}[!ht]
\epsfxsize=3.4in
\begin{center}
\epsffile{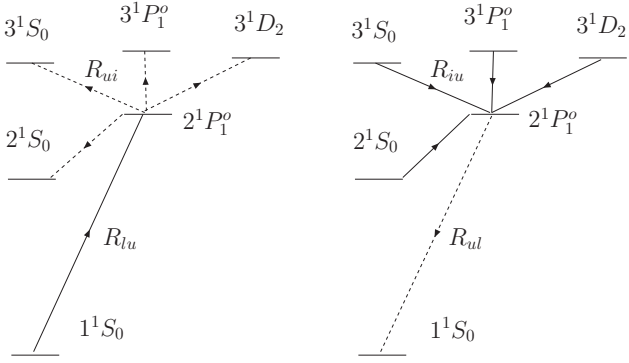}
\end{center}
\caption{The transition \LyaHe\ interpreted as a two-photon process with $2^1P^o_1$ as an intermediate resonant state.  In an incoherent scattering through 
\HeI~~$1^1S_0 \leftrightarrow 2^1P^o_1$, the incoming photon excites $1^1S_0 \rightarrow 2^1P^o_1$.  The atom absorbs a second photon and explores several 
intermediate states before decaying through \LyaHe, emitting a \LyaHe\ photon with complete redistribution.  In coherent scattering, only $1^1S_0$ and 
$2^1P^o_1$ are be involved, and the outgoing photon is emitted with the same energy as the incoming photon in the rest frame of the atom.} 
\label{figs:incoherentdiagram}
\end{figure}

An atom in level $u$ has three possible fates.  It could decay to the ground level $l$ with the rate $A_{u\rightarrow l}$ (we neglect stimulated 
emission 
in the \LyaHe\ line under assumption 1).  It may also decay to another level $a$ with lower energy $E_a<E_u$, with rate 
$A_{u\rightarrow a}[1+\pha(\nu_{ua})]$.  A third possibility is that an atom in level $u$ could absorb a photon and transit 
to an even higher level $b$ with rate $A_{b\rightarrow u}(g_b/g_u)\pha(\nu_{bu})$.  (Note that $b$ could be a continuum level.)  The overall width of the level $u$ is then
\beqa
\Gamma_u \!\!&=&\!\! A_{u\rightarrow l} + \sum_{a<u,a\ne l} A_{u\rightarrow a}[1+\pha(\nu_{ua})]
\nonumber \\ && \!\!
 + \sum_{b>u} A_{b\rightarrow u}\frac{g_b}{g_u}\pha(\nu_{bu}).
\label{eq:gammau}
\eeqa
(Note that the summation over $b$ includes an implied integration over continuum states.)  It is convenient to define the rates
\beq
R_{ui} = \left\{\begin{array}{lcl}
A_{u\rightarrow i}[1+\pha(\nu_{ui})] & & E_i<E_u \\
A_{i\rightarrow u}(g_i/g_u)\pha(\nu_{iu}) & & E_i>E_u
\end{array}\right.
\eeq
so that $\Gamma_u = A_{u\rightarrow l} + \sum_iR_{ui,i \ne l}$.  The fractional contribution to this width from a final level will be denoted 
$f_i=R_{ui}/\Gamma_u$, where we identify $R_{ul}=A_{u\rightarrow l}$.  This should be interpreted as the probability that $u$ will transit
to $i$, given all of its options.  The radiative rate for one-photon radiative transitions from some level $i\ne l$ 
to $u$ is
\beq
R_{iu} = \left\{\begin{array}{lcl}
A_{u\rightarrow i}(g_u/g_i)\pha(\nu_{ui}) & & E_i<E_u \\
A_{i\rightarrow u}[1+\pha(\nu_{iu})] & & E_i>E_u
\end{array}\right..
\eeq
The radiative rate for one-photon excitation from $l$ to $u$ is similar, except that one must average the photon phase space density over the line 
profile because the phase space density may be strongly frequency-dependent (unlike the $u\leftrightarrow e$ transitions where one is dealing with a blackbody 
radiation field):
\beq
R_{lu} = A_{u\rightarrow l}\frac{g_u}{g_l}\bar\pha.
\label{eq:rlu}
\eeq
Figure \ref{figs:incoherentdiagram} summarizes the rates $R_{lu}$, $R_{lu}$, $R_{lu}$, and $R_{lu}$ for the lower excitation states of helium.

It is straightforward to write down the net rate of production (or destruction) of each level via two-photon processes involving the excited state 
$u$, in terms of the rates (Eqs.~\ref{eq:gammau}--\ref{eq:rlu}).  It is simply the difference of destruction rate of level $i$ via transition to $u$, 
and the rate of production of $i$ from all modes involving $u$:
\beqa
\dot x_i|_u &=& -R_{iu}x_i + \sum_{j\neq u} x_jR_{ju}f_i
\nonumber \\
&=& -R_{iu}x_i + R_{ui}\Gamma_u^{-1}\sum_{j\neq u} x_jR_{ju}.
\label{eq:xiu}
\eeqa
If we knew exactly the rate coefficients $R_{ul}$ and $R_{lu}$, Eq.~(\ref{eq:xiu}) would be easy to incorporate into the level codes.  That is because 
we can simply write the steady-state population of $u$ as its production rate multiplied by its lifetime $\Gamma_u^{-1}$,
\beq
x_u = \Gamma_u^{-1}\sum_{j\neq u} x_jR_{ju}.
\label{eq:xu}
\eeq
Then for $i\neq l,u$, Eq.~(\ref{eq:xiu}) is exactly equivalent to the standard rate equation
(Eq.~\ref{eqn:sobolevapprox}).  

The rate equation for the ground level is modified to become
\beq
\dot x_l|_u = -R_{lu}x_l + R_{ul}x_u,
\label{eq:xlu}
\eeq
and that for $u$ becomes
\beq
0 = -\sum_{i\neq u}R_{ui}x_u + \sum_{i\neq u}R_{iu}x_i.
\label{eq:xuu}
\eeq
These two are also similar to Eq.~(\ref{eqn:sobolevapprox}) except that the left-hand side of Eq.~(\ref{eq:xuu}) is zero, as appropriate for a 
steady-state solution, and the rates $R_{ul}$ and $R_{lu}$ may differ from the Sobolev values.  Indeed, the {\em only} modification needed in the 
level code is that these rates need to be replaced with the values $A_{u\rightarrow l}$ and Eq.~(\ref{eq:rlu}).  The only nontrivial part of the calculation is 
to compute $\bar\pha$ appearing in Eq.~(\ref{eq:rlu}).  The calculation of $\bar\pha$ will occupy the rest of this section.

In some parts of this section we will define the additional branching fractions $f_{\rm coh}\equiv f_l$ and $f_{\rm inc}\equiv 1-f_l$.  
These represent the branching fractions for absorption of a line photon to result in coherent scattering ($f_{\rm coh}$) or incoherent 
scattering ($f_{\rm inc}$).  We also use the portions of the Sobolev optical depth due to these processes, $\taucoh\equiv \tauS 
f_l$ and $\tauinc\equiv \tauS(1-f_l)$.

\subsection{The equation of radiative transfer}
\label{ss:eqnradtrans}

In order to compute $\bar\pha$, we need to construct and solve the radiative transfer equation for the photon phase space density in the 
vicinity of the line, $\pha(\nu)$.  The line profile evolves according to four effects: the Hubble redshifting of the photons; absorption and emission 
in the \HI\ continuum; coherent scattering; and incoherent emission/absorption processes (i.e. resonant Raman scattering and two-photon 
emission/absorption, whose line profiles do not depend on the incident radiation field).  Schematically, we write
\beq
\dot\pha = \dot\pha_{\rm Hubble}
+ \dot\pha_{\rm cont}
+ \dot\pha_{\rm coh}
+ \dot\pha_{\rm inc}.
\eeq
The continuum contribution was solved in Sec.~\ref{sec:anacompletecont}:
\beq
\dot\pha_{\rm cont}=-H\nu_{ul}\etaC(\pha-\phaC).
\label{eq:r1a}
\eeq
A photon has a probability per unit time of undergoing a coherent scatter given by $H\nu_{ul}\taucoh\phi(\nu)$, where $\taucoh=\tauS f_l$, the Sobolev optical depth to any type of absorption times $f_l$, where $f_l$ is the fraction of photon absorptions that are followed directly by emission, $\phi(\nu)$ is the line profile in units of fraction of the integrated profile traversed per unit time (s$^{-1}$), and $H\nu_{ul}$ converts this to the fraction of the integrated profile traversed per unit frequency (Hz$^{-1}$).  Thus we have
\beqa
\dot\pha_{\rm coh}(\nu) \!\! &=& \!\! -H\nu_{ul}\taucoh\phi(\nu)\pha(\nu)
\nonumber \\ && \!\! + H\nu_{ul}\taucoh\int \phi(\nu')\pha(\nu')p(\nu|\nu')\,d\nu'.
\label{eq:r-coh}
\eeqa
Here $p(\nu|\nu')$ is the probability distribution for the outgoing frequency $\nu$ of a coherently scattered photon conditioned on its 
ingoing frequency $\nu'$.  (This is commonly known as a redistribution function, and the relevant case here is of the $R_{\rm II}$ type 
\cite{1962MNRAS.125...21H}.)  It satisfies $\int p(\nu|\nu') \,d\nu=1$.

Finally we come to incoherent processes, $\dot\pha_{\rm inc}$.  The probability per unit time of incoherent 
scattering (i.e. excitation of an atom to $u$ followed by transit to a state other than $l$) is $H\nu_{ul}\tauinc\phi(\nu)$.  The 
rate of incoherent emission processes (two-photon emission or resonant Raman scattering, $i\rightarrow u\rightarrow 
l$, with $i\neq l$) per H nucleus per unit time is $\sum_i x_i R_{iu}f_l$.  Thus we have
\beqa
\dot\pha_{\rm inc}(\nu) &=& -H\nu_{ul}\tauinc\phi(\nu)\pha(\nu)
\nonumber \\ &&
+ \frac{\nH c^3}{8\pi\nu_{ul}^2}\sum_{i\neq l} x_i R_{iu}f_l \phi(\nu).
\label{eq:temp-inc}
\eeqa
(Here $\nH c^3/8\pi\nu_{ul}^2$ is a conversion factor: it is the number of H nuclei in a volume containing one photon mode per unit frequency at 
$\nu_{ul}$.)  We note that $f_l=A_{u\rightarrow l}/\Gamma_u$, and using $x_u\ll x_l$, we may write
\beqa
\frac{\nH c^3}{8\pi\nu_{ul}^2}f_l &=& \frac{\nH c^3 A_{u\rightarrow l}}{8\pi\nu_{ul}^2}\Gamma_u^{-1}
= \frac{H\nu_{ul}\tauS}{x_l(g_u/g_l)-x_u}\Gamma_u^{-1}
\nonumber \\
&\approx&\frac{H\nu_{ul}g_l\tauS}{g_ux_l\Gamma_u}.
\eeqa
This turns Eq.~(\ref{eq:temp-inc}) into
\beq
\dot\pha_{\rm inc}(\nu) = H\nu_{ul}\tauinc\phi(\nu)[\phaLO-\pha(\nu)],
\label{eq:r2}
\eeq
where we have defined
\beq
\phaLO = \frac{g_l}{g_ux_l\Gamma_uf_{\rm inc}} \sum_{i\neq l} x_i R_{iu}.
\label{eq:phalo-def}
\eeq
Note that $\phaLO$ is not necessarily equal to the phase space density of photons in equilibrium with the line, which would be 
$\phaL=g_lx_u/g_ux_l$ for $x_u\ll x_l$.  However, the two are related.

In steady state one may combine Eqs.~(\ref{eq:r1}), (\ref{eq:r1a}), (\ref{eq:r-coh}) and (\ref{eq:r2}) and set $\dot\pha\rightarrow0$ to yield
\beqa
\dphadnu &=& 
\etaC[\pha(\nu)-\phaC]
\nonumber \\ &&
+\taucoh\left[\phi(\nu)\pha(\nu)-\int \phi(\nu')\pha(\nu')p(\nu|\nu')\,d\nu'\right]
\nonumber \\ &&
+\tauinc\phi(\nu)[\pha(\nu)-\phaLO].
\label{eq:r3}
\eeqa
The parameters $\phaC$ and $\phaLO$ may be eliminated from this equation via the introduction of the variable
\beq
\xi(\nu) = \frac{\pha(\nu)-\phaC}{\phaLO-\phaC},
\label{eq:xi-pha}
\eeq
or
\beq
\pha(\nu) = \phaC + (\phaLO-\phaC)\xi(\nu).
\label{eq:pha-xi}
\eeq
Expressing Eq.~(\ref{eq:r3}) in terms of $\xi(\nu)$, we find
\beqa
\frac d{d\nu}\xi(\nu) &=&
\taucoh\left[\phi(\nu)\xi(\nu)-\int \phi(\nu')\xi(\nu')p(\nu|\nu')\,d\nu'\right]
\nonumber \\ &&
+ \etaC\xi(\nu) + \tauinc\phi(\nu)[\xi(\nu)-1].
\label{eq:r4}
\eeqa
The boundary condition is that due to formally infinite continuum optical depth if we integrate to $\nu=+\infty$, we have $\xi_+=0$ (where again $\xi_+$ is the value of $\xi$ on the blue side of the line).  This may fail if the optical depth between lines $\tauLL$ is not much larger than 1; the correction for this is discussed at the end of Sec.~\ref{ss:relationphalo}.

We will concern ourselves with solving Eq.~(\ref{eq:r4}) in Sec.~\ref{ss:mcmethods} using a Monte Carlo method.  
However before we do this we will investigate the 
implications of the solution by relating the value of $\xi(\nu)$ averaged across the line,
\beq
\bar\xi\equiv\int\xi(\nu)\phi(\nu)\,d\nu,
\label{eq:barxi}
\eeq
to the value of $\phaLO$ and to the net decay rate $\dot x_l|_u$, which is the quantity we need to know in order to 
solve helium recombination.

\subsection{Relation of $\phaLO$ and $\bar\xi$ to decay rate} 
\label{ss:relationphalo}

The objective of this section is to determine the rate $\dot x_l|_u$ in terms of quantities such as $x_l$, $x_u$, and $f_l$ that can be 
computed easily in the level code, and the quantity $\bar\xi$ that emerges from the solution to Eq.~(\ref{eq:r4}).

As noted at the end of Sec.~\ref{ss:setup}, the upward and downward transition rates in the $l\leftrightarrow u$ line are 
$R_{lu}=(g_u/g_l)A_{u\rightarrow l}\bar\pha$ 
and $R_{ul}=A_{u\rightarrow l}$ respectively.  Averaging Eq.~(\ref{eq:pha-xi}) over the line profile, we find that
\beq
R_{lu} = \frac{g_u}{g_l}A_{u\rightarrow l}[\phaC + (\phaLO-\phaC)\bar\xi].
\eeq
It follows from Eq.~(\ref{eq:xu}) that
\beqa
x_u &=& \Gamma_u^{-1}\Biggl\{ \sum_{i\neq l,u}x_iR_{iu}
\nonumber \\ && + \frac{g_u}{g_l}x_lA_{u\rightarrow l}[\phaC + (\phaLO-\phaC)\bar\xi]
\Biggr\}.
\eeqa
The term $\sum_{i\neq l,u}x_iR_{iu}$ also appears in Eq.~(\ref{eq:phalo-def}), so we can replace it with $\phaLO$.  Noting also that 
$A_{u\rightarrow l}/\Gamma_u=f_l=f_{\rm coh}$, we may write
\beq
x_u = \frac{g_ux_l}{g_l}\left\{
f_{\rm inc}\phaLO + f_{\rm coh}[\phaC + (\phaLO-\phaC)\bar\xi]
\right\}.
\eeq
The quantity in braces is $\phaL=g_lx_u/g_ux_l$, i.e. the phase space density of photons that would be in equilibrium with the $l$ and 
$u$ level populations (again assuming $x_u\ll x_l$):
\beq
\phaL = f_{\rm inc}\phaLO + f_{\rm coh}[\phaC + (\phaLO-\phaC)\bar\xi].
\eeq
If however $x_l$ and $x_u$ are known (they are directly available in the level code), we may rearrange this equation to solve for 
$\phaLO$ in terms of $\phaL$, $\phaC$, $f_{\rm coh}$, and $\bar\xi$.  After some algebra, we find
\beq
\phaLO = \frac{\phaL - f_{\rm coh}(1-\bar\xi)\phaC}{1 - f_{\rm coh}(1-\bar\xi)}.
\eeq

In order to determine the upward transition rate, we need to calculate $\bar\pha$, which is
\beq
\bar\pha = \phaC + (\phaLO-\phaC)\bar\xi
= \phaC + \frac{(\phaL-\phaC)\bar\xi}{1 - f_{\rm coh}(1-\bar\xi)}.
\label{eq:bpha}
\eeq
Then we have
\beqa
\dot x_l|_u \!\! &=& \!\! -R_{lu}x_l + R_{ul}x_u
= A_{u\rightarrow l}\left( -\frac{g_u}{g_l}x_l\bar\pha +x_u\right)
\nonumber \\
&=& \!\! \frac{g_u}{g_l}A_{u\rightarrow l}x_l(-\bar\pha+\phaL). 
\eeqa
One can eliminate $\bar\pha$ using Eq.~(\ref{eq:bpha}), and algebraic simplification then yields
\beq
\dot x_l|_u = \frac{g_u}{g_l}A_{u\rightarrow l}x_l(\phaL-\phaC)\left( 1-\frac{\bar\xi}{1 - f_{\rm coh}(1-\bar\xi)}\right).
\eeq
We now recall that $\phaC$ is very nearly equal to the blackbody function since hydrogen is in Saha equilibrium during \HeI\ 
recombination.
\beq
\dot x_l|_u = A_{u\rightarrow l}\left( x_u-\frac{g_u}{g_l}x_l\phaC\right)\Pprd,
\label{eq:downward0}
\eeq
where
\beq
\Pprd= 1-\frac{\bar\xi}{1 - f_{\rm coh}(1-\bar\xi)} = \frac{(1-\bar\xi)f_{\rm inc}}{1 - f_{\rm coh}(1-\bar\xi)}.
\label{eq:pprd}
\eeq

Before continuing, we wish to generalize this equation to remove one assumption we have made about continuum opacity.  Recall that we have assumed a finite continuum optical depth per unit frequency, which when integrated to $\nu\rightarrow+\infty$ gives formally infinite optical depth and hence drives $\pha_+$ to its equilibrium value.  In practice there are times early in helium recombination where the continuum opacity within a line is unimportant, and the continuum optical depth between the line of interest and the next higher-frequency line is not $\gg 1$.  In these cases $\pha_+$ may differ from $\pha_{\rm Pl}$ due to feedback.  We may trivially correct for this by noting that if feedback is important then the line-line optical depth $\tauLL$ is of order 1 or less.  In this case, then the continuum opacity within an individual \HeI\ line is negligible, so in Eq.~(\ref{eq:r3}) we have $\etaC\approx 0$.  In this case, we are free to replace $\phaC$ in Eq.~(\ref{eq:r3}) with $\pha_+$ without consequence.  Making this replacement in the subsequent equations, in particular Eq.~(\ref{eq:xi-pha}), we find that $\xi$ on the blue side of the line is zero since $\pha=\pha_+$ there.  Hence the boundary condition $\xi_+=0$ that we were using earlier applies, and Eq.~(\ref{eq:downward0}) is also valid, except that we need to replace $\phaC\rightarrow\pha_+$:
\beq
\dot x_l|_u = A_{u\rightarrow l}\left( x_u-\frac{g_u}{g_l}x_l\pha_+\right)\Pprd,
\label{eq:downward}
\eeq
This equation is very similar to the Sobolev rate, Eq.~(\ref{eqn:sobolevapprox}).  The only difference (aside from the factor of $1+\pha_+$, which is irrelevant since $\pha_+\ll 1$ for these lines) is the Sobolev escape probability has been replaced with $\Pprd$, the escape probability including partial redistribution.

\subsection{Monte Carlo method: theory}
\label{ss:mcmethods}

So far in this section, we have constructed an equation of radiative transfer (Eq.~\ref{eq:r4}) for the \HeI\ \LyHe\ lines, and related 
the effective line escape probability to its solution (Eq.~\ref{eq:pprd}).  There is only one major step left: to solve Eq.~(\ref{eq:r4}) for the full partial redistribution plus continuous opacity problem.

A variety of approaches have been taken in the literature for solving line radiative transfer equations including coherent scattering 
terms.  One approach is the diffusive, Fokker-Planck approximation \cite{1992ApJ...387..248H} which replaces the redistribution integral 
(Eq.~\ref{eq:r-coh}) with a second-order differential operator.  This results in a second-order ODE instead of integro-differential 
equation, which is a substantial improvement for most numerical techniques.  The other possibilities are conversion of the equation of 
radiative transfer into a linear algebra problem or solution through Monte Carlo methods \cite{2006MNRAS.367..259H, 2002ApJ...578...33Z,
1979ApJ...233..649B,1979A&A....73...67B,1972ApJ...176..439C, 1968ApJ...153..783A,1968ApJ...152..493A}.  The latter two have the advantage of 
being usable in the Doppler core of the lines, which we expect to be important since for e.g. \HeI\ \LyaHe\ lines, the width of the line 
that is optically thick to incoherent processes $\Delta\nu_{\rm line}$ is only $\sim 30\Delta\nu_{\rm D}$ during most of \HeI\ 
recombination.  Therefore we have not used the Fokker-Planck approach, which we believe is better suited for studying the far damping 
wings of very optically thick lines such as \HI\ Ly$\alpha$ (The Fokker-Planck operator assumes that many scattering events transport a 
photon over a region where the line shape varies slowly; yet in the core of the line, single scatterings can transport a photon over the width of the core.)
We have chosen the Monte Carlo approach here, mainly because we had a pre-existing code that was capable of handling the problem with minor 
modifications \cite{2006MNRAS.367..259H}.

The basic plan of the Monte Carlo simulation is as follows: we begin by injecting a photon with frequency distribution drawn from the 
Voigt line profile, $\phi(\nu)$.  We simulate its fate by assuming that it redshifts at the rate $\dot\nu=-H\nu_{ul}$; undergoes coherent 
scattering with probability per unit time $H\nu_{ul}\tauS f_l\phi(\nu)$; undergoes continuum absorption with probability per unit time 
$H\nu_{ul}\etaC$; and undergoes incoherent absorption with probability per unit time $H\nu_{ul}\tauS(1-f_l)\phi(\nu)$.  The simulation 
is terminated if the photon is absorbed in the \HI\ continuum, if it redshifts out of the line, or if it undergoes incoherent absorption 
in \HeI.  Note that within the idealized conditions $\etaC=$~constant of the simulation, a photon that redshifts out of the line will 
eventually be absorbed by the continuum so long as $\etaC>0$ (though in reality the photon would eventually reach other \HeI\ lines or 
redshift to below $13.6\,$eV).  Therefore only the total probability of these two results is meaningful.  We thus denote by $\PMC$ the 
probability that a photon in the Monte Carlo is terminated by redshifting out of the line or by continuum absorption, and let $1-\PMC$ 
denote the probability that the photon is terminated by incoherent absorption.  The implementation of the Monte Carlo is described in 
Appendix~\ref{sec:mcimplementation}; the rest of this section will be devoted to the problem of extracting $\bar\xi$ from the Monte 
Carlo.  In particular, we will prove that $\bar\xi=1-\PMC$.

Our proof goes as follows.  We begin by considering the probability distribution $\Pi(\nu)$ of the photon frequency in the Monte Carlo 
at any specified time.  [Note that this is not the same as the histogram of frequencies at which the photon scatters, which is 
$\propto\Pi(\nu)\phi(\nu)$ because the scattering rate is proportional to $\phi(\nu)$.]  Now from elementary considerations the 
probability distribution satisfies
\beqa
\dot\Pi(\nu) &=& H\nu_{ul}\Biggl\{
\frac{\partial\Pi}{\partial\nu}
- \etaC\Pi(\nu)
- \tauinc\phi(\nu)\Pi(\nu)
\nonumber \\ &&
- \taucoh\left[\phi(\nu)\Pi(\nu)-
\int\phi(\nu')\Pi(\nu')p(\nu|\nu')\,d\nu'
\right]
\Biggr\}
\nonumber \\ &&
+ \Gamma_{\rm inj}\phi(\nu),
\eeqa
where $\Gamma_{\rm inj}$ is the rate of injection of photons in the Monte Carlo.
Then the steady-state distribution satisfies
\beqa
\frac{\partial\Pi}{\partial\nu} &=& \etaC\Pi(\nu)+\tauinc\phi(\nu)\Pi(\nu)
\nonumber \\ &&
+\taucoh\left[\phi(\nu)\Pi(\nu)-
\int\phi(\nu')\Pi(\nu')p(\nu|\nu')\,d\nu'
\right]
\nonumber \\ &&
- \frac{\Gamma_{\rm inj}}{H\nu_{ul}}\phi(\nu).
\label{eq:pi-e}
\eeqa
This equation has a striking resemblance to Eq.~(\ref{eq:r4}).  They are both linear inhomogeneous integro-differential equations, and 
they obey the same boundary condition, namely that $\xi$ and $\Pi$ go to zero at high frequency (the photons are in equilibrium with 
\HI\ on the blue side of the line).  The only difference is in the source terms: the source term in Eq.~(\ref{eq:r4}) is 
$-\tauinc\phi(\nu)$, whereas the source term in Eq.~(\ref{eq:pi-e}) is $-(\Gamma_{\rm inj}/H\nu_{ul})\phi(\nu)$.  These source 
terms differ only by a constant scaling factor, and hence the solutions are scaled versions of each other with the same scaling factor:
\beq
\Pi(\nu) = \frac{\Gamma_{\rm inj}}{H\nu_{ul}\tauinc}\xi(\nu).
\label{eq:pixi}
\eeq
We now note that in the Monte Carlo, photons are injected (the simulation is restarted with a new photon) when the photon is absorbed
in either an incoherent process or by \HI.  Practically, the frequency span of the simulation is finite, yet photons that 
eventually redshift through the simulation boundaries will eventually be absorbed in a continuum event (assuming the opacity is 
non-zero).  Therefore $\Gamma_{\rm inj} = \Gamma_{\rm inc} + \Gamma_{\rm cont}$,
where $\Gamma_{\rm inc}$ and $\Gamma_{\rm cont}$ are the rates of removal of photons by incoherent scattering and continuum absorption 
respectively.
Of these, $\Gamma_{\rm inc}$ is obtained by averaging the incoherent scattering rate $\tauinc\phi(\nu)$ over the 
probability distribution, which yields
\beq
\Gamma_{\rm inc}
= H\nu_{ul}\tauinc\bar\Pi.
\label{eq:gi}
\eeq
[Here $\bar\Pi$ denotes averaging of $\Pi(\nu)$ over the line profile $\phi(\nu)$, which is equal to the average of the line profile 
over the photon probability distribution $\Pi(\nu)$.]
The rate of removal via continuum opacity is simply
\beq
\Gamma_{\rm cont}=H \nu_{ul} \etaC.
\label{eq:gc}
\eeq
Therefore we may write Eq.~(\ref{eq:pixi}) as
\beq
\Pi(\nu) = \left( \bar\Pi + \frac{\etaC}{\tauinc} \right)\xi(\nu).
\eeq
Multiplying both sides by $\phi(\nu)$ and integrating yields
\beq
\bar\Pi = \left(\bar\Pi + \frac{\etaC}{\tauinc}\right)\bar\xi,
\eeq
from which we may find
\beq
\bar\xi = \frac{\bar\Pi}{\bar\Pi + \etaC/\tauinc}.
\eeq
Now recalling Eqs.~(\ref{eq:gi}) and (\ref{eq:gc}), we see that
\beq
\bar\xi = \frac{\Gamma_{\rm inc}}{\Gamma_{\rm inc}+\Gamma_{\rm cont}}
= \frac{\Gamma_{\rm inc}}{\Gamma_{\rm inj}} = 1-\frac{\Gamma_{\rm cont}}{\Gamma_{\rm inj}} = 1-\PMC.
\eeq
This result connects the Monte Carlo algorithm to the parameter $\bar\xi$ needed in the level code.  Written directly in terms of 
$\PMC$, the modified escape probability (Eq.~\ref{eq:downward}) is
\beq
\Pprd = \frac{\PMC f_{\rm inc}}{1-f_{\rm coh}\PMC}.
\label{eq:downward-new}
\eeq

Before continuing, we note several properties and limiting cases of Eq.~(\ref{eq:downward-new}).  Since it is obvious that $\PMC$ and 
$f_{\rm coh}$, being probabilities, are in the range from 0 to 1, and we recall that $f_{\rm coh}+f_{\rm inc}=1$, it is easily seen that 
$\Pprd$ is also in the range from 0 to 1 (hence its interpretation as an ``escape probability'').  In the limit where there is no 
coherent scattering, we have, unsurprisingly, $\Pprd=\PMC$: the effective escape probability $\Pprd$ appearing in the rate equations is 
exactly the probability of line escape or continuum absorption in the Monte Carlo.  In an optically thin line where essentially all photons in the Monte Carlo escape, $\PMC=1$, then we find $\Pprd=1$.  Finally, if $\PMC\ll 1$ so that almost all photons 
emitted in the line undergo incoherent re-absorption, we have $\Pprd=\PMC f_{\rm inc}$: the escape probability in the rate equations 
becomes the escape probability in the Monte Carlo times the fraction of the Sobolev optical depth due to incoherent processes.

The Monte Carlo method described here raises two theoretical questions: first, does the escape probability in Eq.~(\ref{eq:downward-new}) coincide with 
that calculated in Sec.~\ref{sec:anacompletecont}, as one would expect; and second, is the standard Sobolev result recovered in the absence of 
continuum absorption?  In Appendix~\ref{sec:mcsobolev}, we show that the answer to the first question is in the affirmative.  As for the second 
question, we find (also in Appendix~\ref{sec:mcsobolev}) that by setting $\etaC=0$ the standard Sobolev escape probability is recovered {\em if} the 
line is sufficiently optically thick, i.e. if $\PMC\ll 1$ and if the probability ${\cal P}_+$ that a photon entering the line from the blue side will 
escape to the red side without undergoing any incoherent scattering satisfies ${\cal P}_+\ll 1$.

\subsection{Incorporation in the level code}
\label{ss:levelcoderes}

\begin{figure}[!ht]
\epsfxsize=3.4in
\begin{center}
\epsffile{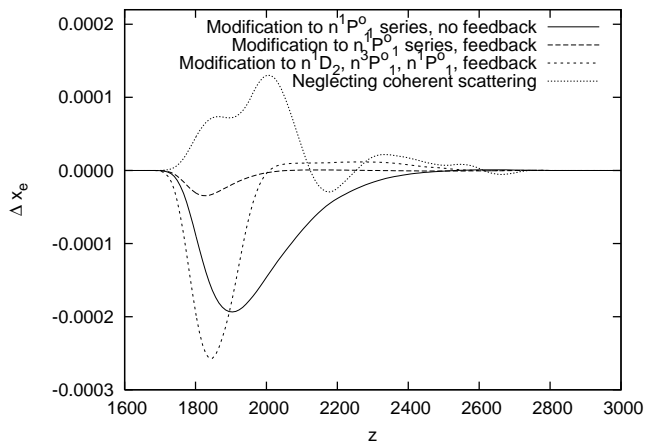}
\end{center}
\caption{Modification to the \HeI\ recombination history due to continuum opacity in the \QuadHe, \InterHe, and \LyHe\ lines (and where coherent scattering through \LyaHe\ is neglected) relative to a model where only continuum opacity in the \LyaHe\ line is accounted for.  The total effect on $x_e$ is at the level of $10^{-4}$.  Without feedback, \LyHe\ slightly speeds recombination relative to just \LyaHe--the allowed rates from $n>2$ to the ground state are accelerated.  \comment{Once the \LybHe\ rate is acclerated and feedback is considered, a larger spectral distortion reaches \LyaHe\, thus slowing down \HeI\ recombination overall.  The effect of feedback wins out until the distortion from \LybHe\ is almost completely absorbed, while rates within \LyaHe\ are increasingly affected by the continuum opacity, around $z<2000$.  Here, these higher order effects still speed \HeI\ recombination overall.}  Coherent scattering through \LyaHe\ is also seen to be negligible for the recombination history.} 
\label{figs:mcmodfdconvergence}
\end{figure}

\begin{figure}[!ht]
\epsfxsize=3.4in
\begin{center}
\epsffile{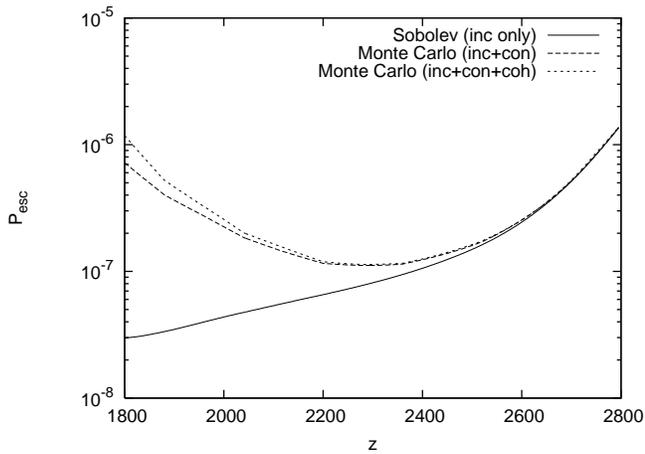}
\end{center}
\caption{The modified escape probability from \Hefour\ \LyaHe\ during \HeI\ recombination, comparing the results of the standard Sobolev approximation and the modification 
due to continuous opacity with and without coherent scattering.  Once coherent scattering is introduced, photon diffusion effects increase the escape 
probability by increasing the span of frequencies traversed by the scattering photon before it escapes.  These probabilities are log-interpolated over 
the grid of $x_{\rm HeI}$ and $z$ used in the level code.  We find that the effect of doubling the grid resolution and with it the smoothness of the 
interpolated probability is negligible.} \label{figs:escape_mc}
\end{figure}

In principle, modifications to transport in the entire \LyHe\ series, the quadrupole series (\QuadHe), and intercombination series 
(\InterHe) of \HeI\ could lead to acceleration of \HeI\ recombination.  To make the problem computationally tractable, the hierarchy of 
higher order $n$ contributions needs to be cut off where there are diminishing returns.  The integral solution to the transport 
equations with complete redistribution (Sec.~\ref{sec:anacompletecont}), while not accurate for \LyHe\ rates, gives a quick estimate of 
whether, for example, continuum effects become more and more important for higher $n$, or die away.  We evaluate the probabilities at 
$z=1606$, at the end of \HeI\ recombination where the continuum effects are expected to be largest, using the integral 
Eq.~(\ref{eqn:absorptionprob}).  It is found that in the intercombination lines, continuum effects become small for $n>4$ (the ratio of 
the probabilities with and without continuous opacity approaches $1$).  For quadrupole transitions (\QuadHe), the corrections are 
significant out to moderate $n$, (modifying the escape probability by a factor of $\sim 2$ at $n=8$).  In the \LyHe\ series, continuum 
effects lead to significant modifications to the escape probability (by a factor of $\sim 100$ at $n=9$ toward the end of \HeI\ recombination)--the 
entire \LyHe\ series sees significant corrections.  (One mitigating factor that makes the calculation practical is that only the \LyaHe\ 
line dominates recombination rates.)  Based on these estimates, we calculate corrections to transport within the line up to $n=6$ in the 
\LyHe\ series, up to $n=4$ in the intercombination lines (\InterHe), and $n=6$ in the quadrupole lines (\QuadHe).  The \LyHe\ series is treated via the 
Monte Carlo technique, while the others are treated here using the integral method of Sec.~\ref{sec:anacompletecont}.  (Fig.~\ref{figs:mcmodfdconvergence} shows 
that higher allowed transitions are negligible.  In Paper III we will show 
results where the Monte Carlo method was used for intercombination and quadrupole lines as well.)  Fig.~\ref{figs:escape_mc} shows the 
the escape probabilities subject to \HI\ opacity (with and without coherent scattering) for \LyaHe\ derived from the Monte Carlo, relative 
to the ordinary Sobolev results.

Both the Monte Carlo and analytic integral methods of finding the escape probability are prohibitively slow to run in real-time with the 
level code.  In the Monte Carlo method, computing the large number of coherent scatters for one photon trajectory in the \HeI\ \LyHe\ 
series is very time-consuming.  In the analytic integral method developed for complete redistribution, the variety of scales in the line 
profile gives, generally, integrands with large dynamic ranges that need to be known to high accuracies.  For this reason, we generate 
tables of the escape probability over a range of parameters and log-interpolate to find the probabilities in the level code.

The escape probability is a function of the coherent, incoherent, and continuum optical depths, and the matter temperature, which sets the Gaussian 
width for the line,
\beq
\Pesc(\{ \tau_{\rm coh}, \tau_{\rm inc}, \etaC, \Tm, \Gamma_{\rm line} \})
\eeq
The wing width is set by $\Gamma_{\rm line}$.  
We will project to a more natural parameter space for the level code,
\beq
\left \{ \tau_{\rm coh}, \tau_{\rm inc}, \etaC, \Tm, \Gamma_{\rm line} 
\right \} \rightarrow \left \{ z, \frac{n_{\rm HI} }{ n_{\rm HI,Saha}}, x_{\rm HeI} \right \}, \label{eqn:mcparmeters}
\eeq 
since for a given cosmology the latter parameter set determines the former. We calculate tables for $11$ linearly-spaced $z$ values between 1400 and 
3000, inclusive, and $21$ logarithmically-spaced $x_{\rm HeI}$ values from $2\times 10^{-5}$ to $0.08$.  When we double the fineness of the probability grid over the 
parameters, the change in free electron density $|\Delta x_e|$ has a maximum of $\sim 5\times 10^{-4}$ at $z \sim 1900$.  The neutral hydrogen 
population is taken to be the evolution determined by the reference level code.  This is very nearly Saha until the end of \HeI\ recombination, by 
which time we find that \HeI\ has been relaxed to equilibrium.  (That is, whether you assume the \HI\ population is Saha or evolves through in the full 
recombination treatment should matter little for the evolution of the \HeI\ ground state: even before neutral hydrogen departs significantly from 
equilbrium, \HeI\ is already relaxed into equilibrium, by that point.)  This set of probabilities required $\sim 4$ days to evaluate over $50\times 
3.2\,$GHz computer nodes.  The convergence criterion is described in Appendix~\ref{sec:mcimplementation}, and shown to converge to $1.25\%$ fractional 
error in probability with negligible bias.  Doubling the number of Monte Carlo trials led to a maximum change of $|\Delta x_e| < 10^{-4}$, and 
resampling with a different random generator (Numerical Recipes {\tt ran2} instead of {\tt ran3} \cite{1992nrca.book.....P}) gives a change $|\Delta 
x_e| < 1.5\times 10^{-4}$.  (More convergence tests are described in Paper III.)

\begin{figure}[!ht]
\epsfxsize=3.4in
\begin{center}
\epsffile{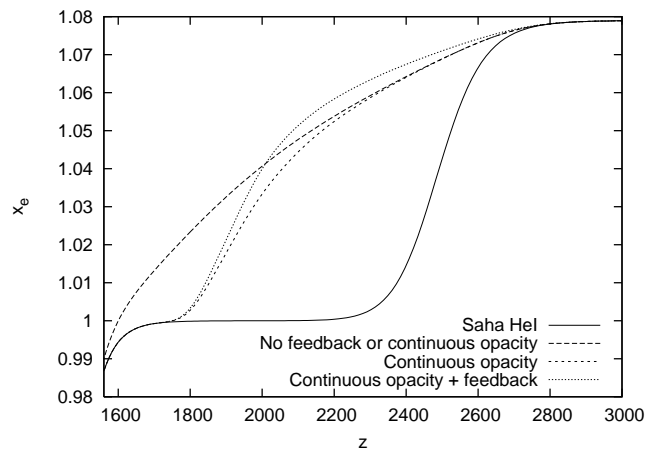}
\end{center}
\caption{Modifications to the \HeI\ recombination history from the inclusion of feedback and continuum opacity in \HeI\ lines, compared 
to a Saha \HeI\ recombination history and the standard \HeI\ recombination history.  Continuous opacity starts to become important at $z \sim 2100$ (as 
suggested by Fig.~\ref{figs:hydrogen_depth}), and pushes the \HeI\ evolution to a Saha by $z \sim 1800$.  The beginning of \HI\ recombination is visible 
here starting at $z\sim 1700$, and for later times $x_e$ drops precipitously.  The modifications to \HeI\ recombination suggested here are twofold: 
at early times during \HeI\ recombination, feedback slows recombination, and at later times, continuous opacity accelerates recombination relative to 
standard models.}
\label{figs:mcmodified}
\end{figure}

The result of including these in the level code is shown in Fig.~\ref{figs:mcmodified}.  Here, we can test the diminishing-returns set 
of modified lines by running a level code with: 1) just \LyaHe\ modifications, 2) modifications to the \LyHe\ series up to $n=6$, 3) 
\LyaHe\ and the 3 intercombination lines, and 4) the \LyHe\ series up to $n=6$, plus the three intercombination lines, shown in 
Fig.~\ref{figs:mcmodfdconvergence}.  It is apparent that most of the effect is due to modification to the \LyaHe\ escape probability, 
leading to relaxation of the `$n=2$' bottleneck and further that coherent scattering is only a small contribution to the recombination history.
This greatly improves prospects for including continuum effects in a practical level code through modifications to \LyaHe\ under the approximations 
developed in Sec.~\ref{sec:anacompletecont} (without coherent scattering).

\section{conclusion}
\label{sec:conclusion}

The efforts of a new generation of precision small-scale CMB temperature anisotropy experiments need to be complemented by confidence that the 
underlying theory is well-understood.  Recent results \cite{2006A&A...446...39C, 2005AstL...31..359D, 2004MNRAS.349..632L} have brought up the 
possibility that new effects may modify the \HI\ and \HeI\ recombination histories.  Here we have examined and extended several of 
these recently proposed effects: the matter temperature evolution and the intercombination and quadrupole lines of \HeI.  We also introduce two new 
effects: feedback of radiation between lines (non-local radiative transport), and the effect of continuum opacity from photoionization of neutral 
hydrogen on transport within and between lines.  The effect of feedback between lines, which has previously been neglected is also shown to be a 
significant effect at the $\sim 1\%$ level by using an iterative method to include the effect.  (Paper II and III will treat two-photon processes, 
electron and \Hethree\ scattering, and rare decays.)

The striking new effect has been the introduction of continuum opacity due to \HI\ photoionization in the \HeI\ \LyaHe\ transition.  We presented two 
methods of understanding this physics, an integral solution to the radiative transport equations for complete redistribution, and a Monte Carlo method 
for partial redistribution (to include strong coherent scattering effects).  This is found to dramatically increase the escape probability and speed up 
\HeI\ recombination once continuum effects in the line become important.  These effects, in combination with the inclusion of the intercombination line 
as suggested by \cite{2005AstL...31..359D}, paint a very different picture of \HeI\ recombination -- one where the recombination accelerates after 
$z\sim 2200$ leading to very little overlap with the \HI\ recombination.  The speed-up is a generic feature of any effects that facilitate the 
formation of ground state \HeI, though of course the details depend on the kinetics of the specific mechanism.

Now that the interpretation of many important cosmological parameters depend on slight shifts (such as evidence for inflation from the primordial slope 
$n_s$), it will be crucial to understand even sub-percent corrections to the recombination history from various slow processes.  We have examined 
several corrections here and more are considered in Paper II and Paper III. Once a set of corrections to hydrogen and helium recombination are 
solidified, the next step will be to parameterize a set of corrections to a recombination model in a fast, easy-to-use package that can be plugged in 
to CMB calculations and parameter estimations.  Distilling the modified escape probabilities for a set of cosmological parameters into a new 
recombination code will be the subject of later work.

\begin{acknowledgments}

E.S. acknowledges the support from grants NASA LTSAA03-000-0090 and NSF PHY-0355328.  C.H. is a John Bahcall Fellow in Astrophysics at the Institute 
for Advanced Study.  We acknowledge useful conversations with Jens Chluba, Bruce Draine, Jim Peebles, Doug Scott, Uro\v s Seljak, and Rashid Sunyaev.  
We also thank Joanna Dunkley for critical readings and comments prior to publication.

\end{acknowledgments}

\appendix 
\section{Atomic levels and radiative rates}
\label{sec:atomicappendix}

\subsection{\HI\ model atom}
\label{ss:hydrogenmodel}

The full model \HI\ atom has 445 levels up to a maximum principal quantum 
number $n_{\rm max}=400$.  The individual $l$-sublevels are resolved for 
$n\le 10$.  The energies of each level are determined from the exact 
nonrelativistic formula $E_{nl}=-R_{\rm H}/n^2$, where the hydrogen Rydberg 
constant is $R_{\rm H}/h=3.2881\,$PHz.  The degeneracy factors are 
$g_{nl}=2(2l+1)$ for the $l$-resolved levels and $g_n=2n^2$ for the 
unresolved levels.  Throughout, we have used a smaller set of $245$ levels,
identical to the full model, except truncated at $n_{\rm max}=200$.  Inclusion of the remaining 
levels up to $n_{\rm max} = 400$ will be important for late-stage \HI\ recombination,
but negligible for \HeI\ recombination.  

The bound-bound electric dipole Einstein coefficients between $l$-resolved 
levels are determined by integration of the exact wave functions.  For 
cases where the upper sublevel is unresolved ($n_u>10$), we compute a 
weighted average of the Einstein coefficient, with the weights determined 
by the degeneracy factors.  This gives the appropriate decay rate in the 
limit where the sublevels are populated according to their statistical 
ratios.  For transitions with $n_l>10$ we used the Gaunt factor 
approximation of Ref.~\cite{1935MNRAS..96...77M}, Eq.~(1.38), with the 
correction described in the Appendix of Ref.~\cite{1958MNRAS.118..477B}.  
Direct summation of the Einstein coefficients shows that this yields an 
error of 1.2\% for the worst case ($n_l=11$, $n_u-n_l=1$).  We have not 
included the \HI\ electric quadrupole lines ($1s$--$nd$, $n\ge 3$) because they are at the same frequency as the Lyman lines 
$1s$--$np$ and hence do not yield acceleration of the \HI\ recombination 
process.  (This is not the case for \HeI.)

The bound-free cross sections for $n\le 10$ are obtained from TOPbase 
\cite{1993A&A...275L...5C}; for $n>10$, the $l$-sublevels are unresolved 
and we have used the Gaunt factor approximation.

The \HI\ atom posesses a metastable $2s$ level, which decays by emission 
of two electric dipole photons.  We have used the two-photon frequency 
distribution and the spontaneous lifetime estimate $\Lambda_{\rm 
HI}=8.2249\,$s$^{-1}$ from Ref.~\cite{1984A&A...138..495N}.

\subsection{\HeI\ model atom}
\label{ss:heliummodel}

The \HeI\ model atom has 289 levels up to a maximum principal quantum 
number $n_{\rm max}=100$.  For all values of $n$ we resolve singlet 
(``para-\HeI'') versus triplet (``ortho-\HeI'') levels.  The $l$-sublevels 
are resolved for $n\le 10$.  We include only configurations of the form 
$1s\;nl$ as the doubly excited configurations are all unbound.  The energy 
levels for the $l$-resolved levels in \HeI\ are taken from the NIST Atomic 
Spectral Database, based on Ref.~\cite{1987PhRvA..36.3575M}, except for 
$n=9,10$, $l>6$, for which hydrogenic values were used.  For the $n>10$ 
levels in which the value of $l$ is unresolved, we have used the Rydberg 
formula (which produces as accurate a value as can be considered 
meaningful given that the various $l$ levels are not strictly degenerate).  
The degeneracy factors are $g_{nLS}=(2L+1)(2S+1)$ for $L$-resolved levels 
and $g_{nS}=n^2(2S+1)$ for unresolved levels.  Note that in ortho-\HeI,
there are multiple allowed values of $J$ for given $L$ and $S$, which 
correspond to fine structure levels; we do not resolve these.

For the allowed bound-bound transition rates, we combined data for several 
sources.  For transitions between two $l$-resolved levels, the electric 
dipole oscillator strengths for allowed $S-P^o$ and $P^o-D$ transitions were 
obtained from Ref.~\cite{1984PhRvA..29.2981K} for upper levels with $n\le 
9$.  The $1^1S-10^1P^o$ transition rate is from 
Ref.~\cite{1988ApJ...329..493K} and other $S-P$ and $P-D$ rates with upper 
level $n=10$ are from TOPbase \cite{1993A&A...275L...5C}.  For $D-F^o$, 
$F^o-G$, etc. transitions, we used the hydrogenic rates since the $l\ge 2$ 
states are well approximated as an electron orbiting a pointlike He$^+$ 
ion.  For transitions between an $l$-resolved lower level and an 
unresolved upper level ($n>10$), we first compute the fully $l$-resolved 
Einstein coefficients and then average.  These $l$-resolved rates are 
taken as hydrogenic for $P-D$, $D-F$, etc. transitions.  For $1^1S-n^1P^o$ 
transitions ($n<10$), we use the asymptotic formula of 
Ref.~\cite{1989ApJ...336..504K}.  (This formula is only 1.4\% different 
from TOPbase at $n=10$.)  The Coulomb approximation 
\cite{1949RSPTA.242..101B} is used for all other $S-P^o$ transitions.  For 
transitions between two unresolved levels ($n_l>10$), we have used the 
hydrogenic Gaunt factor approximation.

In \HeI\ recombination, the intercombination and forbidden lines can be 
competitive with allowed lines as a mechanism of decay to the ground state 
because the allowed lines become very optically thick.  In particular we 
consider the series of intercombination lines \HeI] $1^1S-n^3P^o$ ($n\ge 
2$) and forbidden lines [\HeI] $1^1S-n^1D$ ($n\ge 2$).  For the \HeI] 
$1^1S_0-n^3P_1^o$ lines with $n=2,3$ we have used the Einstein coefficients 
estimated by Ref.~\cite{1978JPhB...11L.391L}; these are extrapolated as 
$A\propto n^{-3}$ for larger $n$.  (Note that the Einstein coefficients 
are usually quoted for the fine structure-resolved $n^3P^o_1$ level.  
Since our code does not resolve the fine structure these coefficients must 
be divided by 3 since only 3 of the 9 $n^3P^o$ states have $J=1$.)  For 
the [\HeI] $1^1S-n^1D$ lines with $3\le n\le 6$, we have used the 
oscillator strengths calculated by Ref.~\cite{2002JPhB...35..421C}.  For 
$n>6$ the oscillator strengths have been extrapolated with the asymptotic 
expansion $f\propto n^{-3}$.

Like \HI, the \HeI\ atom posesses a metastable singlet state $2^1S_0$ that 
decays by emission of two electric dipole photons.  The two-photon rate 
used here is $\Lambda_{\rm HeI}=50.94\,$s$^{-1}$, as determined by 
Ref.~\cite{1986PhRvA..34.2871D}.  The two-photon spectrum of \HeI\ was 
determined by the following formula, which was fitted to the results of 
Ref.~\cite{1986PhRvA..34.2871D}:
\begin{equation}
\phi(\nu) = \frac{19.602}{\nu_{1^1S_0-2^1S_0}}
\frac{\zeta^3(1.742-7.2\zeta+12.8\zeta^2)}{(\zeta+0.03)^2},
\end{equation}
where $\nu_{1^1S_0-2^1S_0}=4.9849\,$PHz is the frequency difference 
between the ground and metastable states, and 
$\zeta=\nu(\nu_{1^1S_0-2^1S_0}-\nu)/\nu_{1^1S_0-2^1S_0}^2$ is between $0$ 
and $1/4$.

\subsection{\HeII\ model ion}
\label{ss;helium2model}

The \HeII\ ion has one electron and hence in nonrelativistic theory all 
rates can be obtained by scaling of \HI.  The energy levels are re-scaled 
in proportion to $Z^2\mu$, where $Z$ is the atomic number and $\mu$ is the 
reduced mass.  The ratio of these factors for \HeII\ versus \HI\ is 
$4.001787$.  The bound-bound rate coefficients scale as $(Z^2\mu)^2$ and 
the photoionization cross sections scale as $(Z^2\mu)^{-1}$.  The 
two-photon decay rate $\Lambda$ scales as $(Z^2\mu)^3$.  The \HeII\ model 
atom is $l$-resolved up to $n=10$ and has a maximum principal quantum 
number $n_{\rm max}=100$, for a total of 145 levels.

\section{Integration of the transport equation for complete redistribution and continuous opacity} 
\label{ss:integraldetails}

In this Appendix we discuss how the transport equation with continuous opacity is integrated to find the continuum absorption probability, 
Eq.~(\ref{eqn:absorptionprob}).  Rybicki and Hummer \cite{1985ApJ...293..258H} develop a convenient analytic approximation to the transport solution 
in the case of a Gaussian profile with high optical depth.  This gives good intuition for the combinations of factors that contribute to the escape 
probability.  For example, it suggests that if the escape probability for a line $i$ is known, then an approximate scaling (in their limits) can be 
used to find the escape probability from $j$,
\beq P_{\rm esc,j} \sim \left ( \frac{P_{\rm esc,i} \tau_{\rm inc,i} }{ \sigma_{\rm D,i} \eta_{\rm C,i} } \right )  \frac{ \sigma_{\rm D,j} \eta_{\rm C,j} }{ \tau_{\rm inc,j} }. \eeq
There are three reasons why we would like a more general solution.  First, the high optical depth limit is not applicable to intercombination or 
quadrupole lines during recombination, though they have nearly Gaussian profiles.  Second, it is desirable to be able to use the macroscopic transport 
equations as a cross-check for the photon Monte Carlo methods developed in the next section for the \LyHe\ series, before the effect of coherent 
scattering is ``turned on''.  This requires solving Eq.~(\ref{eqn:absorptionprob})  with the Voigt profile.  (It also shows directly why complete 
redistribution is a dangerous assumption in the \LyHe\ series.)  Third, \HeI\ recombination rates are sensitive to the probabilities set by 
Eq.~(\ref{eqn:absorptionprob}), so a high-precision numerical result is necessary.  These constraints are met by directly integrating 
Eq.~(\ref{eqn:absorptionprob}) with the Voigt profile.

The Voigt profile presents two characteristic frequency scales: slowly varying functions in wings, and rapidly-varying functions in the core.  In the 
Doppler core, we can follow an analysis similar to \cite{1985ApJ...293..258H} and note that in the inner integrand,
\beq \exp \left \{ -\tau_{\rm inc} \int_\nu^{\tilde \nu} \phi(y) dy \right \},
\label{eqn:rybickiintegralapprox1} \eeq
the contribution is small unless the integral in the exponent is less than $\tauinc^{-1}$.
Yet, for the small differences between $\nu$ and $\tilde \nu$, the integrand is nearly linear,
\beq \int_\nu^{\tilde \nu} \phi(\nu') d \nu' \approx \phi(\nu) (\tilde \nu - \nu). \label{eqn:rybickiintegralapprox2} \eeq

Thus, in the Doppler core, the integral depends on evaluations of the integral over the Voigt profile on very small scales.  The Voigt function and its 
integral are time-consuming to evaluate.  Further, a linear interpolation lookup table must have sufficient resolution in the core, and breadth in the wings 
for Eq.~(\ref{eqn:absorptionprob}) to be evaluated accurately.  We use the Gubner's series \cite{1994JPhA...27L.745G} in the core region and switch to a fourth 
order asymptotic expansion of the Voigt profile in the wings.  (In the boundary between the two regimes, we estimate the error in the asymptotic expansion 
by using the next higher order and ensure that differences between the two approaches are negligible.)  The Voigt function and its integral are 
tabulated 
out to 10000 Doppler widths with 50 values per doppler width, exploiting the symmetry of the profile and its integral.  These are interpolated using a cubic 
spline and Eq.~(\ref{eqn:absorptionprob}) is integrated using a 61-point Gauss-Kronrod adaptive integration scheme.  (This is necessary because the 
integrand of Eq.~(\ref{eqn:absorptionprob}) carries its support over a widely varying range of $\tilde \nu$.)  To apply the numerical results of this 
integral to the recombination setting, $\PC$ is pre-calculated for a range of $\tau_{\rm inc}$ and $\etaC \Delta\nu_{\rm D}$, and then 
log-interpolated.   

\section{Monte Carlo method: implementation}
\label{sec:mcimplementation}

Monte Carlo radiative transfer methods have been used extensively in the literature \cite{2006MNRAS.367..259H, 2002ApJ...578...33Z, 
1979ApJ...233..649B,1979A&A....73...67B,1972ApJ...176..439C,1968ApJ...153..783A,1968ApJ...152..493A}.  We follow an approach similar to 
Hirata \cite{2006MNRAS.367..259H}, and use an algorithm to draw atomic velocity distributions that is discussed clearly in Lee 
\cite{1977ApJ...218..857L,1982ApJ...255..303L}.  A general atomic scattering process can be represented by the joint probability 
function of a photon with $\nuin $ scattering against the atom coherently to produce a photon of $\nuout $ through the scattering 
angle $\chi$, $P(\nuout , \chi|\nuin)$.

In coherent scattering, the photon is emitted with the same frequency it was absorbed in the atom's rest frame, and its emission 
direction relative to the incoming photon is drawn from the angular distribution associated with the transition.  The Doppler shift formula
gives an expression for the frequency shift between incoming and outgoing photons in the lab frame,
\beq \Delta \nu = (f_\parallel + \alpha)(1-\cos\chi)-f_\perp \sin\chi, \label{eqn:cohfreqshift} \eeq
where ${\mathbf f}= \nu_{0} {\mathbf v}/c$, for the transition frequency $\nu_{0}$, and $\alpha = h \nu_{0}^2/(m_p c^2)$.  Thoughout, 
$\parallel$ labels the component of a vector in the direction from which the photon came, and $\perp$ labels the component 
perpendicular to this direction and in the plane of scattering.  Thus, the change in frequency between the input and output states in 
the lab frame is uniquely specified through 
$\chi$, $f_\parallel$ and $f_\perp$.  These three quantities are stochastic, where $f_\parallel$ and $f_\perp$ depend on the 
thermodynamics of the gas, and $\chi$ is determined by the quantum mechanical scattering distribution.  The distribution of the angle 
$\chi$ between outgoing states and incoming states for dipole scattering is given by (over the range $0 \leq \chi \leq \pi$) 
\cite{2006MNRAS.367..259H},
\beq P(\chi) d\chi = \frac{3 }{ 8}( 1 + \cos^2 \chi) \sin\chi\, d\chi. \label{eqn:angulardist} \eeq
We can scale the atomic velocity relative to the other velocity scale in the problem, the characteristic thermal velocity, $u=v 
\sqrt{m/(2\kB T)}$.  This gives the convenient expression $f_\parallel = (\Delta \nuD) u_\parallel = \sqrt{2} \sigma_D u_\parallel$, 
where $\Delta \nuD$ is the Doppler width in standard notation, and $\sigma_D^2$ is the variance of the Doppler Gaussian.  In the 
perpendicular direction, the distribution of atomic velocities is thermal, $\propto e^{-u_\perp^2}$.  Bayes' rule gives the distribution 
of $u_\parallel$ based on known distributions as
\begin{eqnarray}
P(u_\parallel | x_{in}) &=& \frac{ P(x_{in} | u_\parallel) P(u_\parallel) }{ P(x_{in}) } \\ \nonumber 
&=& \frac{ a }{\pi H(a,x_{in})}  \frac{ e^{-u_\parallel^2} }{ a^2 + (x_{in} - u_\parallel)^2 }.
\label{eqn:uparallel}
\end{eqnarray}
Here $H(a,x)$ is the Voigt profile,
\beq H(a,x) = \frac{a }{ \pi} \int_{-\infty}^{\infty} \frac{ e^{-y^2} dy }{ a^2 + (x-y)^2 }, \label{eqn:voigtintegral} \eeq
$x = (\nu - \nu_{0})/\Delta \nuD$, and $a$ is the Voigt width parameter:
\beq a = \frac{ \Gamma_{\rm line} }{ 4 \pi \Delta \nuD }.\label{eq:voigt} \eeq
Typically $a\ll 1$, indicating that the Lorentzian width of the line is very small compared to the Doppler width.
Lee \cite{1977ApJ...218..857L, 1982ApJ...255..303L} presents a rejection method to randomly draw from this distribution.  

When a photon is emitted by an incoherent process -- i.e. it reaches an excited level $u$ either by a radiative transition from another excited level, 
or by recombination -- the outgoing photon has no knowledge of the phase space distribution of pre-existing line photons, so the probability 
distribution function for an outgoing photon is just the Voigt profile.  Because the Voigt profile is the convolution of the Doppler Gaussian 
distribution and the Cauchy distribution, a Voigt-distributed random number is easily implemented as the sum of random numbers drawn from each 
distribution.

The distance between scatters for small frequency shifts is given by $\ell \approx c H^{-1}(z) \nu_0^{-1} \Delta \nu$ (for some central 
frequency $\nu_0$ and shift $\Delta \nu$).  Over the travel times associated with the escape of one photon, the universe has expanded 
very little, and the line depth parameters are effectively constant.

In standard Sobolev theory, where the fate of the photon is either that it escapes or is absorbed in an incoherent process, it is clear what is meant 
by an escape probability.  The transport problem for \HeI\ with continuum opacity is not as clear-cut.  Coherent, incoherent, continuum and redshift 
processes are all active, so there are several choices about what ``escape'' and ``scatter'' mean.  As shown in the main text however, there is one 
specific number we need to know: given a photon was just emitted in the line through an incoherent process, what is the probability $\PMC$ that it will 
escape through redshifting or continuum absorption before it is absorbed in an incoherent process?  
The easiest way to measure $\PMC$ with the Monte Carlo is to inject many photons whose initial frequency distribution is the Voigt profile, and follow 
them until they escape (i.e. redshift out of the line or get absorbed by \HI) or are re-absorbed in an incoherent process.  Here the range of 
frequencies simulated is 
$\pm 360$~THz with a bin size of $0.36$~GHz.  (Using a smaller span or less resolution leads to escape probabilities that are biased high at early times 
during \HeI\ recombination, because absorption can occur far in 
the wings at early times when the incoherent depth is high and the continuum depth is small.  Doubling the boundary and halving the frequency step 
size does not improve the results, within these tolerances.)  The basic steps in the Monte Carlo are:
\newcounter{mcsteps}
\begin{list}{\arabic{mcsteps}. }{\usecounter{mcsteps}}
\item Draw a photon from the Voigt distribution, representing emission from an incoherent process.
\item Draw the optical depth that the photon traverses before scatting from the exponential distribution, $P(\delta \tau) d(\delta 
\tau) = e^{- \delta \tau} d(\delta \tau)$.  The next frequency where the photon scatters is given implicitly by
\beqa
\tilde \tau &=& \int_{\nu_{start}}^\nu \left [ (\tau_{\rm inc} + \tau_{\rm coh}) \phi(\tilde \nu) + \frac{ d\tau }{ d \nu}(\tilde \nu) 
\right ] d \tilde \nu 
\nonumber \\
&\equiv& \int_{\nu_{start}}^\nu d\tilde \nu \eta(\tilde \nu),
\label{eqn:MCsteppingcondition} \eeqa
where we have identified the integrand $d \tau_{tot} / d\nu$ as $I(\nu)$. This is implemented numerically by choosing frequency bins 
($\Delta \nu = 0.36$~GHz) and determining whether a photon crosses that bin, or is absorbed.  If the bins are chosen to be small enough, 
then the integrand is nearly linear over the bin,
\beq \eta(\nu)|_{i} = \frac{\eta_{i+1} - \eta_i}{\Delta \nu} + \eta_i. \label{eqn:linearmcintegrand} \eeq
Integrating this gives a quadratic expression for the fraction $x$ of the bin that the photon traversed:
\beq \frac{ \eta_{i+1} - \eta_i }{ 2\eta_i } x^2 + x = \frac{ \delta \tau }{ (\Delta \nu) \eta_i } \label{eqn:linearmcintegral} 
\eeq
This can either be solved quadratically, or by working to first order in small bins.  We will use the latter approach for small bins,
\beq x_0 = \frac{\delta \tau }{ (\Delta \nu) \eta_i } \mathrm{~~~and~~~} x = x_0 \left ( 1 - x_0 \frac{ \eta_{i+1} - \eta_i }{ 2\eta_i } 
\right ). \label{eqn:linearmcsoln} \eeq
The fraction of the bin traversed indicates whether the photon scattered in the bin, or leaves.  If it escaped the bin, then we move the 
photon to the start of the next bin and goes to step 2.  If it traversed less than the whole bin, we find the frequency in the bin where 
it scatters $\nu_{scatter}$, and go to step 3.
\item Draw a uniform random number between zero and $\frac{d\tau_{\rm tot} }{ d \nu}(\tilde \nu_{\rm scatter})$ and determine the type of 
event (incoherent scatter/absorption, coherent scatter, or \HI\ continuum absorption).
\item If the photon coherently scatters, draw the scattered atom's velocity and angle between incoming and 
out-going states and use Eq.~(\ref{eqn:cohfreqshift}) to find the photon energy after scattering, and go to step 2.  If the photon is incoherently scattered by \HeI, undergoes photoelectric absorption by \HI, or redshifs through a pre-defined simulation boundary, start a new photon in step 1.
\end{list}

The Monte Carlo procedure leaves one issue open, namely the convergence criterion.  We repeat the Monte Carlo until 6400 photons escape from the line, 
which should enable $\PMC$ to be determined to a fractional error of $6400^{-1/2}=0.0125$. A possible concern with this procedure (or any other in 
which the number of photons simulated is not fixed before running the Monte Carlo) is that the result could be biased if the convergence criterion 
depends on the results of the simulation.  We addressed this question by replacing the function that decides whether the photon escapes with a random 
number generator that returns ``escape'' a known fraction $\PMC$ of the time and ``no escape'' (incoherent scatter) a fraction $1-\PMC$ of the time.  
This produces a very fast code that allows us to map the distribution of the estimated escape probability $\hat P_{\rm MC}$ as a function of the true 
$\PMC$.  Across the relevant range of $\PMC$ (down to $10^{-6}$) we find that $\hat P_{\rm MC}$ has a bias $\langle\hat P_{\rm MC}\rangle/\PMC-1$ whose 
absolute value is $<0.1\%$, and a standard deviation $\sigma(\hat P_{\rm MC})/\PMC\approx 0.0125$.

\section{Relation of escape probability methods}
\label{sec:mcsobolev}

In this Appendix, we consider the relation of the Monte Carlo method to the analytic solution of Sec.~\ref{sec:anacompletecont} and to the usual 
Sobolev escape method.  We show explicitly that the solution to the transport equations considered here reduce to those of 
Sec.~\ref{sec:anacompletecont} if $f_{\rm coh}\rightarrow 0$.  We then proceed to investigate the conditions under which the Monte Carlo solution is 
equivalent to the Sobolev escape method.

\subsection{Relation to case of no coherent scattering}

Here we connect the formalism developed in Secs.~\ref{ss:eqnradtrans} and \ref{ss:relationphalo} to the analysis of Sec.~\ref{sec:anacompletecont}, which neglected coherent scattering.

The transport equation for line processes and the continuum, ignoring the coherent scattering diffusion term gives
\beq \dphadnu = \etaC[\pha(\nu)-\phaC]+\tauinc\phi(\nu)[\pha(\nu)-\phaLO].
\label{eq:usualsobolevtrans} \eeq
This can be solved for $\bar \pha$ analogously to the derivation in Sec.~\ref{sec:anacompletecont},
\beq \bar \pha = \phaLO + \PescIV (\phaC - \phaLO), \eeq
where $\PescIV\equiv \PS-\Delta\bar I_L$ is the escape probability derived in Sec.~\ref{sec:anacompletecont} (c.f. Eq.~\ref{eqn:barphaprob}).  Comparing this to the analysis of Sec.~\ref{ss:relationphalo}, in particular Eq.~(\ref{eq:xi-pha}) we find the correspondance
\beq
\bar\xi = 1-\PescIV.
\eeq
Plugging in to the transport result, Eq.~(\ref{eq:pprd}), one finds that $\Pprd$ (which modulates the rate) is now
\beq \Pprd = \frac{\PescIV f_{\rm inc} }{ 1 - f_{\rm coh} \PescIV}. \label{eqn:prdfornocoh}\eeq
This reduces to $\Pprd=\PescIV$ in the limit that scattering in the line is purely incoherent, as expected.  

\subsection{Relation of Monte Carlo to Sobolev method}

Next we consider the relation between this escape probability and the traditional Sobolev probability $\PS$ used in recombination codes in the {\em absence} of coherent scattering. The Sobolev escape calculation in its usual form assumes complete redistribution over the steady-state line profile and neglects continuum opacity \cite{1994ApJ...427..603R,1992ApJ...387..248H}.  Complete redistribution is an accurate assumption in pure-resonance rate problems where the radiation field is forced into equilibrium with the line over most of its central frequency extent by the high incoherent scattering rate (typically when the wing is optically thick to incoherent scattering).  Also it is trivially valid if the scattering optical depth is dominated by incoherent scattering.  Diffusion from Doppler shifts in repeated coherent scattering events has been studied extensively through its redistribution function, the Fokker-Planck approximation (accurate away from the Doppler core), and Monte Carlo methods \cite{1977ApJ...218..857L, 1992ApJ...387..248H, 1994ApJ...427..603R, 2006MNRAS.367..259H}.  This diffusion tends to broaden the jump in the radiation phase space density distribution on the blue side of the line, but does not significantly suppress (or enhance) $\bar \pha$ near line center or modify the atomic transitions rates.  Thus, for practical 
purposes in a recombination level code with pure resonant line processes and no continuous opacity, there is very little loss in assuming 
complete redistribution and steady-state radiation fields.

During recombination, there are two general categories of lines connecting to the ground state to consider: (1) $f_{\rm inc} \approx 1$ so that 
$\tau_{\rm inc} \approx \tauS$ (intercombination and forbidden lines) or (2) $f_{\rm inc}$ is small but $\tau_{\rm inc}$ is large, i.e. $\tau_{\rm 
coh}\gg\tau_{\rm inc}\gg1$ (allowed lines such as \LyHe\ in \HeI\ or the Lyman series in \HI).  We consider what happens in each type of line with 
continuum opacity off ($\etaC=0$).

For case (1), $f_{\rm coh}\ll 1$ but $f_{\rm inc}\approx 1$; then it is immediate that $\Pprd \approx P_{\rm S}$.
The treatment of case (2) is more complicated.  We begin by integrating Eq.~(\ref{eq:r4}) over frequency with $\etaC=0$ to get
\beqa
\xi_+-\xi_- \!\!&=&\!\! \tau_{\rm coh}
\biggl[\int\phi(\nu)\xi(\nu)\,d\nu
\nonumber \\ &&
 - \int\int\phi(\nu')\xi(\nu')p(\nu|\nu')\,d\nu'\,d\nu\biggr]
\nonumber \\ &&
\!\!+\tau_{\rm inc}\int\phi(\nu)[\xi(\nu)-1]\,d\nu.
\eeqa
Here $\xi_\pm$ is the value of $\xi$ on the blue ($+$) or red ($-$) side of the line.  The term multiplying $\tau_{\rm coh}$ vanishes since the redistribution probability $p$ integrates to unity.  Also $\xi_+=0$ because of boundary conditions.  The last term simplifies to $\tau_{\rm inc}(1-\bar\xi)$ or $\tau_{\rm inc}\PMC$.  If we turn off continuum opacity ($\etaC=0$), this then simplifies to
\beq
\PMC = \frac{\xi_-}{\tau_{\rm inc}},
\eeq
which is much less than 1.  Then from Eq.~(\ref{eq:downward-new}):
\beq
\Pprd \approx \frac{(\xi_-/\tau_{\rm inc})f_{\rm inc}}{1-f_{\rm coh}\PMC} \approx \frac{\xi_-f_{\rm inc}}{\tau_{\rm inc}} = \frac{\xi_-}{\tauS}.
\label{eq:temp-01}
\eeq

In order to proceed we must understand the behavior of $\xi_-$ in the case with no continuum opacity.  This can be done by considering a thought 
experiment in which we inject photons into the Monte Carlo on the far-blue side of the line instead of using the Voigt profile.  In this modified Monte 
Carlo, the photon probability distribution $\Pi^{\rm mod}$ satisfies
\beqa
\frac{\partial\Pi^{\rm mod}}{\partial\nu} &=& \tauinc\phi(\nu)\Pi^{\rm mod}(\nu)
\nonumber \\ &&
+\taucoh\Biggl[\phi(\nu)\Pi^{\rm mod}(\nu)
\nonumber \\ &&
-\int\phi(\nu')\Pi^{\rm mod}(\nu')p(\nu|\nu')\,d\nu'
\Biggr],
\label{eq:pi-e-mod}
\eeqa
with the boundary condition $\Pi^{\rm mod}(\nu)=\Gamma_{\rm inj}/H\nu_{ul}$ in the blue wing since in the absence of the line photons redshift at a 
rate $\dot\nu=-H\nu_{ul}$.  Defining the quantity
\beq
X(\nu) = 1 - \frac{H\nu_{ul}}{\Gamma_{\rm inj}}\Pi^{\rm mod}(\nu),
\label{eq:xnu}
\eeq
we see that $X(+\infty)=0$ and that $X(\nu)$ satisfies
\beqa
\frac d{d\nu}X(\nu) &=&
\taucoh\left[\phi(\nu)X(\nu)-\int \phi(\nu')X(\nu')p(\nu|\nu')\,d\nu'\right]
\nonumber \\ &&
+ \tauinc\phi(\nu)[X(\nu)-1],
\label{eq:r4-mod}
\eeqa
where we have used the symmetry relation $\int \phi(\nu')p(\nu|\nu')\,d\nu'=\phi(\nu)$ to eliminate the terms that come from the $1$ term in 
Eq.~(\ref{eq:xnu}).  Now $X(\nu)$ satisfies the same integro-differential equation as $\xi(\nu)$ in Eq.~(\ref{eq:r4}), and has the same boundary 
condition $X(+\infty)=0$.  Therefore we have $X(\nu)=\xi(\nu)$ and $\xi_-=X_-$.  This means that the photon frequency distribution on the red side of 
the line is
\beq
\Pi^{\rm mod}_- = \frac{\Gamma_{\rm inj}}{H\nu_{ul}}(1-\xi_-).
\eeq
Therefore on the red side of the line, the frequency distribution of the photons is suppressed by a factor of $1-\xi_-$.  Therefore, in the absence of 
continuum opacity, the probability $P_{\rm trans}$ that a photon injected on the blue side of the line manages to redshift through the line without any 
incoherent scattering/absorption is $1-\xi_-$.  Conversely, $\xi_- = 1-P_{\rm trans}$ is the probability that such a photon is incoherently 
scattered/absorbed in the line.  This provides a simple physical interpretation of $\xi_-$.

In the case where the line is very optically thick into the damping wings, we expect the line transmission probability $P_{\rm trans}\ll 1$ and hence 
$\xi_-\approx 1$.  It follows that $\Pprd\approx\tauS^{-1}$.  This is in agreement with the Sobolev value for $\tauS\gg 1$. Thus, in both of the 
limiting cases (1) and (2) where the continuum opacity is small, $\Pprd$ is well-approximated by $P_{\rm S}$. The approximation $\Pprd \approx P_{\rm 
S}$ can fail if the coherent depth is very large, while the incoherent depth is optically thin.  This could occur at very early times during 
recombination in the \LyHe\ series, for example.  Yet, for \LyaHe, the difference between $f_{\rm inc} P_{\rm inc}$ and $P_S$ is $<10^{-4}$ (likewise for 
the \LyHe\ series considered here).  The distinction is thus negligible in the overall recombination history.

\section{Photons in the \HeI\ continuum}
\label{ss:heic}

Recombination to the ground state is usually ignored because the photon that is generated immediately re-ionizes another bound atom.  This is a 
single-species picture.  When bound-free rates in helium and hydrogen interact, the recombination photons that would have ionized \HeI\ can instead be 
absorbed by \HI.  This accelerates \HeI\ recombination by allowing recombinations directly to $1^1S$.  (It also forces hydrogen slightly out of 
equilibrium.)  The direct $1^1S$ recombination rate in \HeI\ should roughly scale as the recombination rate to $1^1S_0$ times the probability that 
the ionizing photon is absorbed by hydrogen instead of by \HeI.  In this appendix, we develop this effect in more detail and show that it can be 
neglected.

It is tedious to solve the full transport problem for $\pha(\nu)$ subject to bound-free processes in \HeI\ and \HI, and then integrate to find the 
photoionization and recombination rates.  We will instead track the total number of photons with energies above $24.6$~eV (from recombination to \HeI), 
and develop a method to calculate an effective cross section and photoionization rate for both \HeI\ and \HI\ by photons from region.  Recombination 
rates can be calculated through detailed balancing of the photoionzation rate.

A similar mechanism is active in \HeII, and in principle one should also consider simulating the possibility of direct recombination to \HI\ 1s.  In 
\HI, there is no coupling between species and the answer is straightforward.  In \HeII, recombination proceeds too close to equilibrium for this effect 
to matter.

We lump the radiation field into frequency regions for photons above threshold for each species and consider the loss of photons in a given lumped 
region due to bound-free processes, and redshift.  Note that the integral of the radiation phase space density over a region is not enough to entirely 
specify the behavior.  The number of photons that redshift out of the region depends on the phase space density at its red boundary.  Because the 
spectrum is black-body dominated, we will approximate the radiation phase space density at the boundary by its black-body value there.

If the perturbations to the radiation field caused by bound-free processes are over frequency scales which are large compared to the exponential 
fall-off of the blackbody radiation field and the radiation field is close to equilibrium, then an effective constant cross-section (near threshold) in 
each lumped frequency region provides a good first approximation.
 
Define the number of photons per hydrogen nucleon above the ionization threshold of species $s$ as 
\beqa X_{\rm th,~s}^>  &=& \int_{E_{{\rm th},s}}^\infty \frac{n_\gamma(E)}{n_H} dE \\ \nonumber &=& \frac{h^3}{n_H} 
\int_{\nu_{\rm th,s}}^\infty \pha(\nu) \left ( \frac{8 \pi \nu^2}{c^3} \right ) d \nu. \eeqa  

(Throughout, we will use the subscript $s$ to denote the species.  Unless stated otherwise in this appendix, all occupation variables, photoionization 
cross sections, and photoionization or recombination rates are taken from the ground state of that species.)  To a good approximation, at these 
energies stimulated recombinations can be neglected because the photons are from well into the tail of the blackbody distribution.  One can then write 
the photoionization rate from the ground state due to these photons as
\beq \beta_s \approx \sigma_{\rm eff,s} \frac{X_{\rm th,~s}^{>} c n_H}{h^3},
\label{eq:beta-app}
\eeq
where $\sigma_{\rm eff,s}$ is an effective cross section.
By the principle of detailed balance the recombination rate directly to the ground state is
\beq \alpha_s \equiv \sigma_{\rm eff,s} \left ( \frac{n_s}{n_e n_c} \right )^{\rm LTE} \frac{X_{\rm th,~s}^{>,{\rm~bb}} c n_{\rm H}}{h^3}. \eeq
The effective cross sections are determined by equating Eq.~(\ref{eq:beta-app}) to Eq.~(\ref{eqn:photorate}) for a thermal distribution of photons 
$\pha = e^{-h\nu/\kB T}$, i.e.
\beqa
\sigma_{\rm eff,s} &=& \frac{8\pi c^{-2}\int_{\nu_{\rm min}}^\infty \nu^2\sigma(\nu)e^{-h\nu/\kB T}\,d\nu}
{X_{\rm th,~s}^{>}}
\nonumber \\
&\approx& \sigma(\nu_{\rm min}) + \frac{\kB T}h\sigma'(\nu_{\rm min}),
\eeqa
where $\sigma'$ denotes the derivative of the cross section with respect to frequency.
Note that this assumes the distribution of radiation above the threshold $\nu_{\rm min}$ is distributed as $\pha\propto e^{-h\nu/\kB T}$.  This is 
obviously true in equilibrium if $h\nu\gg\kB T$ but may be violated in the real universe.  Physically we do not expect large deviations from this 
proportionality if the main source of opacity is photoionization, because the photons above ionization threshold should develop a phase space 
distribution $\pha=e^{-(h\nu-\mu_\gamma)/\kB T}$, where the photon chemical potential $\mu_\gamma$ is determined by the non-Saha behavior of the 
ionization stages, e.g.
\beq
\mu_\gamma = \mu_{\rm HII} + \mu_e - \mu_{\rm HI} = \kB T\ln\frac{n_en_{\rm HII}/n_{\rm HI}}{(n_en_{\rm HII}/n_{\rm HI})_{\rm Saha}}.
\eeq
if the source of opacity is \HI.  This may not happen if the photon sees opacity from two competing species (e.g. \HI\ and \HeI) whose relative 
deviations from Saha equilibrium are different.  The most severe case would occur if the cross sections vary rapidly with frequency such that e.g. \HI\ 
opacity dominates at some frequencies and \HeI\ at others.  However the \HI\ and \HeI\ cross sections are smooth functions of frequency in the regime 
of interest (i.e. far below the \HeI\ double-excitation resonance region), so we expect that the current ``one group'' approach is adequate for 
assessing whether continuum photons are important.

Then, the total bound-free rate to the ground state is
\beq \frac{d x_s}{dt} = \frac{c n_H}{h^3} \sigma_{\rm eff,s} \left[ n_e x_c \left ( \frac{n_s}{n_e n_c} \right )^{\rm LTE} X_{\rm th,~s}^{>,{\rm~bb}} -
x_s X_{\rm th,~s}^{>} \right]. \label{eqn:totalbfwithopacity}\eeq
In this picture, in addition to the state occupations evolving, the lumped radiation fields $X_{\rm th,~s}^{>}$ evolve, and can be included as additional ``states" in the level code.  The total number of photons in a frequency region is not conserved.  Thus the evolution equations must account for photons that redshift out of the region, as well as those that are injected or removed by bound-free processes.  Taking the derivative gives a loss rate from the region,
\beqa \frac{d X_{\rm th,~s}^{>}}{dt} &=& \frac{1}{n_H} \frac{d (X_{\rm th,~s}^{>} n_H)}{dt} + 3 H X_{\rm th,~s}^{>} \nonumber \\ 
&=& \frac{8 \pi}{c^3} {h^3 H \over n_H}
\nu^3_{\rm th} \pha(\nu_{\rm th}). \eeqa
Here $\nu_{\rm th}$ is the ionization threshold frequency of the species.  Assuming an $\pha\propto e^{-h\nu/\kB T}$ spectrum, the rate of redshifting 
of photons is (under the considerations developed for \HeI\ recombination),
\beqa \frac{1}{X_{\rm th,~s}^{>}} \frac{d X_{\rm th,~s}^{>}}{dt}  \biggl |_{bb} &=& \left ( \frac{h \nu}{k_B T_{\rm r}} \right )^3 \\ \nonumber 
&\times& \left [ \frac{h \nu}{k_B T_{\rm r}} 
\left ( \frac{h \nu}{k_B T_{\rm r}} + 2 \right ) + 2 \right ]^{-1} H. \eeqa

\begin{figure}
\includegraphics[width=3.2in]{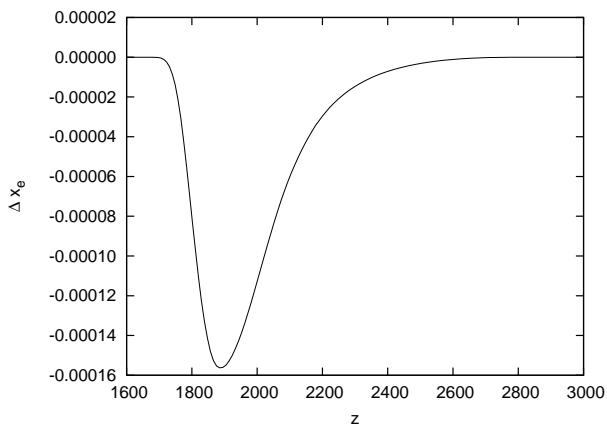}
\caption{\label{figs:bf_opacity}   Recombination directly to \HeI\ $1^1S$ can occur because \HeI\ recombination photons can ionize neutral hydrogen, before they ionize \HeI.  This would tend to accelerate recombination.  Here we show that this effect is leads to a maximum fractional deficit of electrons of around $10^{-4}$.}
\end{figure}

With this choice of variables and tracked states, we can accommodate the interaction between \HeI\ and \HI\ bound-free rates--the overall effect on the 
free electron occupation fraction is shown in Fig.~\ref{figs:bf_opacity}.

\bibliography{recombination}

\begin{thebibliography}{88}
\expandafter\ifx\csname natexlab\endcsname\relax\def\natexlab#1{#1}\fi
\expandafter\ifx\csname bibnamefont\endcsname\relax
  \def\bibnamefont#1{#1}\fi
\expandafter\ifx\csname bibfnamefont\endcsname\relax
  \def\bibfnamefont#1{#1}\fi
\expandafter\ifx\csname citenamefont\endcsname\relax
  \def\citenamefont#1{#1}\fi
\expandafter\ifx\csname url\endcsname\relax
  \def\url#1{\texttt{#1}}\fi
\expandafter\ifx\csname urlprefix\endcsname\relax\def\urlprefix{URL }\fi
\providecommand{\bibinfo}[2]{#2}
\providecommand{\eprint}[2][]{\url{#2}}

\bibitem[{\citenamefont{{Spergel} et~al.}(2006)\citenamefont{{Spergel}, {Bean},
  {Dore'}, {Nolta}, {Bennett}, {Hinshaw}, {Jarosik}, {Komatsu}, {Page},
  {Peiris} et~al.}}]{2006astro.ph..3449S}
\bibinfo{author}{\bibfnamefont{D.~N.} \bibnamefont{{Spergel}}},
  \bibinfo{author}{\bibfnamefont{R.}~\bibnamefont{{Bean}}},
  \bibinfo{author}{\bibfnamefont{O.}~\bibnamefont{{Dore'}}},
  \bibinfo{author}{\bibfnamefont{M.~R.} \bibnamefont{{Nolta}}},
  \bibinfo{author}{\bibfnamefont{C.~L.} \bibnamefont{{Bennett}}},
  \bibinfo{author}{\bibfnamefont{G.}~\bibnamefont{{Hinshaw}}},
  \bibinfo{author}{\bibfnamefont{N.}~\bibnamefont{{Jarosik}}},
  \bibinfo{author}{\bibfnamefont{E.}~\bibnamefont{{Komatsu}}},
  \bibinfo{author}{\bibfnamefont{L.}~\bibnamefont{{Page}}},
  \bibinfo{author}{\bibfnamefont{H.~V.} \bibnamefont{{Peiris}}},
  \bibnamefont{et~al.}, \bibinfo{journal}{ArXiv Astrophysics e-prints}
  (\bibinfo{year}{2006}), \eprint{arXiv:astro-ph/0603449}.

\bibitem[{\citenamefont{{Tegmark} et~al.}(2004)\citenamefont{{Tegmark},
  {Strauss}, {Blanton}, {Abazajian}, {Dodelson}, {Sandvik}, {Wang}, {Weinberg},
  {Zehavi}, {Bahcall} et~al.}}]{2004PhRvD..69j3501T}
\bibinfo{author}{\bibfnamefont{M.}~\bibnamefont{{Tegmark}}},
  \bibinfo{author}{\bibfnamefont{M.~A.} \bibnamefont{{Strauss}}},
  \bibinfo{author}{\bibfnamefont{M.~R.} \bibnamefont{{Blanton}}},
  \bibinfo{author}{\bibfnamefont{K.}~\bibnamefont{{Abazajian}}},
  \bibinfo{author}{\bibfnamefont{S.}~\bibnamefont{{Dodelson}}},
  \bibinfo{author}{\bibfnamefont{H.}~\bibnamefont{{Sandvik}}},
  \bibinfo{author}{\bibfnamefont{X.}~\bibnamefont{{Wang}}},
  \bibinfo{author}{\bibfnamefont{D.~H.} \bibnamefont{{Weinberg}}},
  \bibinfo{author}{\bibfnamefont{I.}~\bibnamefont{{Zehavi}}},
  \bibinfo{author}{\bibfnamefont{N.~A.} \bibnamefont{{Bahcall}}},
  \bibnamefont{et~al.}, \bibinfo{journal}{\prd} \textbf{\bibinfo{volume}{69}},
  \bibinfo{pages}{103501} (\bibinfo{year}{2004}).

\bibitem[{\citenamefont{{Lange} et~al.}(2001)\citenamefont{{Lange}, {Ade},
  {Bock}, {Bond}, {Borrill}, {Boscaleri}, {Coble}, {Crill}, {de Bernardis},
  {Farese} et~al.}}]{2001PhRvD..63d2001L}
\bibinfo{author}{\bibfnamefont{A.~E.} \bibnamefont{{Lange}}},
  \bibinfo{author}{\bibfnamefont{P.~A.} \bibnamefont{{Ade}}},
  \bibinfo{author}{\bibfnamefont{J.~J.} \bibnamefont{{Bock}}},
  \bibinfo{author}{\bibfnamefont{J.~R.} \bibnamefont{{Bond}}},
  \bibinfo{author}{\bibfnamefont{J.}~\bibnamefont{{Borrill}}},
  \bibinfo{author}{\bibfnamefont{A.}~\bibnamefont{{Boscaleri}}},
  \bibinfo{author}{\bibfnamefont{K.}~\bibnamefont{{Coble}}},
  \bibinfo{author}{\bibfnamefont{B.~P.} \bibnamefont{{Crill}}},
  \bibinfo{author}{\bibfnamefont{P.}~\bibnamefont{{de Bernardis}}},
  \bibinfo{author}{\bibfnamefont{P.}~\bibnamefont{{Farese}}},
  \bibnamefont{et~al.}, \bibinfo{journal}{\prd} \textbf{\bibinfo{volume}{63}},
  \bibinfo{pages}{042001} (\bibinfo{year}{2001}).

\bibitem[{\citenamefont{{Jaffe} et~al.}(2001)\citenamefont{{Jaffe}, {Ade},
  {Balbi}, {Bock}, {Bond}, {Borrill}, {Boscaleri}, {Coble}, {Crill}, {de
  Bernardis} et~al.}}]{2001PhRvL..86.3475J}
\bibinfo{author}{\bibfnamefont{A.~H.} \bibnamefont{{Jaffe}}},
  \bibinfo{author}{\bibfnamefont{P.~A.} \bibnamefont{{Ade}}},
  \bibinfo{author}{\bibfnamefont{A.}~\bibnamefont{{Balbi}}},
  \bibinfo{author}{\bibfnamefont{J.~J.} \bibnamefont{{Bock}}},
  \bibinfo{author}{\bibfnamefont{J.~R.} \bibnamefont{{Bond}}},
  \bibinfo{author}{\bibfnamefont{J.}~\bibnamefont{{Borrill}}},
  \bibinfo{author}{\bibfnamefont{A.}~\bibnamefont{{Boscaleri}}},
  \bibinfo{author}{\bibfnamefont{K.}~\bibnamefont{{Coble}}},
  \bibinfo{author}{\bibfnamefont{B.~P.} \bibnamefont{{Crill}}},
  \bibinfo{author}{\bibfnamefont{P.}~\bibnamefont{{de Bernardis}}},
  \bibnamefont{et~al.}, \bibinfo{journal}{Physical Review Letters}
  \textbf{\bibinfo{volume}{86}}, \bibinfo{pages}{3475} (\bibinfo{year}{2001}).

\bibitem[{\citenamefont{{Pryke} et~al.}(2002)\citenamefont{{Pryke},
  {Halverson}, {Leitch}, {Kovac}, {Carlstrom}, {Holzapfel}, and
  {Dragovan}}}]{2002ApJ...568...46P}
\bibinfo{author}{\bibfnamefont{C.}~\bibnamefont{{Pryke}}},
  \bibinfo{author}{\bibfnamefont{N.~W.} \bibnamefont{{Halverson}}},
  \bibinfo{author}{\bibfnamefont{E.~M.} \bibnamefont{{Leitch}}},
  \bibinfo{author}{\bibfnamefont{J.}~\bibnamefont{{Kovac}}},
  \bibinfo{author}{\bibfnamefont{J.~E.} \bibnamefont{{Carlstrom}}},
  \bibinfo{author}{\bibfnamefont{W.~L.} \bibnamefont{{Holzapfel}}},
  \bibnamefont{and}
  \bibinfo{author}{\bibfnamefont{M.}~\bibnamefont{{Dragovan}}},
  \bibinfo{journal}{\apj} \textbf{\bibinfo{volume}{568}}, \bibinfo{pages}{46}
  (\bibinfo{year}{2002}).

\bibitem[{\citenamefont{{Abroe} et~al.}(2002)\citenamefont{{Abroe}, {Balbi},
  {Borrill}, {Bunn}, {Hanany}, {Ferreira}, {Jaffe}, {Lee}, {Olive}, {Rabii}
  et~al.}}]{2002MNRAS.334...11A}
\bibinfo{author}{\bibfnamefont{M.~E.} \bibnamefont{{Abroe}}},
  \bibinfo{author}{\bibfnamefont{A.}~\bibnamefont{{Balbi}}},
  \bibinfo{author}{\bibfnamefont{J.}~\bibnamefont{{Borrill}}},
  \bibinfo{author}{\bibfnamefont{E.~F.} \bibnamefont{{Bunn}}},
  \bibinfo{author}{\bibfnamefont{S.}~\bibnamefont{{Hanany}}},
  \bibinfo{author}{\bibfnamefont{P.~G.} \bibnamefont{{Ferreira}}},
  \bibinfo{author}{\bibfnamefont{A.~H.} \bibnamefont{{Jaffe}}},
  \bibinfo{author}{\bibfnamefont{A.~T.} \bibnamefont{{Lee}}},
  \bibinfo{author}{\bibfnamefont{K.~A.} \bibnamefont{{Olive}}},
  \bibinfo{author}{\bibfnamefont{B.}~\bibnamefont{{Rabii}}},
  \bibnamefont{et~al.}, \bibinfo{journal}{\mnras}
  \textbf{\bibinfo{volume}{334}}, \bibinfo{pages}{11} (\bibinfo{year}{2002}).

\bibitem[{\citenamefont{{Tauber}}(2004)}]{2004AdSpR..34..491T}
\bibinfo{author}{\bibfnamefont{J.~A.} \bibnamefont{{Tauber}}},
  \bibinfo{journal}{Advances in Space Research} \textbf{\bibinfo{volume}{34}},
  \bibinfo{pages}{491} (\bibinfo{year}{2004}).

\bibitem[{\citenamefont{{Kosowsky}}(2003)}]{2003NewAR..47..939K}
\bibinfo{author}{\bibfnamefont{A.}~\bibnamefont{{Kosowsky}}},
  \bibinfo{journal}{New Astronomy Review} \textbf{\bibinfo{volume}{47}},
  \bibinfo{pages}{939} (\bibinfo{year}{2003}).

\bibitem[{\citenamefont{{Ruhl} et~al.}(2004)\citenamefont{{Ruhl}, {Ade},
  {Carlstrom}, {Cho}, {Crawford}, {Dobbs}, {Greer}, {Halverson}, {Holzapfel},
  {Lanting} et~al.}}]{2004SPIE.5498...11R}
\bibinfo{author}{\bibfnamefont{J.}~\bibnamefont{{Ruhl}}},
  \bibinfo{author}{\bibfnamefont{P.~A.~R.} \bibnamefont{{Ade}}},
  \bibinfo{author}{\bibfnamefont{J.~E.} \bibnamefont{{Carlstrom}}},
  \bibinfo{author}{\bibfnamefont{H.-M.} \bibnamefont{{Cho}}},
  \bibinfo{author}{\bibfnamefont{T.}~\bibnamefont{{Crawford}}},
  \bibinfo{author}{\bibfnamefont{M.}~\bibnamefont{{Dobbs}}},
  \bibinfo{author}{\bibfnamefont{C.~H.} \bibnamefont{{Greer}}},
  \bibinfo{author}{\bibfnamefont{N.~w.} \bibnamefont{{Halverson}}},
  \bibinfo{author}{\bibfnamefont{W.~L.} \bibnamefont{{Holzapfel}}},
  \bibinfo{author}{\bibfnamefont{T.~M.} \bibnamefont{{Lanting}}},
  \bibnamefont{et~al.}, in \emph{\bibinfo{booktitle}{Astronomical Structures
  and Mechanisms Technology. Edited by Antebi, Joseph; Lemke, Dietrich.
  Proceedings of the SPIE, Volume 5498, pp. 11-29 (2004).}}, edited by
  \bibinfo{editor}{\bibfnamefont{J.}~\bibnamefont{{Zmuidzinas}}},
  \bibinfo{editor}{\bibfnamefont{W.~S.} \bibnamefont{{Holland}}},
  \bibnamefont{and}
  \bibinfo{editor}{\bibfnamefont{S.}~\bibnamefont{{Withington}}}
  (\bibinfo{year}{2004}), pp. \bibinfo{pages}{11--29}.

\bibitem[{\citenamefont{{Kuo} et~al.}(2004)\citenamefont{{Kuo}, {Ade}, {Bock},
  {Cantalupo}, {Daub}, {Goldstein}, {Holzapfel}, {Lange}, {Lueker}, {Newcomb}
  et~al.}}]{2004ApJ...600...32K}
\bibinfo{author}{\bibfnamefont{C.~L.} \bibnamefont{{Kuo}}},
  \bibinfo{author}{\bibfnamefont{P.~A.~R.} \bibnamefont{{Ade}}},
  \bibinfo{author}{\bibfnamefont{J.~J.} \bibnamefont{{Bock}}},
  \bibinfo{author}{\bibfnamefont{C.}~\bibnamefont{{Cantalupo}}},
  \bibinfo{author}{\bibfnamefont{M.~D.} \bibnamefont{{Daub}}},
  \bibinfo{author}{\bibfnamefont{J.}~\bibnamefont{{Goldstein}}},
  \bibinfo{author}{\bibfnamefont{W.~L.} \bibnamefont{{Holzapfel}}},
  \bibinfo{author}{\bibfnamefont{A.~E.} \bibnamefont{{Lange}}},
  \bibinfo{author}{\bibfnamefont{M.}~\bibnamefont{{Lueker}}},
  \bibinfo{author}{\bibfnamefont{M.}~\bibnamefont{{Newcomb}}},
  \bibnamefont{et~al.}, \bibinfo{journal}{\apj} \textbf{\bibinfo{volume}{600}},
  \bibinfo{pages}{32} (\bibinfo{year}{2004}).

\bibitem[{\citenamefont{{Stompor} et~al.}(2001)\citenamefont{{Stompor},
  {Abroe}, {Ade}, {Balbi}, {Barbosa}, {Bock}, {Borrill}, {Boscaleri}, {de
  Bernardis}, {Ferreira} et~al.}}]{2001ApJ...561L...7S}
\bibinfo{author}{\bibfnamefont{R.}~\bibnamefont{{Stompor}}},
  \bibinfo{author}{\bibfnamefont{M.}~\bibnamefont{{Abroe}}},
  \bibinfo{author}{\bibfnamefont{P.}~\bibnamefont{{Ade}}},
  \bibinfo{author}{\bibfnamefont{A.}~\bibnamefont{{Balbi}}},
  \bibinfo{author}{\bibfnamefont{D.}~\bibnamefont{{Barbosa}}},
  \bibinfo{author}{\bibfnamefont{J.}~\bibnamefont{{Bock}}},
  \bibinfo{author}{\bibfnamefont{J.}~\bibnamefont{{Borrill}}},
  \bibinfo{author}{\bibfnamefont{A.}~\bibnamefont{{Boscaleri}}},
  \bibinfo{author}{\bibfnamefont{P.}~\bibnamefont{{de Bernardis}}},
  \bibinfo{author}{\bibfnamefont{P.~G.} \bibnamefont{{Ferreira}}},
  \bibnamefont{et~al.}, \bibinfo{journal}{\apjl}
  \textbf{\bibinfo{volume}{561}}, \bibinfo{pages}{L7} (\bibinfo{year}{2001}).

\bibitem[{\citenamefont{{Netterfield} et~al.}(2002)\citenamefont{{Netterfield},
  {Ade}, {Bock}, {Bond}, {Borrill}, {Boscaleri}, {Coble}, {Contaldi}, {Crill},
  {de Bernardis} et~al.}}]{2002ApJ...571..604N}
\bibinfo{author}{\bibfnamefont{C.~B.} \bibnamefont{{Netterfield}}},
  \bibinfo{author}{\bibfnamefont{P.~A.~R.} \bibnamefont{{Ade}}},
  \bibinfo{author}{\bibfnamefont{J.~J.} \bibnamefont{{Bock}}},
  \bibinfo{author}{\bibfnamefont{J.~R.} \bibnamefont{{Bond}}},
  \bibinfo{author}{\bibfnamefont{J.}~\bibnamefont{{Borrill}}},
  \bibinfo{author}{\bibfnamefont{A.}~\bibnamefont{{Boscaleri}}},
  \bibinfo{author}{\bibfnamefont{K.}~\bibnamefont{{Coble}}},
  \bibinfo{author}{\bibfnamefont{C.~R.} \bibnamefont{{Contaldi}}},
  \bibinfo{author}{\bibfnamefont{B.~P.} \bibnamefont{{Crill}}},
  \bibinfo{author}{\bibfnamefont{P.}~\bibnamefont{{de Bernardis}}},
  \bibnamefont{et~al.}, \bibinfo{journal}{\apj} \textbf{\bibinfo{volume}{571}},
  \bibinfo{pages}{604} (\bibinfo{year}{2002}).

\bibitem[{\citenamefont{{Halverson} et~al.}(2002)\citenamefont{{Halverson},
  {Leitch}, {Pryke}, {Kovac}, {Carlstrom}, {Holzapfel}, {Dragovan},
  {Cartwright}, {Mason}, {Padin} et~al.}}]{2002ApJ...568...38H}
\bibinfo{author}{\bibfnamefont{N.~W.} \bibnamefont{{Halverson}}},
  \bibinfo{author}{\bibfnamefont{E.~M.} \bibnamefont{{Leitch}}},
  \bibinfo{author}{\bibfnamefont{C.}~\bibnamefont{{Pryke}}},
  \bibinfo{author}{\bibfnamefont{J.}~\bibnamefont{{Kovac}}},
  \bibinfo{author}{\bibfnamefont{J.~E.} \bibnamefont{{Carlstrom}}},
  \bibinfo{author}{\bibfnamefont{W.~L.} \bibnamefont{{Holzapfel}}},
  \bibinfo{author}{\bibfnamefont{M.}~\bibnamefont{{Dragovan}}},
  \bibinfo{author}{\bibfnamefont{J.~K.} \bibnamefont{{Cartwright}}},
  \bibinfo{author}{\bibfnamefont{B.~S.} \bibnamefont{{Mason}}},
  \bibinfo{author}{\bibfnamefont{S.}~\bibnamefont{{Padin}}},
  \bibnamefont{et~al.}, \bibinfo{journal}{\apj} \textbf{\bibinfo{volume}{568}},
  \bibinfo{pages}{38} (\bibinfo{year}{2002}).

\bibitem[{\citenamefont{{Pearson} et~al.}(2003)\citenamefont{{Pearson},
  {Mason}, {Readhead}, {Shepherd}, {Sievers}, {Udomprasert}, {Cartwright},
  {Farmer}, {Padin}, {Myers} et~al.}}]{2003ApJ...591..556P}
\bibinfo{author}{\bibfnamefont{T.~J.} \bibnamefont{{Pearson}}},
  \bibinfo{author}{\bibfnamefont{B.~S.} \bibnamefont{{Mason}}},
  \bibinfo{author}{\bibfnamefont{A.~C.~S.} \bibnamefont{{Readhead}}},
  \bibinfo{author}{\bibfnamefont{M.~C.} \bibnamefont{{Shepherd}}},
  \bibinfo{author}{\bibfnamefont{J.~L.} \bibnamefont{{Sievers}}},
  \bibinfo{author}{\bibfnamefont{P.~S.} \bibnamefont{{Udomprasert}}},
  \bibinfo{author}{\bibfnamefont{J.~K.} \bibnamefont{{Cartwright}}},
  \bibinfo{author}{\bibfnamefont{A.~J.} \bibnamefont{{Farmer}}},
  \bibinfo{author}{\bibfnamefont{S.}~\bibnamefont{{Padin}}},
  \bibinfo{author}{\bibfnamefont{S.~T.} \bibnamefont{{Myers}}},
  \bibnamefont{et~al.}, \bibinfo{journal}{\apj} \textbf{\bibinfo{volume}{591}},
  \bibinfo{pages}{556} (\bibinfo{year}{2003}).

\bibitem[{\citenamefont{{Grainge} et~al.}(2002)\citenamefont{{Grainge},
  {Grainger}, {Jones}, {Kneissl}, {Pooley}, and
  {Saunders}}}]{2002MNRAS.329..890G}
\bibinfo{author}{\bibfnamefont{K.}~\bibnamefont{{Grainge}}},
  \bibinfo{author}{\bibfnamefont{W.~F.} \bibnamefont{{Grainger}}},
  \bibinfo{author}{\bibfnamefont{M.~E.} \bibnamefont{{Jones}}},
  \bibinfo{author}{\bibfnamefont{R.}~\bibnamefont{{Kneissl}}},
  \bibinfo{author}{\bibfnamefont{G.~G.} \bibnamefont{{Pooley}}},
  \bibnamefont{and}
  \bibinfo{author}{\bibfnamefont{R.}~\bibnamefont{{Saunders}}},
  \bibinfo{journal}{\mnras} \textbf{\bibinfo{volume}{329}},
  \bibinfo{pages}{890} (\bibinfo{year}{2002}).

\bibitem[{\citenamefont{{G{\'e}nova-Santos}
  et~al.}(2005)\citenamefont{{G{\'e}nova-Santos}, {Rubi{\~n}o-Mart{\'{\i}}n},
  {Rebolo}, {Cleary}, {Davies}, {Davis}, {Dickinson}, {Falc{\'o}n}, {Grainge},
  {Guti{\'e}rrez} et~al.}}]{2005MNRAS.363...79G}
\bibinfo{author}{\bibfnamefont{R.}~\bibnamefont{{G{\'e}nova-Santos}}},
  \bibinfo{author}{\bibfnamefont{J.~A.}
  \bibnamefont{{Rubi{\~n}o-Mart{\'{\i}}n}}},
  \bibinfo{author}{\bibfnamefont{R.}~\bibnamefont{{Rebolo}}},
  \bibinfo{author}{\bibfnamefont{K.}~\bibnamefont{{Cleary}}},
  \bibinfo{author}{\bibfnamefont{R.~D.} \bibnamefont{{Davies}}},
  \bibinfo{author}{\bibfnamefont{R.~J.} \bibnamefont{{Davis}}},
  \bibinfo{author}{\bibfnamefont{C.}~\bibnamefont{{Dickinson}}},
  \bibinfo{author}{\bibfnamefont{N.}~\bibnamefont{{Falc{\'o}n}}},
  \bibinfo{author}{\bibfnamefont{K.}~\bibnamefont{{Grainge}}},
  \bibinfo{author}{\bibfnamefont{C.~M.} \bibnamefont{{Guti{\'e}rrez}}},
  \bibnamefont{et~al.}, \bibinfo{journal}{\mnras}
  \textbf{\bibinfo{volume}{363}}, \bibinfo{pages}{79} (\bibinfo{year}{2005}).

\bibitem[{\citenamefont{{White}}(2001)}]{2001ApJ...555...88W}
\bibinfo{author}{\bibfnamefont{M.}~\bibnamefont{{White}}},
  \bibinfo{journal}{\apj} \textbf{\bibinfo{volume}{555}}, \bibinfo{pages}{88}
  (\bibinfo{year}{2001}).

\bibitem[{\citenamefont{{Dodelson} et~al.}(1997)\citenamefont{{Dodelson},
  {Kinney}, and {Kolb}}}]{1997PhRvD..56.3207D}
\bibinfo{author}{\bibfnamefont{S.}~\bibnamefont{{Dodelson}}},
  \bibinfo{author}{\bibfnamefont{W.~H.} \bibnamefont{{Kinney}}},
  \bibnamefont{and} \bibinfo{author}{\bibfnamefont{E.~W.}
  \bibnamefont{{Kolb}}}, \bibinfo{journal}{\prd} \textbf{\bibinfo{volume}{56}},
  \bibinfo{pages}{3207} (\bibinfo{year}{1997}).

\bibitem[{\citenamefont{{Peiris} et~al.}(2003)\citenamefont{{Peiris},
  {Komatsu}, {Verde}, {Spergel}, {Bennett}, {Halpern}, {Hinshaw}, {Jarosik},
  {Kogut}, {Limon} et~al.}}]{2003ApJS..148..213P}
\bibinfo{author}{\bibfnamefont{H.~V.} \bibnamefont{{Peiris}}},
  \bibinfo{author}{\bibfnamefont{E.}~\bibnamefont{{Komatsu}}},
  \bibinfo{author}{\bibfnamefont{L.}~\bibnamefont{{Verde}}},
  \bibinfo{author}{\bibfnamefont{D.~N.} \bibnamefont{{Spergel}}},
  \bibinfo{author}{\bibfnamefont{C.~L.} \bibnamefont{{Bennett}}},
  \bibinfo{author}{\bibfnamefont{M.}~\bibnamefont{{Halpern}}},
  \bibinfo{author}{\bibfnamefont{G.}~\bibnamefont{{Hinshaw}}},
  \bibinfo{author}{\bibfnamefont{N.}~\bibnamefont{{Jarosik}}},
  \bibinfo{author}{\bibfnamefont{A.}~\bibnamefont{{Kogut}}},
  \bibinfo{author}{\bibfnamefont{M.}~\bibnamefont{{Limon}}},
  \bibnamefont{et~al.}, \bibinfo{journal}{\apjs}
  \textbf{\bibinfo{volume}{148}}, \bibinfo{pages}{213} (\bibinfo{year}{2003}).

\bibitem[{\citenamefont{{Sunyaev} and {Zeldovich}}(1970)}]{1970Ap&SS...7....3S}
\bibinfo{author}{\bibfnamefont{R.~A.} \bibnamefont{{Sunyaev}}}
  \bibnamefont{and} \bibinfo{author}{\bibfnamefont{Y.~B.}
  \bibnamefont{{Zeldovich}}}, \bibinfo{journal}{\apss}
  \textbf{\bibinfo{volume}{7}}, \bibinfo{pages}{3} (\bibinfo{year}{1970}).

\bibitem[{\citenamefont{{Peebles} and {Yu}}(1970)}]{1970ApJ...162..815P}
\bibinfo{author}{\bibfnamefont{P.~J.~E.} \bibnamefont{{Peebles}}}
  \bibnamefont{and} \bibinfo{author}{\bibfnamefont{J.~T.} \bibnamefont{{Yu}}},
  \bibinfo{journal}{\apj} \textbf{\bibinfo{volume}{162}}, \bibinfo{pages}{815}
  (\bibinfo{year}{1970}).

\bibitem[{\citenamefont{{Bond} and {Efstathiou}}(1987)}]{1987MNRAS.226..655B}
\bibinfo{author}{\bibfnamefont{J.~R.} \bibnamefont{{Bond}}} \bibnamefont{and}
  \bibinfo{author}{\bibfnamefont{G.}~\bibnamefont{{Efstathiou}}},
  \bibinfo{journal}{\mnras} \textbf{\bibinfo{volume}{226}},
  \bibinfo{pages}{655} (\bibinfo{year}{1987}).

\bibitem[{\citenamefont{{Hu} et~al.}(1997)\citenamefont{{Hu}, {Sugiyama}, and
  {Silk}}}]{1997Natur.386...37H}
\bibinfo{author}{\bibfnamefont{W.}~\bibnamefont{{Hu}}},
  \bibinfo{author}{\bibfnamefont{N.}~\bibnamefont{{Sugiyama}}},
  \bibnamefont{and} \bibinfo{author}{\bibfnamefont{J.}~\bibnamefont{{Silk}}},
  \bibinfo{journal}{\nat} \textbf{\bibinfo{volume}{386}}, \bibinfo{pages}{37}
  (\bibinfo{year}{1997}).

\bibitem[{\citenamefont{{Silk}}(1968)}]{1968ApJ...151..459S}
\bibinfo{author}{\bibfnamefont{J.}~\bibnamefont{{Silk}}},
  \bibinfo{journal}{\apj} \textbf{\bibinfo{volume}{151}}, \bibinfo{pages}{459}
  (\bibinfo{year}{1968}).

\bibitem[{\citenamefont{{Hu} and {White}}(1997)}]{1997ApJ...479..568H}
\bibinfo{author}{\bibfnamefont{W.}~\bibnamefont{{Hu}}} \bibnamefont{and}
  \bibinfo{author}{\bibfnamefont{M.}~\bibnamefont{{White}}},
  \bibinfo{journal}{\apj} \textbf{\bibinfo{volume}{479}}, \bibinfo{pages}{568}
  (\bibinfo{year}{1997}).

\bibitem[{\citenamefont{{Lewis} et~al.}(2006)\citenamefont{{Lewis}, {Weller},
  and {Battye}}}]{2006MNRAS.373..561L}
\bibinfo{author}{\bibfnamefont{A.}~\bibnamefont{{Lewis}}},
  \bibinfo{author}{\bibfnamefont{J.}~\bibnamefont{{Weller}}}, \bibnamefont{and}
  \bibinfo{author}{\bibfnamefont{R.}~\bibnamefont{{Battye}}},
  \bibinfo{journal}{\mnras} \textbf{\bibinfo{volume}{373}},
  \bibinfo{pages}{561} (\bibinfo{year}{2006}), \eprint{astro-ph/0606552}.

\bibitem[{\citenamefont{{Seager} et~al.}(2000)\citenamefont{{Seager},
  {Sasselov}, and {Scott}}}]{2000ApJS..128..407S}
\bibinfo{author}{\bibfnamefont{S.}~\bibnamefont{{Seager}}},
  \bibinfo{author}{\bibfnamefont{D.~D.} \bibnamefont{{Sasselov}}},
  \bibnamefont{and} \bibinfo{author}{\bibfnamefont{D.}~\bibnamefont{{Scott}}},
  \bibinfo{journal}{\apjs} \textbf{\bibinfo{volume}{128}}, \bibinfo{pages}{407}
  (\bibinfo{year}{2000}).

\bibitem[{\citenamefont{{Peebles}}(1968)}]{1968ApJ...153....1P}
\bibinfo{author}{\bibfnamefont{P.~J.~E.} \bibnamefont{{Peebles}}},
  \bibinfo{journal}{\apj} \textbf{\bibinfo{volume}{153}}, \bibinfo{pages}{1}
  (\bibinfo{year}{1968}).

\bibitem[{\citenamefont{{Zeldovich} et~al.}(1968)\citenamefont{{Zeldovich},
  {Kurt}, and {Syunyaev}}}]{1968ZhETF..55..278Z}
\bibinfo{author}{\bibfnamefont{Y.~B.} \bibnamefont{{Zeldovich}}},
  \bibinfo{author}{\bibfnamefont{V.~G.} \bibnamefont{{Kurt}}},
  \bibnamefont{and} \bibinfo{author}{\bibfnamefont{R.~A.}
  \bibnamefont{{Syunyaev}}}, \bibinfo{journal}{Zhurnal Eksperimental noi i
  Teoreticheskoi Fiziki} \textbf{\bibinfo{volume}{55}}, \bibinfo{pages}{278}
  (\bibinfo{year}{1968}).

\bibitem[{\citenamefont{{Matsuda} et~al.}(1969)\citenamefont{{Matsuda}, {Sat{\=
  o}}, and {Takeda}}}]{1969PThPh..42..219M}
\bibinfo{author}{\bibfnamefont{T.}~\bibnamefont{{Matsuda}}},
  \bibinfo{author}{\bibfnamefont{H.}~\bibnamefont{{Sat{\= o}}}},
  \bibnamefont{and} \bibinfo{author}{\bibfnamefont{H.}~\bibnamefont{{Takeda}}},
  \bibinfo{journal}{Progress of Theoretical Physics}
  \textbf{\bibinfo{volume}{42}}, \bibinfo{pages}{219} (\bibinfo{year}{1969}).

\bibitem[{\citenamefont{{Matsuda} et~al.}(1971)\citenamefont{{Matsuda}, {Sato},
  and {Takeda}}}]{1971PThPh..46..416M}
\bibinfo{author}{\bibfnamefont{T.}~\bibnamefont{{Matsuda}}},
  \bibinfo{author}{\bibfnamefont{H.}~\bibnamefont{{Sato}}}, \bibnamefont{and}
  \bibinfo{author}{\bibfnamefont{H.}~\bibnamefont{{Takeda}}},
  \bibinfo{journal}{Progress of Theoretical Physics}
  \textbf{\bibinfo{volume}{46}}, \bibinfo{pages}{416} (\bibinfo{year}{1971}).

\bibitem[{\citenamefont{{Liubarskii} and
  {Sunyaev}}(1983)}]{1983A&A...123..171L}
\bibinfo{author}{\bibfnamefont{I.~E.} \bibnamefont{{Liubarskii}}}
  \bibnamefont{and} \bibinfo{author}{\bibfnamefont{R.~A.}
  \bibnamefont{{Sunyaev}}}, \bibinfo{journal}{\aap}
  \textbf{\bibinfo{volume}{123}}, \bibinfo{pages}{171} (\bibinfo{year}{1983}).

\bibitem[{\citenamefont{{Lepp} and {Shull}}(1984)}]{1984ApJ...280..465L}
\bibinfo{author}{\bibfnamefont{S.}~\bibnamefont{{Lepp}}} \bibnamefont{and}
  \bibinfo{author}{\bibfnamefont{J.~M.} \bibnamefont{{Shull}}},
  \bibinfo{journal}{\apj} \textbf{\bibinfo{volume}{280}}, \bibinfo{pages}{465}
  (\bibinfo{year}{1984}).

\bibitem[{\citenamefont{{Fahr} and {Loch}}(1991)}]{1991A&A...246....1F}
\bibinfo{author}{\bibfnamefont{H.~J.} \bibnamefont{{Fahr}}} \bibnamefont{and}
  \bibinfo{author}{\bibfnamefont{R.}~\bibnamefont{{Loch}}},
  \bibinfo{journal}{\aap} \textbf{\bibinfo{volume}{246}}, \bibinfo{pages}{1}
  (\bibinfo{year}{1991}).

\bibitem[{\citenamefont{{Galli} and {Palla}}(1998)}]{1998A&A...335..403G}
\bibinfo{author}{\bibfnamefont{D.}~\bibnamefont{{Galli}}} \bibnamefont{and}
  \bibinfo{author}{\bibfnamefont{F.}~\bibnamefont{{Palla}}},
  \bibinfo{journal}{\aap} \textbf{\bibinfo{volume}{335}}, \bibinfo{pages}{403}
  (\bibinfo{year}{1998}).

\bibitem[{\citenamefont{{Seager} et~al.}(1999)\citenamefont{{Seager},
  {Sasselov}, and {Scott}}}]{1999ApJ...523L...1S}
\bibinfo{author}{\bibfnamefont{S.}~\bibnamefont{{Seager}}},
  \bibinfo{author}{\bibfnamefont{D.~D.} \bibnamefont{{Sasselov}}},
  \bibnamefont{and} \bibinfo{author}{\bibfnamefont{D.}~\bibnamefont{{Scott}}},
  \bibinfo{journal}{\apjl} \textbf{\bibinfo{volume}{523}}, \bibinfo{pages}{L1}
  (\bibinfo{year}{1999}).

\bibitem[{\citenamefont{{Chluba} and {Sunyaev}}(2006)}]{2006A&A...446...39C}
\bibinfo{author}{\bibfnamefont{J.}~\bibnamefont{{Chluba}}} \bibnamefont{and}
  \bibinfo{author}{\bibfnamefont{R.~A.} \bibnamefont{{Sunyaev}}},
  \bibinfo{journal}{\aap} \textbf{\bibinfo{volume}{446}}, \bibinfo{pages}{39}
  (\bibinfo{year}{2006}).

\bibitem[{\citenamefont{{Dubrovich} and {Grachev}}(2005)}]{2005AstL...31..359D}
\bibinfo{author}{\bibfnamefont{V.~K.} \bibnamefont{{Dubrovich}}}
  \bibnamefont{and} \bibinfo{author}{\bibfnamefont{S.~I.}
  \bibnamefont{{Grachev}}}, \bibinfo{journal}{Astronomy Letters}
  \textbf{\bibinfo{volume}{31}}, \bibinfo{pages}{359} (\bibinfo{year}{2005}).

\bibitem[{\citenamefont{{Leung} et~al.}(2004)\citenamefont{{Leung}, {Chan}, and
  {Chu}}}]{2004MNRAS.349..632L}
\bibinfo{author}{\bibfnamefont{P.~K.} \bibnamefont{{Leung}}},
  \bibinfo{author}{\bibfnamefont{C.~W.} \bibnamefont{{Chan}}},
  \bibnamefont{and} \bibinfo{author}{\bibfnamefont{M.-C.} \bibnamefont{{Chu}}},
  \bibinfo{journal}{\mnras} \textbf{\bibinfo{volume}{349}},
  \bibinfo{pages}{632} (\bibinfo{year}{2004}).

\bibitem[{\citenamefont{{Wong} and
  {Scott}}(2006{\natexlab{a}})}]{2006astro.ph.10691W}
\bibinfo{author}{\bibfnamefont{W.~Y.} \bibnamefont{{Wong}}} \bibnamefont{and}
  \bibinfo{author}{\bibfnamefont{D.}~\bibnamefont{{Scott}}},
  \bibinfo{journal}{ArXiv Astrophysics e-prints}
  (\bibinfo{year}{2006}{\natexlab{a}}), \eprint{astro-ph/0610691}.

\bibitem[{\citenamefont{{Kholupenko} and
  {Ivanchik}}(2006)}]{2006AstL...32..795K}
\bibinfo{author}{\bibfnamefont{E.~E.} \bibnamefont{{Kholupenko}}}
  \bibnamefont{and} \bibinfo{author}{\bibfnamefont{A.~V.}
  \bibnamefont{{Ivanchik}}}, \bibinfo{journal}{Astronomy Letters}
  \textbf{\bibinfo{volume}{32}}, \bibinfo{pages}{795} (\bibinfo{year}{2006}),
  \eprint{astro-ph/0611395}.

\bibitem[{\citenamefont{{Wong} and
  {Scott}}(2006{\natexlab{b}})}]{2006astro.ph.12322W}
\bibinfo{author}{\bibfnamefont{W.~Y.} \bibnamefont{{Wong}}} \bibnamefont{and}
  \bibinfo{author}{\bibfnamefont{D.}~\bibnamefont{{Scott}}},
  \bibinfo{journal}{ArXiv Astrophysics e-prints}
  (\bibinfo{year}{2006}{\natexlab{b}}), \eprint{astro-ph/0612322}.

\bibitem[{\citenamefont{{Chluba} et~al.}(2007)\citenamefont{{Chluba},
  {Rubi{\~n}o-Mart{\'{\i}}n}, and {Sunyaev}}}]{2007MNRAS.374.1310C}
\bibinfo{author}{\bibfnamefont{J.}~\bibnamefont{{Chluba}}},
  \bibinfo{author}{\bibfnamefont{J.~A.}
  \bibnamefont{{Rubi{\~n}o-Mart{\'{\i}}n}}}, \bibnamefont{and}
  \bibinfo{author}{\bibfnamefont{R.~A.} \bibnamefont{{Sunyaev}}},
  \bibinfo{journal}{\mnras} \textbf{\bibinfo{volume}{374}},
  \bibinfo{pages}{1310} (\bibinfo{year}{2007}), \eprint{astro-ph/0608242}.

\bibitem[{\citenamefont{{Hu} et~al.}(1995)\citenamefont{{Hu}, {Scott},
  {Sugiyama}, and {White}}}]{1995PhRvD..52.5498H}
\bibinfo{author}{\bibfnamefont{W.}~\bibnamefont{{Hu}}},
  \bibinfo{author}{\bibfnamefont{D.}~\bibnamefont{{Scott}}},
  \bibinfo{author}{\bibfnamefont{N.}~\bibnamefont{{Sugiyama}}},
  \bibnamefont{and} \bibinfo{author}{\bibfnamefont{M.}~\bibnamefont{{White}}},
  \bibinfo{journal}{\prd} \textbf{\bibinfo{volume}{52}}, \bibinfo{pages}{5498}
  (\bibinfo{year}{1995}), \eprint{astro-ph/9505043}.

\bibitem[{\citenamefont{{Gopal} and {Sethi}}(2005)}]{2005MNRAS.363..521G}
\bibinfo{author}{\bibfnamefont{R.}~\bibnamefont{{Gopal}}} \bibnamefont{and}
  \bibinfo{author}{\bibfnamefont{S.~K.} \bibnamefont{{Sethi}}},
  \bibinfo{journal}{\mnras} \textbf{\bibinfo{volume}{363}},
  \bibinfo{pages}{521} (\bibinfo{year}{2005}).

\bibitem[{\citenamefont{{Battye} et~al.}(2001)\citenamefont{{Battye},
  {Crittenden}, and {Weller}}}]{2001PhRvD..63d3505B}
\bibinfo{author}{\bibfnamefont{R.~A.} \bibnamefont{{Battye}}},
  \bibinfo{author}{\bibfnamefont{R.}~\bibnamefont{{Crittenden}}},
  \bibnamefont{and} \bibinfo{author}{\bibfnamefont{J.}~\bibnamefont{{Weller}}},
  \bibinfo{journal}{\prd} \textbf{\bibinfo{volume}{63}},
  \bibinfo{pages}{043505} (\bibinfo{year}{2001}).

\bibitem[{\citenamefont{{Mapelli} and {Ferrara}}(2005)}]{2005MNRAS.364....2M}
\bibinfo{author}{\bibfnamefont{M.}~\bibnamefont{{Mapelli}}} \bibnamefont{and}
  \bibinfo{author}{\bibfnamefont{A.}~\bibnamefont{{Ferrara}}},
  \bibinfo{journal}{\mnras} \textbf{\bibinfo{volume}{364}}, \bibinfo{pages}{2}
  (\bibinfo{year}{2005}), \eprint{astro-ph/0508413}.

\bibitem[{\citenamefont{{Padmanabhan} and
  {Finkbeiner}}(2005)}]{2005PhRvD..72b3508P}
\bibinfo{author}{\bibfnamefont{N.}~\bibnamefont{{Padmanabhan}}}
  \bibnamefont{and} \bibinfo{author}{\bibfnamefont{D.~P.}
  \bibnamefont{{Finkbeiner}}}, \bibinfo{journal}{\prd}
  \textbf{\bibinfo{volume}{72}}, \bibinfo{pages}{023508}
  (\bibinfo{year}{2005}), \eprint{astro-ph/0503486}.

\bibitem[{\citenamefont{{Peebles} et~al.}(2000)\citenamefont{{Peebles},
  {Seager}, and {Hu}}}]{2000ApJ...539L...1P}
\bibinfo{author}{\bibfnamefont{P.~J.~E.} \bibnamefont{{Peebles}}},
  \bibinfo{author}{\bibfnamefont{S.}~\bibnamefont{{Seager}}}, \bibnamefont{and}
  \bibinfo{author}{\bibfnamefont{W.}~\bibnamefont{{Hu}}},
  \bibinfo{journal}{\apjl} \textbf{\bibinfo{volume}{539}}, \bibinfo{pages}{L1}
  (\bibinfo{year}{2000}), \eprint{astro-ph/0004389}.

\bibitem[{\citenamefont{{Mapelli} et~al.}(2006)\citenamefont{{Mapelli},
  {Ferrara}, and {Pierpaoli}}}]{2006MNRAS.369.1719M}
\bibinfo{author}{\bibfnamefont{M.}~\bibnamefont{{Mapelli}}},
  \bibinfo{author}{\bibfnamefont{A.}~\bibnamefont{{Ferrara}}},
  \bibnamefont{and}
  \bibinfo{author}{\bibfnamefont{E.}~\bibnamefont{{Pierpaoli}}},
  \bibinfo{journal}{\mnras} \textbf{\bibinfo{volume}{369}},
  \bibinfo{pages}{1719} (\bibinfo{year}{2006}), \eprint{astro-ph/0603237}.

\bibitem[{\citenamefont{{Hannestad}}(2001)}]{2001NewA....6...17H}
\bibinfo{author}{\bibfnamefont{S.}~\bibnamefont{{Hannestad}}},
  \bibinfo{journal}{New Astronomy} \textbf{\bibinfo{volume}{6}},
  \bibinfo{pages}{17} (\bibinfo{year}{2001}), \eprint{astro-ph/0008452}.

\bibitem[{\citenamefont{{Pierpaoli}}(2004)}]{2004PhRvL..92c1301P}
\bibinfo{author}{\bibfnamefont{E.}~\bibnamefont{{Pierpaoli}}},
  \bibinfo{journal}{Physical Review Letters} \textbf{\bibinfo{volume}{92}},
  \bibinfo{pages}{031301} (\bibinfo{year}{2004}), \eprint{astro-ph/0310375}.

\bibitem[{\citenamefont{{Avelino} et~al.}(2000)\citenamefont{{Avelino},
  {Martins}, {Rocha}, and {Viana}}}]{2000PhRvD..62l3508A}
\bibinfo{author}{\bibfnamefont{P.~P.} \bibnamefont{{Avelino}}},
  \bibinfo{author}{\bibfnamefont{C.~J.~A.~P.} \bibnamefont{{Martins}}},
  \bibinfo{author}{\bibfnamefont{G.}~\bibnamefont{{Rocha}}}, \bibnamefont{and}
  \bibinfo{author}{\bibfnamefont{P.}~\bibnamefont{{Viana}}},
  \bibinfo{journal}{\prd} \textbf{\bibinfo{volume}{62}},
  \bibinfo{pages}{123508} (\bibinfo{year}{2000}), \eprint{astro-ph/0008446}.

\bibitem[{\citenamefont{{Bonometto} and
  {Shouping}}(1986)}]{1986A&A...157L...7B}
\bibinfo{author}{\bibfnamefont{S.~A.} \bibnamefont{{Bonometto}}}
  \bibnamefont{and}
  \bibinfo{author}{\bibfnamefont{X.}~\bibnamefont{{Shouping}}},
  \bibinfo{journal}{\aap} \textbf{\bibinfo{volume}{157}}, \bibinfo{pages}{L7}
  (\bibinfo{year}{1986}).

\bibitem[{\citenamefont{{Rybicki} and {Lightman}}(1986)}]{1986rpa..book.....R}
\bibinfo{author}{\bibfnamefont{G.~B.} \bibnamefont{{Rybicki}}}
  \bibnamefont{and} \bibinfo{author}{\bibfnamefont{A.~P.}
  \bibnamefont{{Lightman}}}, \emph{\bibinfo{title}{{Radiative Processes in
  Astrophysics}}} (\bibinfo{publisher}{Radiative Processes in Astrophysics, by
  George B.~Rybicki, Alan P.~Lightman, pp.~400.~ISBN 0-471-82759-2.~Wiley-VCH ,
  June 1986.}, \bibinfo{year}{1986}).

\bibitem[{\citenamefont{{Sutherland}}(1998)}]{1998MNRAS.300..321S}
\bibinfo{author}{\bibfnamefont{R.~S.} \bibnamefont{{Sutherland}}},
  \bibinfo{journal}{\mnras} \textbf{\bibinfo{volume}{300}},
  \bibinfo{pages}{321} (\bibinfo{year}{1998}).

\bibitem[{\citenamefont{{Chen} and
  {Miralda-Escud{\'e}}}(2004)}]{2004ApJ...602....1C}
\bibinfo{author}{\bibfnamefont{X.}~\bibnamefont{{Chen}}} \bibnamefont{and}
  \bibinfo{author}{\bibfnamefont{J.}~\bibnamefont{{Miralda-Escud{\'e}}}},
  \bibinfo{journal}{\apj} \textbf{\bibinfo{volume}{602}}, \bibinfo{pages}{1}
  (\bibinfo{year}{2004}), \eprint{astro-ph/0208235}.

\bibitem[{\citenamefont{{Press} et~al.}(1992)\citenamefont{{Press},
  {Teukolsky}, {Vetterling}, and {Flannery}}}]{1992nrca.book.....P}
\bibinfo{author}{\bibfnamefont{W.~H.} \bibnamefont{{Press}}},
  \bibinfo{author}{\bibfnamefont{S.~A.} \bibnamefont{{Teukolsky}}},
  \bibinfo{author}{\bibfnamefont{W.~T.} \bibnamefont{{Vetterling}}},
  \bibnamefont{and} \bibinfo{author}{\bibfnamefont{B.~P.}
  \bibnamefont{{Flannery}}}, \emph{\bibinfo{title}{{Numerical recipes in C. The
  art of scientific computing}}} (\bibinfo{publisher}{Cambridge: University
  Press, |c1992, 2nd ed.}, \bibinfo{year}{1992}).

\bibitem[{\citenamefont{{Switzer} and {Hirata}}(2005)}]{2005PhRvD..72h3002S}
\bibinfo{author}{\bibfnamefont{E.~R.} \bibnamefont{{Switzer}}}
  \bibnamefont{and} \bibinfo{author}{\bibfnamefont{C.~M.}
  \bibnamefont{{Hirata}}}, \bibinfo{journal}{\prd}
  \textbf{\bibinfo{volume}{72}}, \bibinfo{pages}{083002}
  (\bibinfo{year}{2005}), \eprint{astro-ph/0507106}.

\bibitem[{\citenamefont{{Baker} and {Menzel}}(1938)}]{1938ApJ....88...52B}
\bibinfo{author}{\bibfnamefont{J.~G.} \bibnamefont{{Baker}}} \bibnamefont{and}
  \bibinfo{author}{\bibfnamefont{D.~H.} \bibnamefont{{Menzel}}},
  \bibinfo{journal}{\apj} \textbf{\bibinfo{volume}{88}}, \bibinfo{pages}{52}
  (\bibinfo{year}{1938}).

\bibitem[{\citenamefont{{Krolik}}(1989)}]{1989ApJ...338..594K}
\bibinfo{author}{\bibfnamefont{J.~H.} \bibnamefont{{Krolik}}},
  \bibinfo{journal}{\apj} \textbf{\bibinfo{volume}{338}}, \bibinfo{pages}{594}
  (\bibinfo{year}{1989}).

\bibitem[{\citenamefont{{Krolik}}(1990)}]{1990ApJ...353...21K}
\bibinfo{author}{\bibfnamefont{J.~H.} \bibnamefont{{Krolik}}},
  \bibinfo{journal}{\apj} \textbf{\bibinfo{volume}{353}}, \bibinfo{pages}{21}
  (\bibinfo{year}{1990}).

\bibitem[{\citenamefont{{Hummer} and {Rybicki}}(1985)}]{1985ApJ...293..258H}
\bibinfo{author}{\bibfnamefont{D.~G.} \bibnamefont{{Hummer}}} \bibnamefont{and}
  \bibinfo{author}{\bibfnamefont{G.~B.} \bibnamefont{{Rybicki}}},
  \bibinfo{journal}{\apj} \textbf{\bibinfo{volume}{293}}, \bibinfo{pages}{258}
  (\bibinfo{year}{1985}).

\bibitem[{\citenamefont{{Hummer} and {Rybicki}}(1992)}]{1992ApJ...387..248H}
\bibinfo{author}{\bibfnamefont{D.~G.} \bibnamefont{{Hummer}}} \bibnamefont{and}
  \bibinfo{author}{\bibfnamefont{G.~B.} \bibnamefont{{Rybicki}}},
  \bibinfo{journal}{\apj} \textbf{\bibinfo{volume}{387}}, \bibinfo{pages}{248}
  (\bibinfo{year}{1992}).

\bibitem[{\citenamefont{{Hummer}}(1962)}]{1962MNRAS.125...21H}
\bibinfo{author}{\bibfnamefont{D.~G.} \bibnamefont{{Hummer}}},
  \bibinfo{journal}{\mnras} \textbf{\bibinfo{volume}{125}}, \bibinfo{pages}{21}
  (\bibinfo{year}{1962}).

\bibitem[{\citenamefont{{Hirata}}(2006)}]{2006MNRAS.367..259H}
\bibinfo{author}{\bibfnamefont{C.~M.} \bibnamefont{{Hirata}}},
  \bibinfo{journal}{\mnras} \textbf{\bibinfo{volume}{367}},
  \bibinfo{pages}{259} (\bibinfo{year}{2006}), \eprint{astro-ph/0507102}.

\bibitem[{\citenamefont{{Zheng} and
  {Miralda-Escud{\'e}}}(2002)}]{2002ApJ...578...33Z}
\bibinfo{author}{\bibfnamefont{Z.}~\bibnamefont{{Zheng}}} \bibnamefont{and}
  \bibinfo{author}{\bibfnamefont{J.}~\bibnamefont{{Miralda-Escud{\'e}}}},
  \bibinfo{journal}{\apj} \textbf{\bibinfo{volume}{578}}, \bibinfo{pages}{33}
  (\bibinfo{year}{2002}).

\bibitem[{\citenamefont{{Bonilha} et~al.}(1979)\citenamefont{{Bonilha},
  {Ferch}, {Salpeter}, {Slater}, and {Noerdlinger}}}]{1979ApJ...233..649B}
\bibinfo{author}{\bibfnamefont{J.~R.~M.} \bibnamefont{{Bonilha}}},
  \bibinfo{author}{\bibfnamefont{R.}~\bibnamefont{{Ferch}}},
  \bibinfo{author}{\bibfnamefont{E.~E.} \bibnamefont{{Salpeter}}},
  \bibinfo{author}{\bibfnamefont{G.}~\bibnamefont{{Slater}}}, \bibnamefont{and}
  \bibinfo{author}{\bibfnamefont{P.~D.} \bibnamefont{{Noerdlinger}}},
  \bibinfo{journal}{\apj} \textbf{\bibinfo{volume}{233}}, \bibinfo{pages}{649}
  (\bibinfo{year}{1979}).

\bibitem[{\citenamefont{{Bernes}}(1979)}]{1979A&A....73...67B}
\bibinfo{author}{\bibfnamefont{C.}~\bibnamefont{{Bernes}}},
  \bibinfo{journal}{\aap} \textbf{\bibinfo{volume}{73}}, \bibinfo{pages}{67}
  (\bibinfo{year}{1979}).

\bibitem[{\citenamefont{{Caroff} et~al.}(1972)\citenamefont{{Caroff},
  {Noerdlinger}, and {Scargle}}}]{1972ApJ...176..439C}
\bibinfo{author}{\bibfnamefont{L.~J.} \bibnamefont{{Caroff}}},
  \bibinfo{author}{\bibfnamefont{P.~D.} \bibnamefont{{Noerdlinger}}},
  \bibnamefont{and} \bibinfo{author}{\bibfnamefont{J.~D.}
  \bibnamefont{{Scargle}}}, \bibinfo{journal}{\apj}
  \textbf{\bibinfo{volume}{176}}, \bibinfo{pages}{439} (\bibinfo{year}{1972}).

\bibitem[{\citenamefont{{Auer}}(1968)}]{1968ApJ...153..783A}
\bibinfo{author}{\bibfnamefont{L.~H.} \bibnamefont{{Auer}}},
  \bibinfo{journal}{\apj} \textbf{\bibinfo{volume}{153}}, \bibinfo{pages}{783}
  (\bibinfo{year}{1968}).

\bibitem[{\citenamefont{{Avery} and {House}}(1968)}]{1968ApJ...152..493A}
\bibinfo{author}{\bibfnamefont{L.~W.} \bibnamefont{{Avery}}} \bibnamefont{and}
  \bibinfo{author}{\bibfnamefont{L.~L.} \bibnamefont{{House}}},
  \bibinfo{journal}{\apj} \textbf{\bibinfo{volume}{152}}, \bibinfo{pages}{493}
  (\bibinfo{year}{1968}).

\bibitem[{\citenamefont{{Menzel} and {Pekeris}}(1935)}]{1935MNRAS..96...77M}
\bibinfo{author}{\bibfnamefont{D.~H.} \bibnamefont{{Menzel}}} \bibnamefont{and}
  \bibinfo{author}{\bibfnamefont{C.~L.} \bibnamefont{{Pekeris}}},
  \bibinfo{journal}{\mnras} \textbf{\bibinfo{volume}{96}}, \bibinfo{pages}{77}
  (\bibinfo{year}{1935}).

\bibitem[{\citenamefont{{Burgess}}(1958)}]{1958MNRAS.118..477B}
\bibinfo{author}{\bibfnamefont{A.}~\bibnamefont{{Burgess}}},
  \bibinfo{journal}{\mnras} \textbf{\bibinfo{volume}{118}},
  \bibinfo{pages}{477} (\bibinfo{year}{1958}).

\bibitem[{\citenamefont{{Cunto} et~al.}(1993)\citenamefont{{Cunto}, {Mendoza},
  {Ochsenbein}, and {Zeippen}}}]{1993A&A...275L...5C}
\bibinfo{author}{\bibfnamefont{W.}~\bibnamefont{{Cunto}}},
  \bibinfo{author}{\bibfnamefont{C.}~\bibnamefont{{Mendoza}}},
  \bibinfo{author}{\bibfnamefont{F.}~\bibnamefont{{Ochsenbein}}},
  \bibnamefont{and} \bibinfo{author}{\bibfnamefont{C.~J.}
  \bibnamefont{{Zeippen}}}, \bibinfo{journal}{\aap}
  \textbf{\bibinfo{volume}{275}}, \bibinfo{pages}{L5+} (\bibinfo{year}{1993}).

\bibitem[{\citenamefont{{Nussbaumer} and
  {Schmutz}}(1984)}]{1984A&A...138..495N}
\bibinfo{author}{\bibfnamefont{H.}~\bibnamefont{{Nussbaumer}}}
  \bibnamefont{and}
  \bibinfo{author}{\bibfnamefont{W.}~\bibnamefont{{Schmutz}}},
  \bibinfo{journal}{\aap} \textbf{\bibinfo{volume}{138}}, \bibinfo{pages}{495}
  (\bibinfo{year}{1984}).

\bibitem[{\citenamefont{{Martin}}(1987)}]{1987PhRvA..36.3575M}
\bibinfo{author}{\bibfnamefont{W.~C.} \bibnamefont{{Martin}}},
  \bibinfo{journal}{\pra} \textbf{\bibinfo{volume}{36}}, \bibinfo{pages}{3575}
  (\bibinfo{year}{1987}).

\bibitem[{\citenamefont{{Kono} and {Hattori}}(1984)}]{1984PhRvA..29.2981K}
\bibinfo{author}{\bibfnamefont{A.}~\bibnamefont{{Kono}}} \bibnamefont{and}
  \bibinfo{author}{\bibfnamefont{S.}~\bibnamefont{{Hattori}}},
  \bibinfo{journal}{\pra} \textbf{\bibinfo{volume}{29}}, \bibinfo{pages}{2981}
  (\bibinfo{year}{1984}).

\bibitem[{\citenamefont{{Khan} et~al.}(1988)\citenamefont{{Khan}, {Khandelwal},
  and {Wilson}}}]{1988ApJ...329..493K}
\bibinfo{author}{\bibfnamefont{F.}~\bibnamefont{{Khan}}},
  \bibinfo{author}{\bibfnamefont{G.~S.} \bibnamefont{{Khandelwal}}},
  \bibnamefont{and} \bibinfo{author}{\bibfnamefont{J.~W.}
  \bibnamefont{{Wilson}}}, \bibinfo{journal}{\apj}
  \textbf{\bibinfo{volume}{329}}, \bibinfo{pages}{493} (\bibinfo{year}{1988}).

\bibitem[{\citenamefont{{Khandelwal} et~al.}(1989)\citenamefont{{Khandelwal},
  {Khan}, and {Wilson}}}]{1989ApJ...336..504K}
\bibinfo{author}{\bibfnamefont{G.~S.} \bibnamefont{{Khandelwal}}},
  \bibinfo{author}{\bibfnamefont{F.}~\bibnamefont{{Khan}}}, \bibnamefont{and}
  \bibinfo{author}{\bibfnamefont{J.~W.} \bibnamefont{{Wilson}}},
  \bibinfo{journal}{\apj} \textbf{\bibinfo{volume}{336}}, \bibinfo{pages}{504}
  (\bibinfo{year}{1989}).

\bibitem[{\citenamefont{{Bates} and {Damgaard}}(1949)}]{1949RSPTA.242..101B}
\bibinfo{author}{\bibfnamefont{D.~R.} \bibnamefont{{Bates}}} \bibnamefont{and}
  \bibinfo{author}{\bibfnamefont{A.}~\bibnamefont{{Damgaard}}},
  \bibinfo{journal}{Royal Society of London Philosophical Transactions Series
  A} \textbf{\bibinfo{volume}{242}}, \bibinfo{pages}{101}
  (\bibinfo{year}{1949}).

\bibitem[{\citenamefont{{Laughlin}}(1978)}]{1978JPhB...11L.391L}
\bibinfo{author}{\bibfnamefont{C.}~\bibnamefont{{Laughlin}}},
  \bibinfo{journal}{Journal of Physics B Atomic Molecular Physics}
  \textbf{\bibinfo{volume}{11}}, \bibinfo{pages}{L391} (\bibinfo{year}{1978}).

\bibitem[{\citenamefont{{Cann} and {Thakkar}}(2002)}]{2002JPhB...35..421C}
\bibinfo{author}{\bibfnamefont{N.~M.} \bibnamefont{{Cann}}} \bibnamefont{and}
  \bibinfo{author}{\bibfnamefont{A.~J.} \bibnamefont{{Thakkar}}},
  \bibinfo{journal}{Journal of Physics B Atomic Molecular Physics}
  \textbf{\bibinfo{volume}{35}}, \bibinfo{pages}{421} (\bibinfo{year}{2002}).

\bibitem[{\citenamefont{{Drake}}(1986)}]{1986PhRvA..34.2871D}
\bibinfo{author}{\bibfnamefont{G.~W.~F.} \bibnamefont{{Drake}}},
  \bibinfo{journal}{\pra} \textbf{\bibinfo{volume}{34}}, \bibinfo{pages}{2871}
  (\bibinfo{year}{1986}).

\bibitem[{\citenamefont{{Gubner}}(1994)}]{1994JPhA...27L.745G}
\bibinfo{author}{\bibfnamefont{J.~A.} \bibnamefont{{Gubner}}},
  \bibinfo{journal}{Journal of Physics A Mathematical General}
  \textbf{\bibinfo{volume}{27}}, \bibinfo{pages}{L745} (\bibinfo{year}{1994}).

\bibitem[{\citenamefont{{Lee}}(1977)}]{1977ApJ...218..857L}
\bibinfo{author}{\bibfnamefont{J.-S.} \bibnamefont{{Lee}}},
  \bibinfo{journal}{\apj} \textbf{\bibinfo{volume}{218}}, \bibinfo{pages}{857}
  (\bibinfo{year}{1977}).

\bibitem[{\citenamefont{{Lee}}(1982)}]{1982ApJ...255..303L}
\bibinfo{author}{\bibfnamefont{J.~S.} \bibnamefont{{Lee}}},
  \bibinfo{journal}{\apj} \textbf{\bibinfo{volume}{255}}, \bibinfo{pages}{303}
  (\bibinfo{year}{1982}).

\bibitem[{\citenamefont{{Rybicki} and
  {dell'Antonio}}(1994)}]{1994ApJ...427..603R}
\bibinfo{author}{\bibfnamefont{G.~B.} \bibnamefont{{Rybicki}}}
  \bibnamefont{and} \bibinfo{author}{\bibfnamefont{I.~P.}
  \bibnamefont{{dell'Antonio}}}, \bibinfo{journal}{\apj}
  \textbf{\bibinfo{volume}{427}}, \bibinfo{pages}{603} (\bibinfo{year}{1994}).

\end{thebibliography}

\end{document}